\documentstyle[12pt,aaspp4]{article}

\input epsf
\def\plotfiddle#1#2#3#4#5#6#7{\centering \leavevmode \vbox to#2{\rule{0pt}{#2}} \includegraphics{#1}}

\slugcomment{To appear in \apjs} 

\lefthead{Knauth et al.}
\righthead{Physical Conditions in Foreground PDRs}

\begin{document}

\title{Physical Conditions in the Foreground Gas of Reflection Nebulae:
NGC~2023, vdB~102, and NGC~7023}
\author{David C. Knauth\altaffilmark{1}, S. R. Federman\altaffilmark{1,2}, K.
Pan, and M. Yan}
\affil{Department of Physics and Astronomy, University of Toledo, Toledo, OH
43606}

\and

\author{David L. Lambert}
\affil{Department of Astronomy, University of Texas, Austin, TX 78712}
\altaffiltext{1}{Guest Observer, McDonald Observatory, University of Texas at 
Austin.}
\altaffiltext{2}{Visiting Astronomer, Cerro Tololo Inter-American Observatory, 
National Optical Astronomy Observatories, which is operated by the Association 
of Universities for Research in Astronomy, Inc., under cooperative agreement 
with the National Science Foundation.}
\authoremail{dknauth@astro1.panet.utoledo.edu}
{}
\vfill

\begin{abstract}

High resolution optical spectra of HD~37903 and HD~147009, which illuminate the
reflection nebulae, NGC~2023 and vdB~102, were obtained for comparison with our
results for HD~200775 and NGC~7023.  Ground-based measurements of the molecules,
CH, C$_2$, and CN, and the atoms, Na~{\small I} and K~{\small I}, were
analyzed to extract physical conditions in the foreground cloud.  Estimates
of the gas density, gas temperature and flux of ultraviolet radiation were
derived and were compared with the results from infrared and radio studies of 
the main molecular cloud.  The conditions are similar to those found in studies
of diffuse clouds.  The foreground material is less dense than the gas in the
molecular cloud behind the star(s).  The gas temperature was set at 40 K, the
temperature determined for the foreground gas in NGC~7023.  The flux of
ultraviolet radiation was found to be less intense than in the molecular
material behind the star(s).  The column densities of Na~{\small I} and
K~{\small I} were reproduced reasonably well when the extinction curve for the
specific line of sight was adopted.  We obtained NEWSIPS data from the $IUE$
archive for HD~37903 and HD~200775.  The ultraviolet data on C~{\small I} and CO
allow extraction of the physical conditions by alternate methods.  General
agreement among the various diagnostics was found, leading to self-consistent
pictures of the foreground photodissociation regions.

An Appendix describes checks on the usefulness of $IUE$ NEWSIPS data for
interstellar studies.  Equivalent widths are compared with a previous analysis
of $IUE$ observations for interstellar gas toward 20~Aquilae.  Excellent
agreement is found with NEWSIPS results having smaller errors.  A comparison of
NEWSIPS data for C {\small I} toward X~Per with data acquired with the {\it
Hubble Space Telescope} shows similar levels of agreement in equivalent widths
and that the derived column densities from $IUE$ results are accurate to
better than a factor of 2 for absorption lines of moderate strength.
\end{abstract}


\section{Introduction}

Stars form under the gravitational collapse of gas and dust in interstellar
molecular clouds.  When a newly-formed star interacts with the surrounding
material, the cloud's physical conditions are altered.  For instance, early type
stars create photodissociation regions (PDRs), which are boundary layers between
a source of far ultraviolet (FUV) radiation and the predominantly neutral cloud
(Jaffe et al. 1990).  These boundaries also exist between H~{\small II} regions
produced by O stars and the ambient material surrounding them.  In
photodissociation regions, the FUV radiation dominates the heating and chemical
processes (Tielens \& Hollenbach 1985a, 1985b; Hollenbach \& Tielens 1999,
HT99) and controls which atoms and molecules are observed.  The observed
relative abundances of atomic and molecular species and the use of chemical
models yields the physical conditions in the PDR.  These conditions provide
clues to processes that occurred immediately following star formation.  In this
paper, we describe results for foreground PDR's in three reflection nebulae
based on measurements of absorption at visible and ultraviolet wavelengths
occurring at the velocity of the PDR behind the illuminating source seen via 
emission lines.  While absorption is sometimes seen at other velocities, the 
main components all have velocities consistent with the molecular cloud.  From 
this fact, we infer an association with the material behind the star. 

PDRs in molecular clouds and reflection nebulae have been studied extensively at
all wavelengths.  Analyses have been based on measurements from x-rays to radio
wavelengths.  Specific observations on the nebulae studied here are given below.
In an effort to obtain a comprehensive understanding of the conditions that
prevail in these regions, one dimensional, time-independent models have been
developed (e.g., Tielens \& Hollenbach 1985a, 1985b; van Dishoeck \& Black 1986;
Hollenbach, Takahashi \& Tielens 1991; Draine \& Bertoldi 1996, 1999).  More
realistic PDR models with clumpy structure and time dependent effects have been
attempted as well (Meixner \& Tielens 1993; Hollenbach \& Natta 1995; Bertoldi
\& Draine 1996).  Hollenbach \& Tielens (1999) provide a comprehensive review of
observations and modeling efforts for photodissociation regions.

Little attention has been given to studying PDRs through atomic and molecular
absorption in the visible and ultraviolet portions of the spectrum observed
along the line of sight to the central or illuminating star(s) .  Federman et
al. (1997 - hereafter Paper~I) analyzed visible spectra for the line of sight
toward HD~200775, which illuminates the reflection nebula NGC~7023.  In an
effort to expand upon this work, we obtained high resolution visible spectra
toward NGC~2023 and vdB~102 as well as ultraviolet data on atoms and CO
molecules in the reflection nebulae, NGC~2023 and NGC~7023.  The ultraviolet
data were obtained from the $International$ $Ultraviolet$ $Explorer$ ($IUE$)
archive, which was recalibrated with the New Spectral Image Processing System
(NEWSIPS).  The visible and UV absorption spectra yield information about
the chemical composition and the physical conditions present in the foreground
PDR.  A more complete picture of PDRs emerges when the physical conditions in
front of the star are compared to those inferred from analyses of data obtained
at other wavelengths of the main molecular cloud behind the star.

\subsection{NGC~2023}

NGC~2023 is a reflection nebula embedded in the L1630 molecular cloud (Jaffe et 
al. 1990) in the constellation of Orion.  Infrared images of the nebula (Gatley
et al. 1987; Field et al. 1994, 1998) show that it has a 6$^{\prime}$ extent 
with visible filamentary structures throughout and is approximately centered on 
the star HD~37903.  In a far-infrared study, Harvey, Thronson, \& Gatley (1980)
showed that the molecular cloud lies behind HD~37903, which is located at a
distance of 470 pc (Perryman et al. 1997).  Several Herbig-Haro objects, 
which are associated with pre-main sequence stars, are observed in the 
nebula.  Two Herbig-Haro objects (HH4 and HH5) in the southeastern part 
of the nebula are illuminated by star C (Malin, Ogura \& Walsh 1987).  Star 
C is thought to be a T Tauri star based on spectroscopy and its near infrared
excess.  Many embedded infrared sources have also been detected (Strom et al.
1975).   Wyrowski et al. (2000) detected a ultra-high density clump ($n \sim$
10$^{7}$ cm$^{-3}$) which they propose could be the early stages of a protostar.
Thus, NGC~2023 is an active region of star formation (Strom et al. 1975;
Malin et al. 1987; Lada et al. 1991; Freyberg \& Schmitt 1995).  

The primary illuminating source for this nebula is HD~37903, a B1.5 V star with 
an effective temperature of approximately 22,000 K (Field et al. 1998).  The 
radiation from this star has enough energy at $\lambda$ $\le$ 912 \AA\ to 
form an H~{\small II} region, 0.015 pc in size, detected by radio continuum
emission at the star's position within the nebula (Pankonin \& Walmsley 1976,
1978; Wyrowski et al. 2000).  The photons with $\lambda$ $\ge$ 912 \AA\ create 
a photodissociation region between the H~{\small II} region and the rest of the
neutral cloud.  The proximity of the nebula and its bubble shaped geometry 
(Gatley et al. 1987) make it an ideal target for studying
the interaction of the UV radiation from HD~37903 with the molecular material in
L1630.  tHE reflection nebula and its PDR have been extensively studied via
x-ray (Freyberg \& Schmitt 1995), ultraviolet (Witt, Bohlin, \& Stecher 1984 -
hereafter WBS84; Cardelli, Clayton, \& Mathis 1989 - hereafter CCM89; Buss et
al. 1994), infrared (e.g., Martini, Sellgren, \& DePoy 1999 and references
therein), millimeter (Jaffe et al. 1990), and radio observations (Fuente,
Mart\'{\i}n-Pintado \& Gaume 1995; Lebr\'{o}n \& Rodr\'{\i}guez 1997; Wyrowski
et al. 1997, 2000). The studies relevant to our work will be discussed in detail
later. 

\subsection{vdB~102}
	
The central star of vdB~102 is HD~147009, an A0 V star, with an effective 
temperature of about 9,500 K (de Geus, de Zeeuw \& Lub 1989 - hereafter GZL89).
This nebula was first catalogued and photometrically studied (van den Bergh
1966; Racine 1968) in the late 1960's.  The nebula is 160 pc away (Perryman et
al. 1997) and is one of 10 reflection nebulae in the Sco OB2 Association (van
den Bergh 1966; Vall\'{e}e 1987; Eggen 1998).  The stars in Sco OB2 illuminate
different portions of the same molecular cloud.  Radio observations reveal the
presence of the parent molecular cloud (Kutner et al. 1980; Vall\'{e}e 1987).
Few observational efforts have examined the reflection nebula and its
illuminating star  (Chaffee \& White 1982; Cappa de Nicolau \& P\"{o}ppel 1986;
Vall\'{e}e 1987; Federman et al. 1994).  For this reason, the present study
greatly improves upon our knowledge of PDRs around relatively cool stars.

The remainder of the paper is organized in the following manner.  In \S 2 we
discuss the optical and ultraviolet observations obtained for this study, 
as well as the consistency of our measurements with previous efforts.  We
describe the chemical and atomic analyses performed here and the results of
these analyses as well as a comparison with our earlier results on HD~200775
(Paper I) in \S 3.  We discuss in \S 4 the derived physical conditions and
compare them with the physical conditions obtained through diagnostics at other
wavelengths.  Our conclusions are found in \S 5.  An Appendix describes checks
on the usefulness of NEWSIPS data for interstellar studies such as ours.     

\section{Observations and Data Reduction}
\subsection{Ground-based Observations}
\subsubsection{McDonald 2dcoud\'{e} Data}

High-resolution echelle spectra of HD~37903, HD~147009 and 20~Aql were taken
with the 2.7~m telescope of the University of Texas McDonald Observatory during
June 1996, July 1997, and January 1998.   Observations of 20~Aql were made
since it allows for the comparison of our optical data with others and it is
located behind a relatively isolated cloud for an additional test of
the atomic analysis discussed below. The ``2dcoud\'{e}" spectrograph (Tull
et al. 1995) was used with 2 camera setups and a Tektronix charge coupled device
(CCD) was the detector.  One camera provided high spectral resolution with a
resolving power at 4000 \AA\ of R $\approx$ 180,000 ($\Delta$v $\approx$ 2 km
s$^{-1}$) and covered the wavelength range 3800 \AA\ to 4400 \AA.  This high
resolution mode yielded spectra containing 8 disjoint echelle orders
approximately 15 \AA\ wide, separated by approximately 65 \AA.  This setup
provided spectra on the following interstellar species CH$^+$, CH, CN,
Ca~{\small II}~K, and Ca~{\small I}.  The second camera setting resulted in
moderate spectral resolution with a resolving power at 7000 \AA\ of R $\approx$
52,000 ($\Delta$v $\approx$ 6 km s$^{-1}$) and covered a wavelength range from
3600 \AA\ to 9700 \AA.   The moderate resolution spectra consisted of 59 orders
with an overlap of about 15 \AA\ at the blue end.  The inter-order separation
increased with wavelength until there was approximately 150~\AA\ between orders
at near infrared wavelengths.  In addition to observing the same species
obtained from the high resolution spectra, we also obtained data on
Ca~{\small II}~H, Na~{\small I} D$_1$, K~{\small I}, and C$_2$.  The resolution
was determined from the full width at half maximum (FWHM) of Thorium-Argon
(Th-Ar) lines.   

The data were reduced in a standard way utilizing the {\bf NOAO SUN/IRAF} 
software (Revision 2.10.4).  Dark, bias, and flat lamp exposures were taken each
night to remove any instrumental effects due to the CCD detector.  Comparison 
spectra were taken periodically throughout the night, typically every two hours.
The average bias exposure was subtracted from all raw stellar, comparison
(Th-Ar) and flat images.  The scattered light was fitted by a low order
polynomial, in both the dispersion direction and perpendicular to it for the
multi-order observations, and removed.  The pixel to pixel sensitivity was
removed by dividing the normalized average flat into the stellar spectra.  Next,
the pixels perpendicular to the dispersion were summed in each order for each
stellar and comparison lamp exposure.  The extracted spectra were placed on an
appropriate wavelength scale with the Th-Ar comparison spectra, and
Doppler-corrected.  The spectra were coadded and normalized to unity yielding a
final spectrum with high signal to noise (based on the root-mean-square
deviations in the continuum), typically 100:1.  

\subsubsection{McDonald Observations with the 6-Foot Camera}

The 6-foot camera of the coud\'{e} spectrograph on the 2.7 m telescope was 
used to acquire high-resolution spectra of HD~147009 and 20~Aql.  Observations,
at R $\approx$ 200,000, of interstellar Na {\small I} D
$\lambda$$\lambda$5896, 5890, K {\small I} $\lambda$7699, CH$^+$ $\lambda$4232,
and Ca {\small I} $\lambda$4226 were obtained in June 1990, Ca {\small II} K
$\lambda$3933 observations were made in June 1991 and CH $\lambda$4300
observations were obtained in June 1992.  The TI2 CCD with 15 $\mu$ pixels was
used in all setups.  Since no cross disperser was used, single orders of the
echelle grating were detected in the vicinity of the line of interest.
Appropriate interference filters were placed at the entrance slit to remove
light from unwanted orders.  Scattered light was found to be
negligible since only one order was observed.  Comparison spectra from a Th-Ar
hollow cathode were taken throughout the night, while biases and flat fields
were acquired each night.  Dark frames were also obtained as a check on the
amount of thermal noise from the cooled CCD.  The slit was imaged onto 2.5
pixels for all setups, providing a resolution of 1.6 to 1.8 km s$^{-1}$.  The
data were reduced in the standard manner described above.  The resulting
signal-to-noise ratios ranged from 30 to 200.

\subsubsection{CTIO Observations}

The stars, HD~147009 and 20~Aql, were observed with the Bench Mounted Echelle 
on the 1.5 m telescope in July 1991.  The setup involved the 590 mm long 
camera, the Ritchey Chr\'{e}tien grating from the 4-m telescope as a cross
disperser, the TI1 800 $\times$ 800 pixel CCD, and a CuSO$_4$ blocking filter.
The entrance slit at the end of the 200 $\mu$m fiber optic feed was set at 45
$\mu$m, illuminating 2.6 pixels as determined by the full width at half maximum
of Th-Ar lines.  The resulting instrumental resolving power was about 65,000.  
The setup allowed measurement of lines between 3800 \AA\ and 4400 \AA.  Bias 
frames and flat fields were acquired each night.  The wavelength scale was 
determined from spectra of a Th-Ar hollow cathode taken throughout the 
night.  The data were reduced in the standard manner described above.  The 
extracted spectra had signal-to-noise ratios between 40 and 150.

A sample of the rectified ground-based spectra for atoms and molecules toward
HD~37903 are shown in Figures 1 $-$ 3 and Figures 4 $-$ 6 exhibit spectra of
HD~147009.  The equivalent width, $W_{\lambda}$, for a line was determined by a
Gaussian fit; single Gaussians were adequate for all except the
atomic species Na {\small I} and Ca {\small II} toward HD~37903.  The measured
equivalent widths and $v_{LSR}$ for the ground-based results toward HD~37903 and
HD~147009 are shown in Tables 1 $-$ 3.  Where applicable, measurements taken
from the literature are also shown.  All $W_{\lambda}$'s for a given
line agree within measured uncertainties.  The differences in $v_{LSR}$ are
attributed to differences in spectral resolution and to systematic effects
involving the wavelength solution.  Where no interstellar absorption was
detected, a 3-$\sigma$ upper limit is given.  

Each $W_{\lambda}$ was converted into a column density utilizing curves of
growth.  The column densities for each species were consistent within the errors
from individual measurements.  The final column densities reported in Tables 4
and 5 for HD~37903 and HD~147009, respectively, are the weighted average of all
detections reported in this work and the literature.  For the C$_2$ results
reported in Table 3, the most restrictive upper limits for a rotational
level were summed to get the total upper limit in column density.  
  
The Doppler parameter ({\it b}-value) was set at 1 km s$^{-1}$ for 
all lines except those of Ca~{\small II} and CH$^+$, where a {\it b}-value of
2.5 km s$^{-1}$ was used (e.g., Welty, Morton, \& Hobbs 1996).  For Na~{\small
I} and K~{\small I} absorption associated with the molecular material of the
PDR, {\it b}-values of 2.0 km s$^{-1}$ toward HD~37903 and 1.5 km s$^{-1}$
toward HD~147009 were required by the atomic analysis discussed below.  These
$b$-values are similar to other sightlines studied in Orion and the Sco OB2
association based on high resolution surveys of interstellar Na~{\small I} D$_1$
and K~{\small I} (Welty, Hobbs, \& Kulkarni 1994; Welty \& Hobbs 2000).  The
change in $b$-value affected only the column of Na~{\small I} since the optical
depth at line center is larger for Na~{\small I} than for K~{\small I}.  This
resulted in a 20 \% decrease in $N$(Na~{\small I}) toward HD~37903 and a 10 \%
decrease toward HD~147009; no change in $N$(K~{\small I}) occurred with the 
change in $b$-value.
   
Editor place figures 1-6 here.
Editor place tables 1-5 here.

\subsection{Ultraviolet Data}

High-resolution ultraviolet spectra of HD~37903 and HD~200775 were available 
in the $IUE$ NEWSIPS archive.  Table 6 lists the short-wavelength spectral 
images used in our study.  The data were extracted with the set of programs in 
IUERDAF.  For features that appeared in two orders, the cleaner looking 
spectrum was used for further analysis.  Once extracted, the spectra were 
analyzed with the IRAF package in much the same way as described above for 
the ground-based measurements.  Our focus was on absorption from $^{12}$CO, 
$^{13}$CO, C~{\small I}, S~{\small I}, and Ni~{\small II}; sample spectra are
shown in Figures 7 $-$ 8.  Since C~{\small I} and S~{\small I} represent minor
constituents for their respective elements and have similar ionization
potentials, the S~{\small I} lines provided a check on the C~{\small I} lines.
On the other hand, Ni~{\small II} is the dominant form of Ni in cloud envelopes;
the synthesis of Ni {\small II} lines allowed for a comparison of $b$-values
which could elucidate the relative volumes occupied by dominant versus minor
forms of an element.  

Editor place table 6 here.
Editor place figures 7 - 8 here.

The compilation of $W_{\lambda}$ values appears in Table 7.  
Joseph et al. (1986) reported results for HD~37903 from the same $IUE$ 
spectra, but with an earlier version of the archiving process.  The overall 
agreement between the two sets of $W_{\lambda}$ is good; the NEWSIPS data 
used here produce spectra with higher signal to noise, as 
seen by the smaller uncertainties in $W_{\lambda}$.  We performed several checks
on $W_{\lambda}$'s measured from NEWSIPS spectra; these are described in the
Appendix.  For the features considered in our study, we are confident that they
have been measured reliably.

Profile synthesis was used to extract $b$-values and column densities from 
the spectra.  A chi-squared minimization procedure was adopted for the fits.  
The syntheses were based on the following set of oscillator stengths: 
Morton \& Noreau (1994) for $^{12}$CO and $^{13}$CO; Morton (1991) and 
Zsarg\'{o}, Federman, \& Cardelli (1997) for C {\small I}; Federman \& 
Cardelli (1995) for S {\small I}; and Morton (1991) and Zsarg\'{o} \& Federman 
(1998) for Ni {\small II}.  The resulting column densities appear in 
Table 8.  For the synthesis of the CO bands,  the excitation temperature was set
at 4 K (Lambert et al. 1994) and the $b$-value was found to be  somewhat less
than 1 km s$^{-1}$.  For the atomic lines, the $b$-value was determined by fits
to several lines.  There was no discernible difference in $b$-value between
neutral atoms and ions; we found $b$ $=$ $2.0 \pm 0.5$ km s$^{-1}$.  Since CO is
expected to occupy a more restricted volume than even neutral atoms, a smaller
$b$-value for this species is appropriate.

Editor place tables 7 - 8 here.

\section{Analysis and Results}
\subsection{Chemical Modeling}

The chemistry of the simple molecular species, CH, C$_2$, and CN, is
relatively well understood (van Dishoeck \& Black 1986; Federman et al. 1994)
and relevant reaction rates are known or have been calculated.  The purpose of
obtaining data on several species of carbon-bearing molecules is to glean
information about the physical conditions where these species reside.   In
particular, gas density ({\it n}), gas temperature ({\it T}), and the
enhancement of the ultraviolet flux over the average interstellar value given by
Draine (1978), {\it I}$_{uv}$, can be extracted from the measurements.  

The analysis is based on steady-state chemical rate equations, which were
successful in reproducing data for quiescent diffuse interstellar clouds (e.g.,
Federman et al. 1994) and the PDR toward HD~200775 (Paper I).  In this analysis,
the three quantities, {\it n}, {\it T}, and {\it I}$_{uv}$ are inferred from
observed column densities of CH, C$_2$ and CN.  However, only upper limits to
the column densities are available for the C$_2$ molecule toward HD~37903 and
HD~147009 and for CN toward HD~147009, and as a consequence, the physical
conditions in the foreground gas are not as well constrained as in other clouds
(e.g., Paper I).

The rate equations used here are based on our earlier work.  Federman and
Huntress (1989) developed the chemistry for C$_2$ in diffuse clouds, while
Federman, Danks, and Lambert (1984) and Federman and Lambert (1988) did the 
same for CN.  The incorporation of NH by Federman et al.  (1994) into the 
rate equation for CN was the result of the discovery of NH in cloud envelopes
(Meyer \& Roth 1991).  The respective equations for C$_2$ and CN are: 

\begin{equation}
N({\rm C}_2) = { {k_1x({\rm C}^+)N({\rm CH})n\alpha} \over {G({\rm C}_2) + 
k_2x({\rm O})n + k_3x({\rm N})n}}~{\rm and}
\end{equation}

\begin{equation}
N({\rm CN}) = { {[k_3x({\rm N})N({\rm C}_2) + k_4x({\rm N})N({\rm CH}) + 
k_5x({\rm C}^+)N({\rm NH})\alpha]n} \over {G({\rm CN}) + k_6x({\rm O})n} }.
\end{equation}

\noindent In these equations, the reaction rate constants are denoted by {\it 
k}$_i$ (see Table 9); the abundance of species X relative to the proton
abundance and the column density of species X are given by {\it x}(X) and {\it
N}(X), respectively. The total proton density [{\it n}(H) + 2{\it n}(H$_2$)] is
given by {\it n}; the photodestruction rate of species X is {\it G}(X); 
and $\alpha$ accounts for the conversion of C$^+$ into CO (Federman et al.
1994).  The rate constants for neutral-neutral reactions include a factor,
(T/300)$^{0.5}$, to account for the velocity term in the rate constant.  The
attenuation of radiation by dust is included in the photodissociation rate,
{\it G}(X) = {\it~G}$_o${\it~I}$_{uv}$exp(-$\tau_{uv}$), where {\it G}$_o$ is
the rate at the surface of the cloud, and $\tau_{uv}$ is the optical depth at
1000~\AA.  A wavelength of 1000 \AA\ for $\tau_{uv}$ was chosen because it is in
the middle of the range of wavelengths that leads to photodissociation of C$_2$
and CN (912 \AA\ $\le$ $\lambda$ $\le$ 1200 \AA) (Pouilly et al. 1983; Lavendy,
Robbe, \& Gandara 1987).  The observed values of {\it N}(CH) and {\it N}(C$_2$)
were used in determining the predicted values of {\it N}(C$_2$) and {\it N}(CN).
The value of {\it N}(NH) was set to 0.041{\it N}(CH) since {\it N}(NH) scales
with {\it N}(CH) (Meyer \& Roth 1991; Federman et al. 1994). 

Editor place table 9 here.

The following considerations apply to the parameters, $\tau_{uv}$, $\alpha$, and
$x$(X).  From knowledge of the ratio of total to selective extinction, {\it R}$_v$ = {\it A}$_v$/{\it E}({\it B}~$-$~{\it V}), and the ultraviolet
extinction curve (CCM89; Buss et al. 1994), information on $\tau_{uv}$ can be
derived.  For the direction toward HD~37903, $\tau_{uv}$ was set at 2$A_v$, with
an $R_v$ somewhat larger than average (Buss et al. 1994).  Since the extinction
curve for HD~147009 is similar to that seen for $\rho$ Oph, $\tau_{uv}$ toward
$\rho$ Oph (Federman et al. 1994) was scaled by $E$($B$ $-$ $V$) to obtain the
value for the sight line to HD~147009.  For optical depths greater than 2,
an additional term of the form, $\alpha$ = [1~+~14($\tau_{uv}$~$-$~2)/5]$^{-1}$
(Federman et al. 1994), was added to account for the trends between the
predicted and observed $N$(C$_2$) and $\tau_{uv}$ caused by C$^+$ being
converted into CO (Federman \& Huntress 1989).  Fractional interstellar
abundances of 1.4~$\times$~10$^{-4}$ (Sofia et al. 1997), 7.5~$\times$~10$^{-5}$
(Meyer, Cardelli, \& Sofia 1997), and 3.2~$\times$~10$^{-4}$ (Meyer, Jura, \&
Cardelli 1998) for C$^+$, N, and O, respectively, were adopted from measurements
with the {\it Hubble Space Telescope} ({\it HST}).  These abundances account for
modest depletion of the elements onto interstellar grains in cloud envelopes.
These values are somewhat different than the fractional abundances quoted in
Paper I: while the O and N abundances differ by 10\%, the C$^+$ abundance is
almost a factor-of-2 less.  The lower fractional abundance of carbon requires a
higher density for self-consistent chemical results, mainly because C$_2$ is
formed through reactions involving C$^+$ and CH.  

The column densities derived from the rate equations compare favorably with 
the observed columns for C$_2$ and CN.  The predicted column densities and
inferred physical conditions are found in Table 10.  The Table also displays
predictions for the molecular columns toward HD~200775 based on the atomic
abundances adopted here; the greatest effect was on the extracted density.
Changing the temperature did not significantly alter the predicted column
densities in the chemical analysis above. The temperature used in this analysis
was 40 K, the average value obtained from the C {\small I} excitation analysis
($T$ $\sim$ 50 K) and the temperature determined from CO fractionation ($T$
$\sim$ 30 K).  The average of these two temperatures is reasonable in light of
the fact that C {\small I} is more extended than CO in the cloud.  The value
of {\it I}$_{uv}$/{\it n} is the most important parameter in PDRs (HT99); it
controls the chemistry and structure.  When {\it n} was increased in our model,
{\it I}$_{uv}$ had to be increased in order to obtain the best match to the
observed column densities.  The ratio {\it I}$_{uv}$/{\it n} remained unchanged
(see eqn. 1 and 2).  

Editor place table 10 here.
	
\subsection{Atomic Analysis}

In this analysis, only the columns of Na~{\small I} and K~{\small I} associated
with the molecular material (i.e., at the same velocity as the molecules CH, CN,
and C$_2$) are used.  The atoms Na and K are predominantly singly ionized in
interstellar space.  Through ionization balance, we can analyze the neutral
columns with the equation,

\begin{equation}
N({\rm X}) =   { {N_{tot}({\rm H})A({\rm X})\alpha({\rm X})n_e\alpha_c} \over {G({\rm X})}}.
\end{equation}

\noindent In this equation, {\it A}({\rm X}) is the elemental abundance, 
{\it N}$_{tot}$(H) is the total proton column density, {\it N}(H) + 2{\it N}(H$_2$), $\alpha$(X) is the radiative recombination rate constant
(P\'{e}quinot \& Aldrovandi 1986), {\it G}(X) is the photoionization
rate for species X, and {\it n}$_e$ is the electron density.  {\it A}({\rm X})
includes depletion onto grains at about 25\% for alkali elements (Phillips,
Pettini \& Gondhalekar 1984) and $\alpha_c$ incorporates the conversion of C$^+$
into CO in the molecular material (as opposed to $\alpha$ in the chemical
analysis which is an average for the total amount of neutral material).  This
conversion is a significant effect only when the fraction of molecular gas is
$\ge$ 50\% of the neutral gas.  The form of $\alpha_c$ is $\alpha$$f$ + (1 $-$
$f$), where $f$ is the H$_2$ fraction (2$N$(H$_2$)/$N_{tot}$(H)) and $\alpha$ is
described in \S3.1.  Many of the uncertain factors (e.g., {\it n}$_e$ and depletion) in the above
equation can be eliminated by taking the ratio of {\it N}(Na~{\small I}) to {\it
N}(K~{\small I}).  This ratio can be written as

\begin{equation}
{ \left[ {N({\rm Na\ I}) } \over {N({\rm K\ I})} \right] } = 
{ { \left[ {A({\rm Na})} \over { A({\rm K})} \right] } 
{ \left[ {\alpha({\rm Na\ I}) } \over { \alpha({\rm K\ I}) } \right] }
{ \left[ {G({\rm K\ I}) } \over { G({\rm Na\ I}) } \right]} }.
\end{equation}  

\noindent The appropriate value for {\it A}(Na)/{\it A}(K) is 16 and
$\alpha$(Na~{\small I})/$\alpha$(K~{\small I}) is 1.05.  The latter ratio was
incorrectly given in Paper I; the change results in a 10\% increase in the 
predicted ratio of Na {\small I} and K {\small I} columns.  The predicted
ratio mainly depends on the photoionization rates.  The photoionization rates
for Na~{\small I} and K~{\small I} at the cloud surface were determined from the
measured ionization cross sections of Hudson \& Carter (1967a) and of Hudson 
\& Carter (1965, 1967b), Marr \& Creek (1968) and Sandner et al. (1981),
respectively.  

To complete the atomic analysis, details about the extinction curve are needed.
The main difference for the two sight lines studied here 
involves the effect of extinction from interstellar grains.   The pair method
(WBS84) was used to derive the extinction curve for the central
star in the nebula vdB~102, HD~147009.  Extinction curves for two other
stars, HD~37023 and HD~37903, were computed for comparison with extinction
curves derived by Fitzpatrick \& Massa (1990 - hereafter FM90).  Stellar spectra
of HD~31726, HD~37023, HD~37903, HD~55857, HD~64802, HD~147009 and $\zeta$~Aql 
were obtained from the $IUE$ Archive;  the filenames are listed in Table 6.
The unreddened standard stars used in FM90 are used in this work, with the
exception of $\zeta$~Aql (Witt, Bohlin, \& Stecher 1981), in order to obtain 
the most consistent results.  The extinction curves for HD~37023 and HD~37903
derived by FM90 are similar to those derived by us, demonstrating that the
extinction curve for HD~147009 derived here is reliable. 

The extinction curves for HD~37903 and HD~147009 were found to be anomalous when
compared to the one representing typical Galactic extinction, that for $\zeta$
Per (Savage \& Mathis 1979).  For this analysis, the extinction curve for
$\zeta$~Per was scaled to the same reddening parameter of the observed star. 
Figure 9 displays the resulting extinction curves.  Since the photoionization
rates were determined for $\zeta$~Per (Paper I), the difference between
extinction toward the reddened line of sight compared to the average is used to
derive an attenuated photoionization rate for the reddened star.  Changes in
extinction introduce a multiplicative factor to the photoionization rates of the
form 10$^{\beta}$.  For the wavelengths where the curve for $\zeta$~Per shows
more (less) extinction, the exponent $\beta$ is positive (negative).  Since the
two extinction curves can intersect, for some wavelength intervals the
extinction can be enhanced and diminished for others.  The total rate is the sum
of the rates for each wavelength interval.  

Editor place figure 9 here.

\subsection{Atomic Results}

The atomic spectra for Ca {\small II} H and K and Na~{\small I} D exhibit 6
velocity components along the line of sight toward HD~37903 (see Figure 1).
There is no molecular absorption seen in 5 of them. The atomic gas not
associated with molecular absorption is attributed to Orion's Cloak.  Orion's
Cloak is a rapidly (35 $\le$ {\it v} $\le$ 120 km s$^{-1}$) expanding shell of
gas centered on the Orion OB1 association (Cowie, Songaila \& York 1979).  
While the 5 velocity components toward HD~37903 have lower velocities ($\le$ 30
km s$^{-1}$) than is seen in Orion's Cloak, this velocity structure is observed
toward many other stars in Orion (Cowie et al. 1979; Welty et al. 1994; Welty et
al. 1996) showing the high velocity gas of Orion's Cloak.  Our atomic analysis
involves only absorption associated with the molecular material in the nebula.

The atomic observations yield information on the photoionization
rates present in the gas (see eqn. 4) because eqn. 4 is not dependent on the
physical conditions.  Toward $\zeta$~Per, an observed ratio Na {\small I}/K
{\small I} of 100 $\pm$ 6 was obtained.   For the predicted ratio a refined
analysis was performed.  The difference in grain attenuation between 2400 and
2900 \AA\ ($\Delta$$A_{\lambda}$ = 0.2) was taken into account.  This resulted
in a predicted ratio of 67.  Toward HD~200775, the observed ratio is 54 $\pm$ 9,
which is in excellent agreement with the predicted ratio of 54 when the effects
of the extinction curve are applied. (Note that the sum of all velocity
components of Na {\small I} and K {\small I} are used for this sight line, see
\S4.4.)  

The observed column densities were (1.1 $\pm$ 0.2) $\times$ 10$^{13}$ cm$^{-2}$
and 1.1 $\times$ 10$^{11}$ cm$^{-2}$ for Na~{\small I} and K~{\small I} toward
HD~37903, giving an observed ratio of 100~$\pm$~18.  The inferred Na~{\small I}
and K~{\small I} photoionization rates are 1.2 $\times$ 10$^{-11}$ s$^{-1}$ and
4.4 $\times$ 10$^{-11}$ s$^{-1}$, which yield a predicted ratio of 62.  For the
line of sight to HD~147009, the observed column densities of (1.5~$\pm$~0.3)
$\times$ 10$^{13}$ cm$^{-2}$ and (2.3~$\pm$~0.1) $\times$ 10$^{11}$ cm$^{-2}$
for Na~{\small I} and K~{\small I}, respectively, were determined.  These
columns produce an observed ratio of 65 $\pm$ 13.  The Na~{\small I} and
K~{\small I} photoionization rates for this gas are 2.9 $\times$ 10$^{-10}$
s$^{-1}$ and 9.7 $\times$ 10$^{-10}$ s$^{-1}$, respectively.  These values lead
to a predicted ratio of 56.  As expected the correspondence between observed and
predicted $N$(Na {\small I})/$N$(K {\small I}) ratios is good.

Editor place table 11 here.

Another useful way to study the observed atomic abundances is to compare, for a 
specific atom, the column density obtained toward the program stars to the 
column density obtained toward $\zeta$~Per.  In this way we check for potential 
differences in depletion onto grains and {\it n}$_e$ (see eqn. 3).  Curves 
of growth with {\it b}~=~1.5~km~s$^{-1}$ for the weak Na~{\small I} doublet at
3302 \AA\ (Crutcher 1975) as well as the strong D-lines at $\lambda\lambda$
5889, 5895 (Hobbs 1974a, 1974b; Welty et al. 1994; Andersson, private
communication) and the weak K~{\small I} doublet at $\lambda\lambda$ 4044, 4047
(Knauth et al. 2000) yielded weighted average column densities for Na~{\small I}
of (7.3~$\pm$~0.3) $\times$ 10$^{13}$ cm$^{-2}$ and (7.3~$\pm$~0.3) $\times$
10$^{11}$ cm$^{-2}$ for K~{\small I} toward $\zeta$~Per.  Our precise column
density measurements of K {\small I} $\lambda\lambda$4044, 4047 yield similar
results to the column density obtained from the K {\small I} $\lambda$7699 line.
The photoionization rates of Na~{\small I} and K~{\small I} toward $\zeta$~Per
are 4.5 $\times$ 10$^{-12}$ s$^{-1}$ and 1.8 $\times$ 10$^{-11}$ s$^{-1}$ (Paper
I).  For this analysis we require the total proton density toward each star.
The proton density is (1.6 $\pm$ 0.3) $\times$ 10$^{21}$ cm$^{-2}$ toward $\zeta$~Per (Bohlin, Savage, \& Drake 1978) and is (3.6 $\pm$ 0.7) $\times$
10$^{21}$ cm$^{-2}$ toward HD~37903 (Buss et al. 1994).  Toward HD~147009, the
total proton column density was determined from several sources.
From radio observations of CH, the column density of H$_2$ was found to be
2.5~$\times$~10$^{20}$~cm$^{-2}$ (Vall\'{e}e 1987) using the relationship
derived by Federman (1982) and Danks, Federman, \& Lambert (1984).  A value of
{\it N}(H~{\small I})~=~1.1~$\times$~10$^{21}$~cm$^{-2}$ was determined from the
21-cm results of Cappa de Nicolau \& P\"{o}ppel (1986), indicating
{\it N}$_{tot}$(H)~=~1.6~$\times$~10$^{21}$~cm$^{-2}$.  Also from 21-cm
observations, de Geus \& Burton (1991) derived {\it N}(H~{\small I}) = 1.8
$\times$ 10$^{21}$ cm$^{-2}$, which yields {\it N}$_{tot}$(H) = 2.3 $\times$
10$^{21}$ cm$^{-2}$.  The average proton column density toward HD~147009 is
therefore (1.9 $\pm$ 0.4) $\times$ 10$^{21}$ cm$^{-2}$.  In our analysis, we adopted a 20 \% error in $N_{tot}$(H) for all stars except HD~200775 where the
observed precision was higher.  

Most of the total proton density is assumed to lie in front of HD~147009, since
most of $N$(H {\small I}) resides in one component (Cappa de Nicolau \&
P\"{o}ppel 1986) and the star is at high galactic latitude, $b^{\circ}$ = 20.894
(GZL89).  The relationship of Bohlin et al. (1978),
$N_{tot}$(H)/$E$($B$~$-$~$V$) = 5.8 $\times$ 10$^{21}$ atoms cm$^{-2}$
mag$^{-1}$, supports this assumption.  This relationship yields $N_{tot}$(H) =
2.2 $\times$ 10$^{21}$ cm$^{-2}$, which is similar to the column density
determined above.  Therefore it is reasonable to assume all of the neutral
hydrogen resides in front of the star.  

Estimates for gas density, temperature and UV field strength are needed to 
complete this analysis.  The values of {\it n}, {\it T}, and {\it I}$_{uv}$ 
toward $\zeta$~Per (Federman et al. 1994) were re-derived with the new values
for the fractional abundances of C$^+$, N, and O, resulting in 1200 cm$^{-3}$,
40 K, and 1.  For the gas toward HD~37903 and HD~147009, the adopted values of
{\it n} and {\it T} are from the chemical analysis above for {\it I}$_{uv}$ = 1
(see Table 10).  Here we set {\it n}$_e$ = {\it n}~$x$(C$^+$), since the
electrons mostly come from ionized carbon.  The predicted ratios toward
HD~200775 are 0.05 $\pm$ 0.01 and 0.06 $\pm$ 0.01 for Na~{\small I} and
K~{\small I}, respectively.  The observed ratios are 0.08~$\pm$~0.01 and
0.15~$\pm$~0.02 for Na~{\small I} and K~{\small I}, respectively.  (These ratios
are based on our refined analysis and differ somewhat from the ratios in Paper
I.)  For the case of HD~37903, the effect of $\alpha_c$ needed to be
included since the fraction of molecular gas is $\ge$ 50 \%.  This resulted in 
a predicted ratio {\it N}(Na~{\small I})$_{HD~37903}$/{\it N}(Na~{\small
I})$_{\zeta\ Per}$ of 0.23 $\pm$ 0.06 and 0.21 $\pm$ 0.06 for K~{\small I}.  
The observed ratios are 0.15 $\pm$ 0.03 for Na~{\small I} and 0.15 $\pm$ 0.01
for K~{\small I}.  As for the PDR toward HD~147009, the analysis gives a ratio
of 0.01 $\pm$ 0.01 for both ratios.  The observed column density ratios are 0.21
$\pm$ 0.04 for Na~{\small I} and 0.32 $\pm$ 0.02 for K~{\small I}.  

The results of our analysis are summarized in Table 11.  It can be seen from the
ratios of Na {\small I} and especially K {\small I} that a
modest change in {\it n}$_e$ and/or depletion onto grains is needed to bring 
the predictions and observations for the foreground PDR of NGC~7023 into
agreement.  This change from Paper I is caused by the decrease in $x$(C$^+$).
Since the observed and predicted results for Na {\small I} and K {\small I}
toward HD~37903 with respect to $\zeta$~Per agree within the errors, our estimates for $n_e$ and depletion for the foreground material toward HD~37903  are reasonable.  The situation is more extreme for the foreground PDR of
vdB~102.  A factor-of-20 increase in {\it n}$_e$ is needed to bring the
predictions into agreement with the observations.  While the flat FUV extinction
curve for HD~147009 allows more radiation to permeate the nebula, {\it n}$_e$ is
limited by the elemental abundance of carbon.  Because HD~147009 is an A0 V
star, enhanced ionization in an H {\small II} region is not expected, unless the
enhanced ionization arises from the combined effects of other, earlier type,
stars in Sco OB2.

\subsection{UV Results}

The $IUE$ data on C {\small I} and CO provide additional constraints on the
physical conditions for the foreground PDR.  The relative populations of fine
structure levels in the electronic ground state of C {\small I} arises from a
combination of collisional excitation and de-excitation, radiative decay, and in
enhanced radiation fields, UV pumping.  Analysis of the observed distribution
via statistical equilibrium yields pressure ($n_{tot}$ $T$) and sometimes
$I_{uv}$.  Here, $n_{tot}$ is the density of collision partners, $n$(H) +
$n$(H$_2$).  If the kinetic temperature is known from another diagnostic,
$n_{tot}$ can be obtained and compared with $n$ [$n$(H) + 2 $n$(H$_2$)] inferred
from the chemistry.  The ratio of $N$($^{12}$CO)/$N$($^{13}$CO) is affected by
chemical fractionation.  When the gas is cold ($T$ $\le$ 40 K) and is permeated
by a strong UV radiation field so that the abundance of C$^+$ is high, the ratio
of columns is less than the ratio of ambient $^{12}$C/$^{13}$C (about 60)
because $^{13}$CO has the lower zero-point energy (cf. Watson, Anicich, \&
Huntress 1976).  If the gas in not especially cold, but still affected by UV
photons, selective isotopic photodissociation enhances $N$($^{12}$CO) relative
to $N$($^{13}$CO) (e.g., Bally \& Langer 1982; van Dishoeck \& Black 1988).  The
enhancement is the result of $^{12}$CO self-shielding itself from
photodissociation because its lines are more optically thick.

The analysis of C {\small I} excitation described by Lambert et al. (1994) forms
the basis of our results.  In particular, their collisional rate constants and 
radiative rates are adopted here.  The fractional abundances of collisional
partners are based on the observations of Buss et al. (1994) - see Table 12,
with the added assumption that 10\% of the gas is He.  The analysis also assumes
that absorption from each fine structure level, J, arises from the same parcel
of gas.  Figure 10 shows the results of our calculations.  Here we plot the
density and temperature consistent with the observed column density ratios,
$N$(J $=$ 1)/$N$(J $=$ 0) $-$ dashed curves $-$ and $N$(J $=$ 2)/$N$(J $=$ 0)
$-$ dash-dotted curves.  The upper and lower bounds are based on $\pm$
1-$\sigma$ errors in column density.  For HD~37903, results for $I_{uv}$ of 3
are indistinguishable from those with $I_{uv}$ of 1.  The strongest constraint
comes from $N$(2)/$N$(0): $n_{tot}$ lies between 300 and 500 cm$^{-3}$ and $T$
about 50 K.  Based on the observed columns of H and H$_2$, the inferred value
for $n$ is 400 to 700 cm$^{-3}$.  As for the foreground PDR toward HD~200775,
the lower limits from $N$(2)/$N$(0) and the upper limits from $N$(1)/$N$(0)
indicate that $n_{tot}$ is about 700 cm$^{-3}$ (or $n$ is $\approx$ 850
cm$^{-3}$) for $T$ between 40 and 60 K.  Analysis of C$_2$ excitation from the
data in Paper I is consistent with the C {\small I} results if $I_{ir}$ is 3.
For C$_2$, optical pumping to an excited electronic state involves absorption
of infrared radiation; hence the use of $I_{ir}$.  [Again, the curves for
$I_{uv}$ of 3 are essentially those for $I_{uv}$ of 1, except for the lower
range allowed by $N$(2)/$N$(0).]

Editor place table 12 here.
Editor place figure 10 here.

The inferred values for $n$ agree with the chemical results for the two
foreground PDRs.  The gas containing C {\small I} toward HD~37903 appears to
have a density most consistent with the chemical results for $I_{uv}$ $=$ 1.
Since C {\small I} may occupy a larger volume of material than the molecules
C$_2$ and CN, the inferred average density from C {\small I} excitation could 
be smaller by about a factor of 2 compared to the density in CN-rich gas.  This
seems to be the case for $\zeta$ Oph (Lambert et al. 1994).  In other words, the
chemical results for $I_{uv}$ of 1 $-$ 3 are still consistent with the results
from C {\small I}.  The situation is different for the gas toward HD~200775.
Here, the C {\small I} results are most consistent with the chemical results for
$I_{uv}$ of 3.  Since this comparison also applies to the results for C$_2$ with
$I_{ir}$ of 3, the strengths of the UV and IR fields seem to scale together for
the foreground PDR.

The observations of HD~200775 reveal that $N$($^{12}$CO)/$N$($^{13}$CO) is about
20.  An enhancement in $N$($^{13}$CO) arises when the gas is cold, yet contains
an appreciable fraction of C$^+$.  Lambert et al. (1994) described the amount of
fractionation in terms of the quantity $F_{13}$, the column density ratio 
divided by the ambient $^{12}$C/$^{13}$C ratio.  When the ratio of columns is
smaller than the ambient ratio, $F_{13}$ equals exp(-35/{\it T}).  Since the gas
toward HD~200775 has $F_{13}$ of $\approx$ 0.33, the CO observations indicate a
temperature of 30 $-$ 35 K.  The need for C$^+$ suggests that the UV
radiation field is somewhat higher than in a typical diffuse cloud, consistent
with the results described above.  Since $^{13}$CO was not detected toward HD
37903, no meaningful analysis is possible.

\section{Discussion}
\subsection{Structure of PDRs}

Since {\it I}$_{uv}$/{\it n} determines the structure of PDR's (HT99), we can
estimate the fraction of foreground material in the PDR.  We use eqn.~7 from
Hollenbach \& Tielens (1999), which approximately relates the formation rate of
H$_2$ to the photodissociation rate, for $N_{tot}$(H) $\le$ 10$^{21}$ cm$^{-2}$.
Due to its approximate nature, we assume that the use of this equation for lines
of sight with $N_{tot}$(H) of a few times 10$^{21}$ cm$^{-2}$ to be
valid as well.  This steady state expression is used to reproduce the column of
H$_2$, assuming the observed H$_2$ is completely within the PDR:  

\begin{equation}
N({\rm H}_2) = { {(N_{tot}({\rm H})A)^4N_o} \over {N_o^4}} = { {(nN_{tot}
({\rm H})\gamma_{\rm H_2})^4N_o} \over {(4I_{diss}(0)N_o)^4} }.
\end{equation}

\noindent In this equation, {\it N}$_o$ is the column density at which
self-shielding becomes important ({\it N}$_{o}$ = 1 $\times$ 10$^{14}$
cm$^{-2}$), $\gamma_{H_2}$ is the formation rate of H$_2$ on grains
($\gamma_{H_2}$ = 3 $\times$ 10$^{-17}$ cm$^3$ s$^{-1}$), and {\it
I}$_{diss}$(0) is the photodissociation rate (4 $\times$ 10$^{-11}${\it
I}$_{uv}$ s$^{-1}$).  Table 12 shows that the predicted values for {\it
I}$_{uv}$ are 30 to 60 times greater than the values derived from the chemical
analysis.  The $I_{uv}$ determined from the chemistry must be the same as that
determined from the above equation for H$_2$; these results suggest that only a
small fraction ($\sim$ 3\%) of the foreground material is actually associated
with the PDR in front of the star.

The linear extent of the foreground PDR ({\it L}$_{tr}$) can be compared to that
in the background molecular cloud.  We adopted the expression for {\it L}$_{tr}$
of Federman, Glassgold, \& Kwan (1979):

\begin{equation}
L_{tr} = { {2.93 \times 10^{14} n^{-2.4} (5.6 \times 10^{-11} I_{uv})^{1.4} } \over {(2 \gamma_{\rm H_2})^{1.4}}}~{\rm cm}.
\end{equation}

\noindent For the foreground PDR of NGC~2023, this equation yields an extent of
0.002 $-$ 0.007 pc for densities of 500 $-$ 1500 cm$^{-3}$ and respective
values for {\it I}$_{uv}$ of 1 $-$ 3.  For the background cloud with 
{\it n} of 10$^5$ cm$^{-3}$ and {\it I}$_{uv}$ of 10$^4$ (Rouan et al. 1997), the extent is 0.0086 pc.  For NGC~7023, the range in extent is 0.008 $-$ 0.03 pc
when {\it n} of 300 $-$ 900 cm$^{-3}$ and respective {\it I}$_{uv}$ of 1 $-$ 3
are used for the foreground cloud.  Utilizing the density (10$^4$ cm$^{-3}$) and
{\it I}$_{uv}$ of 10$^3$ from Rogers, Heyer \& Dewdney (1995) for the background
cloud indicates a length scale of 0.086 pc.  The extent of the PDR into the
molecular cloud behind the stars is comparable to its extent in the lower
density foreground gas.  For vdB~102 information is only available for the
foreground gas; an extent of 0.004 $-$ 0.01 pc is found from densities and 
{\it I}$_{uv}$ of 400 $-$ 1200 cm$^{-3}$ and 1 $-$ 3, respectively. 

From the column of neutral hydrogen and the density of the cloud, one can
estimate the size of the neutral cloud in front of the star, assuming all the
hydrogen is associated with the cloud.  Toward HD~37903, a neutral cloud size of
0.8 $-$ 2.3 pc is derived.  A cloud size of 0.5 $-$ 1.5 pc toward
HD~147009 and toward HD~200775 a cloud size of 0.8 $-$ 2.5 pc are obtained.
These simple calculations show that the PDR does indeed reside in a small
fraction ($\sim$ 1 \%) of the neutral cloud.

\subsection{NGC~2023}
                     
In order to arrive at a consistent model of PDR's associated with these
reflection nebulae, it is necessary to compare the results obtained here to the
physical conditions inferred from diagnostics at other wavelengths of the 
background molecular cloud.  As noted in \S1.1, there have been many 
studies of NGC~2023 whose focus was the molecular cloud behind HD~37903.  Here,
we highlight those studies that relate to our measurements.

Infrared H$_2$ observations (Martini et al. 1999 and references therein) of this
nebula show that there is a neutral shell of molecular material 0.13 $-$ 0.20~pc
from the star.  However, this neutral shell is not seen in the H~I data.  The
H~I emission is observed to be centered on HD~37903 (Lebr\'{o}n \&
Rodr\'{\i}guez 1997).  Clumps of material have been detected in the nebula on
less than parsec size scales (Martini et al. 1999).  This implies that there are
two components within the nebula.  Steiman-Cameron et al. (1997) determined that
the densities for the low and high density components are 750 cm$^{-3}$ and
10$^5$ cm$^{-3}$, respectively, with a temperature range of 250 K $\le$ $T$
$\le$ 750 K and a radiation field of 1.5 $\times$ 10$^4$ times the average
interstellar radiation field.  For the gas directly behind HD~37903, Wyrowski et
al. (1997) found a density of 3 $\times$ 10$^{4}$ cm$^{-3}$, a radiation field
$\sim$ 500 times the average interstellar radiation field (ISRF) and a
temperature of 125 K.

Martini et al. (1999) measured the near-infrared emission from H$_2$ in two
regions, 60$^{\prime\prime}$ south and 160$^{\prime\prime}$ north of HD~37903.
Their results show that these regions are of relatively high density, {\it n}
$\sim$ 10$^5$ $-$ 10$^6$ cm$^{-3}$, with a ratio of radiation field to density
{\it I}$_{uv}$/{\it n} $\sim$ 0.1 $-$ 0.01 cm$^{3}$.  These measurements are in
agreement with other studies which find high density clumps (10$^5$ $-$ 10$^7$
cm$^{-3}$) in a slightly less dense interclump medium ($\sim$~10$^4$ $-$
10$^{5}$~cm$^{-3}$) with a enhancement of the UV radiation field over the
average ISRF of 10$^3$ $-$ 10$^4$ for NGC~2023 (Martini et al. 1999, Takami et
al. 2000).  

Our chemical analysis reveals that the radiation field is weaker in the
foreground material, {\it I}$_{uv}$ $\sim$ 1 $-$ 3, than the values determined
in previous work for the molecular cloud behind the star (Field et al. 1994,
1998; Fuente et al. 1995; Steiman-Cameron et al. 1997; Wyrowski et al. 1997).
The density of the foreground gas has a range of a few times 10$^{2}$ to
10$^{3}$ cm$^{-3}$, a factor of 10 to 1000 less dense than the gas in the
molecular cloud behind HD~37903.  The temperature of the foreground gas is a
factor-of-2 less than the material behind the star.  These observations of the
foreground material indicating lower densities support the idea that HD~37903
formed near the edge of the molecular cloud and rapidly dispersed the material
at the edge of the cloud.  NGC~2023 could be a face-on example of star formation
by the blister model (Zuckerman 1973; Israel 1978). 

Burton et al. (1998) indicate clump densities of $\approx$ 10$^4$ cm$^{-3}$ and
a far ultraviolet radiation field about 10$^3$ times the ambient interstellar
value at a projected distance of 0.18 pc.  Buss et al. (1994) measured a stellar
flux at 1050 \AA\ of 1.5~$\times$~10$^{-11}$~erg~cm$^{-2}$~s$^{-1}$~\AA$^{-1}$. 
Using an average interstellar flux of 2.65 $\times$ 10$^{-6}$ ergs cm$^{-2}$
s$^{-1}$ \AA$^{-1}$, $\tau_{uv}$ = 2.9, and a distance to the star of 470 pc
(Perryman et al. 1997), we find a distance to the foreground gas of
2.5~$-$~4.5~pc for our values of {\it I}$_{uv}$ between 1 and 3.  Therefore the
radiation field at a distance of 0.1~$-$~0.2~pc from the star would be 
500 $-$ 2000 times the average radiation field, which is consistent with
Burton et al.'s (1998) value of {\it I}$_{uv}$ $\sim$ 1000 (usually denoted by
{\it G}$_o$) at a distance 0.18 pc from the star.

From maps of CN~$J$~=~1~$\rightarrow$~0~(113.5 GHz) and 
HCN~$J$~=~1~$\rightarrow$~0~(88.6 GHz) emission, Fuente et al. (1995) detected
an enhanced amount of CN in the molecular shell in NGC~2023.  The CN/HCN ratio
inside the H$_2$ shell was found to be 25 times larger than outside.  Outside
the shell this ratio is close to the interstellar value for molecular clouds
($\approx$ 2).  HCN photodissociation plays an important role because it
produces CN and because the photodissociation rate of HCN is greater than that
of CN.  While we previously found an enhancement in the abundance of CN toward
HD~200775 (Paper I), consistent with radio observations of NGC~7023 (Fuente et
al. 1993), our chemical results for the foreground PDR of NGC~2023 show no such
enhancement.  Most of the CN is in the envelope of the molecular cloud, not in
the foreground PDR. 

\subsection{vdB~102}

The chemical analysis for HD~147009 matches well with the observed upper limits
for C$_2$ and CN, but the physical conditions derived in our chemical analysis
could not be constrained well.  Unfortunately, little is known about the
physical conditions in this nebula.  Kutner et al. (1980) mapped
CO emission in reflection nebulae and determined that the peak CO emission is
offset, about 3.3 arcmin to the north-east, from the star.  From radio
observations of CH in the cloud, a column density {\it N}(CH) = (1.0 $\pm$ 0.3)
$\times$ 10$^{13}$ cm$^{-2}$ was obtained (Vall\'{e}e 1987).  This is comparable
to the column density obtained from our measurements, {\it N}(CH) = (1.5 $\pm$ 
0.1) $\times$ 10$^{13}$ cm$^{-2}$.  The peak CH emission reported by Vall\'{e}e
is at the same offset as Kutner's CO observations.  Since there is a 
proportional relationship between {\it N}(CH) and {\it N}(H$_2$) (Federman 1982;
Danks et al.  1984; Mattila 1986), Vall\'{e}e inferred that {\it N}(H$_2$) = 
2.5 $\times$ 10$^{20}$ cm$^{-2}$.  Vall\'{e}e used a length of 0.13~pc to
convert column density to space density.  This length was determined from the
fact that the CH abundance extends over 0.3 magnitudes in blue absorption,
$A_B$.  There could easily be a factor-of-2 error in this estimate.  With this
length, Vall\'{e}e obtained {\it n}(H$_2$) = 600 cm$^{-3}$ and then converted
{\it n}(H$_2$) into a total proton density of {\it n} = 1300 cm$^{-3}$ with the
relation {\it n} = 2.2 {\it n}(H$_2$), assuming 10\% of the gas is He.
Vall\'{e}e's density, which is similar to ours for {\it I}$_{uv}$ = 3, may not
represent the molecular cloud behind the star as he reported, since his CH
abundance is comparable to ours.  The molecular material may in fact reside in
front of HD~147009.  Moreover, this density is similar to densities found in the
diffuse component in the reflection nebulae NGC~2023 (this work) and NGC~7023
(Paper I and this work).  

\subsection{Optical Results for Three Nebulae}

The reflection nebulae NGC~2023 and NGC~7023 are relatively bright extended 
objects, located at the edges of their respective molecular clouds.   For these 
reasons they have been extensively studied at a variety of wavelengths.  
The two nebulae have been shown to have clumpy, filamentary structure from 
infrared observations.  The central stars, HD~37903 and HD~200775, of these 
nebulae have similar spectral types, B1.5~V and B3e~V, and similar 
distances of 470 pc and 430 pc (Perryman et al. 1997), respectively.  The peak
infrared emission is offset from the stars in both nebulae.  As
for vdB~102, the illuminating source, HD~147009, is of later spectral type
(A0~V) and is closer to the Sun at a distance of 160 pc (Perryman et al. 1997).
The extent of filamentary structure is unknown. 

The physical conditions obtained from optical spectra toward the illuminating
stars in the reflection nebulae NGC~2023 and vdB~102 are similar to those
obtained from an earlier study of NGC~7023 (Paper I) and updated here.  The
physical conditions {\it n}, $T$, and {\it I}$_{uv}$ inferred for NGC~7023
are 300 $-$ 900 cm$^{-3}$, 40 K, and 1 $-$ 3, respectively, once the effects of
a lower $x$(C$^+$) are included (Table 10).  These physical conditions are
similar to those obtained from grain scattering observations of NGC~7023 (Walker
et al. 1980; Witt \& Cottrell 1980a, 1980b; Witt et al. 1982).  The physical 
conditions found toward NGC~2023 are 500 $-$ 1500 cm$^{-3}$, 40 K, and 1 $-$ 3 
and toward vdB~102 400 $-$ 1200 cm$^{-3}$, 40 K, and 1 $-$ 3, for $n$, $T$, and 
$I_{uv}$, respectively.  Therefore, the material in front of the stars in the 
reflection nebulae NGC~2023, vdB~102, and NGC~7023 is relatively diffuse. 

In Paper I, an enhancement in the CN abundance over that described in eqn. 2 was
(and is still needed) to explain the observed CN abundance in the foreground PDR
of NGC~7023. This enhancement of CN in the foreground material of NGC~7023 is
the result of a contribution from dark cloud chemistry (Paper I) $-$ HCN
$\rightarrow$ CN through photodissociation.  If we really are seeing an effect
of dark cloud chemistry from the molecular cloud behind the star, the foreground
PDR may be closer to the star than our estimates suggest, even for an
enhancement by a factor-of-3 in the radiation field.  This enhancement is not 
detected toward either HD~37903 or HD~147009, unless the amount of C$_2$ in the 
gas is much less than our current upper limit.  This could be a consequence of 
the different distances between each star and its foreground gas.  

In the case of NGC~7023, an hourglass shaped cavity has been detected
(Lemaire et al. 1996; Fuente et al. 1998) around the star.  CO radio maps reveal
(Gerin et al. 1998) that the $^{13}$CO emission profiles are relatively 
smooth and centered at 2.5 km s$^{-1}$, while the $^{12}$CO profiles indicate
more complicated structure.  This structure shows a gradual shift in the
velocity of the gas from 2.5 km s$^{-1}$ to 1.5 km s$^{-1}$, which is the
velocity of the foreground molecular material reported in Paper I.  This
suggests that there should be a diffuse $^{12}$CO component in the cavity or the
foreground material.  We associate this diffuse component with the CO seen in UV
absorption.  Lemaire et al. (1999) report redshifted H {\small I} with
respect to H$_2$ in the molecular cloud behind the star.  This trend is also
detected in atomic absorption (Paper I) which is blueshifted compared to the
foreground molecular material.  This is consistent with an expanding PDR around
the star (Gerin et al. 1998; Fuente et al. 1998).      

The analysis of atomic data provides insight into the processes taking place in
the foreground gas.  Using the extinction curve for the line of sight in our
model, the observed ratio of Na~{\small I} to K~{\small I} toward NGC~7023 and
vdB~102 matched the predicted ratio.  In the cases for $\zeta$ Per and NGC~2023,
the observed ratio was about twice the predicted value.  Adjustments to the
photoionization rates would bring the predicted ratios into better agreement
with observations.  In particular, a finer grid for computing the differences 
in extinction relative to the extinction for $\zeta$~Per may be more
appropriate; we only considered average differences over wavelength regions
where the curves intersected.  In the comparison of the column density of each
species toward the reflection nebulae and $\zeta$~Per, a small change in $n_e$
or an extra source of depletion is needed toward NGC~7023 to match the observed
and predicted results.  Toward HD~147009, a fairly substantial modification is
required.  A significant increase in {\it n}$_e$, possibly provided by protons
in diffuse ionized gas created by earlier type stars in Sco OB2, is needed to
account for the differences found between the nebula vdB~102 and gas toward
$\zeta$~Per.

The analysis of ultraviolet data provides additional pieces to the picture for
the gas associated with NGC~2023 and NGC~7023.  It appears that the foreground
PDR in NGC~7023 is more homogeneous (because C {\small I} excitation, C$_2$
excitation, CO fractionation, and chemical analysis yield very similar physical
conditions) and is permeated by a slightly stronger radiation field compared to
that for NGC~2023.  

\section{Conclusions}

We presented the first chemical results for foreground material associated 
with the PDRs of NGC~2023 and vdB~102.  The foreground material in NGC~2023 is
more diffuse than the molecular cloud behind HD~37903 and is subjected to a less
intense radiation field.  The results are similar to what we found for NGC~7023
(Paper I).  As for vdB~102, our analyses are the most comprehensive to date and will be the basis for other studies of nebulae illuminated by A-type stars.  

Future studies should incorporate details that extend the treatment presented
here.  The effects of a clumpy distribution of material and a more extensive 
set of chemical reactions in the framework of a time-dependent model need to be considered.  These models have to include the ``anomalous" extinction law found for both stars in this work (see also FM90; WBS84).  The consequences of
NGC~2023 and vdB~102 being located in active star-forming regions is another area for future analysis.  Other absorption measurements of species detected at
ultraviolet wavelengths, such as H$_2$, are needed to place more stringent
constraints on physical conditions.  Observations with the {\it Far Ultraviolet
Spectroscopic Explorer} will soon be obtained.  This combination of
visible and UV data, on the one hand, and more sophisticated modeling efforts,
on the other, will provide a clearer description of the interaction between a
star's radiation field and the surrounding molecular cloud.

\acknowledgments  We thank David Doss for assistance with the instrumental 
setups at McDonald Observatory and Tom Ingerson for help with the setup at 
CTIO.  The McDonald Observatory 2dcoud\'{e} data presented here formed the basis
for David Knauth's Master's Thesis.  Tony Imperial and Marissa Lingen helped in
reducing CTIO data, and Kyle Westfall reduced the NEWSIPS data for X~Per.
Marissa and Kyle participated in the Research Experience for Undergraduates
(REU) Program at the University of Toledo, which was funded by NSF.  We also
thank Yaron Sheffer for the use of his code.  This research made use of the
Simbad database, operated at CDS, Strasbourg, France. This research was
supported in part by NASA LTSA grant NAG5-4957.

\begin{center}
{\bf A. Reliability of NEWSIPS Data}
\end{center}

The $IUE$ archive is now comprised of NEWSIPS data, which yields spectra with 
superior signal to noise.  Techniques were introduced to perform geometric 
and photometric corrections, to improve the ripple correction, and to 
determine the image background (e.g., Nichols 1998).  We examined the
reliability of our data in two ways.  First, we compared our results for another
star, 20~Aql, with those of Hanson, Snow, \& Black (1992) based on an earlier
form of the archive.  Then we compared $W_{\lambda}$'s of C {\small I} lines
seen in the $IUE$ spectra of X~Per with measurements from high-resolution
spectra acquired with ECH-A of the Goddard High Resolution Spectrograph on the
{\it Hubble Space Telescope}.  Our comparisons show that NEWSIPS data provide a
consistent, reliable set of $W_{\lambda}$'s with improved signal to noise for
interstellar studies, thereby confirming the analyses performed by
Gonz\'{a}lez-Riestra et al. (2000).

Tables 13 to 15 provide the basis for comparison.  Table 13 shows our results 
for interstellar gas toward 20~Aql based on NEWSIPS data and the results of 
Hanson et al. (1992).  The agreement in $W_{\lambda}$ for values from 
less than 10 to 150 m\AA\ is excellent, without any discrepancies beyond the 
mutual uncertainties.  The uncertainties associated with the NEWSIPS 
spectra also tend to be smaller.  Sample spectra appear in Figure 11, and 
Table 14 displays the column densities derived from the fits.  In Table 15,
NEWSIPS results for C {\small I} toward X~Per are compared with measurements
from ECH-A spectra.  Again, the correspondence is excellent.  Syntheses of the
ECH-A data give a measure of the precision in column density attainable from
NEWSIPS data.  Since $b$-values can be determined from NEWSIPS data to about 0.5
km s$^{-1}$ and 0.1 km s$^{-1}$ from ECH-A data, column densities from
relatively strong lines in NEWSIPS data are accurate to about 50\%.   This
confirms the measure of uncertainty given to atomic column densities in Table 8.

For completeness, Tables 16 to 18 indicate a synthesis of results from 
ground-based observations of 20~Aql, incorporating data acquired during
observing runs described here.  This compilation includes the highest-resolution
spectra taken of 20~Aql.  Our spectra are displayed in Figures 12-14.  Analyses,
similar to those for NGC~2023 and vdB~102, were applied to the data. The
density, temperature and radiation field are 1400 cm$^{-3}$, 50 K and 1; the
factor-of-2 increase from Federman, Strom, \& Good (1991) in the derived density
is due to the lower fractional abundance of carbon.   With the exception of the
density, the results, derived from the chemical analysis, are much like the
results of Federman et al. (1991) and Hanson et al. (1992).  

The atomic analysis performed above is repeated for data toward 20~Aquilae.
The inferred Na~{\small I} and K~{\small I} photoionization rates are 3.4
$\times$ 10$^{-13}$ s$^{-1}$ and 1.1 $\times$ 10$^{-12}$ s$^{-1}$, which yield
a predicted ratio of 54.  The observed column densities of (1.3~$\pm$~0.1)
$\times$ 10$^{15}$ cm$^{-2}$ and (1.4~$\pm$~0.1) $\times$ 10$^{13}$ cm$^{-2}$
for Na~{\small I} and K~{\small I}, respectively, produce an observed ratio of
93 $\pm$ 10.  The correspondence between observed and predicted $N$(Na {\small
I})/$N$(K {\small I}) ratio is about two to one, similar to that found for gas
toward $\zeta$ Per and HD~37903.

For the comparison of the columns Na {\small I} and K {\small I} of 20~Aquilae 
to those of $\zeta$~Per, the total proton density toward 20~Aquilae is needed.
The proton density is (3.0 $\pm$ 0.6) $\times$ 10$^{21}$ cm$^{-2}$ (Hanson et
al. 1992) with an assumed 20 \% error (see text above).  This results in a
predicted ratio {\it N}(Na~{\small I})$_{20~Aquilae}$/{\it N}(Na {\small
I})$_{\zeta\ Per}$ of 29 $\pm$ 8 and 36 $\pm$ 10 for K~{\small I}.  The observed
ratios are 18 $\pm$ 2 for Na~{\small I} and 19 $\pm$ 2 for K~{\small I}.  The
effect of $\alpha_c$ was not needed in this analysis (see text above). 
Formally, the predicted and observed ratios are consistent with (at most) a
modest change in {\it n}$_e$ and/or depletion to bring the predictions into
better agreement with the observations, much like the results for the PDR toward
HD~200775.

Editor place figures 11 - 14 here.
Editor place tables 13 - 18 here.

\newpage
\clearpage

\begin{figure}[p]
\begin{center}
\plotfiddle{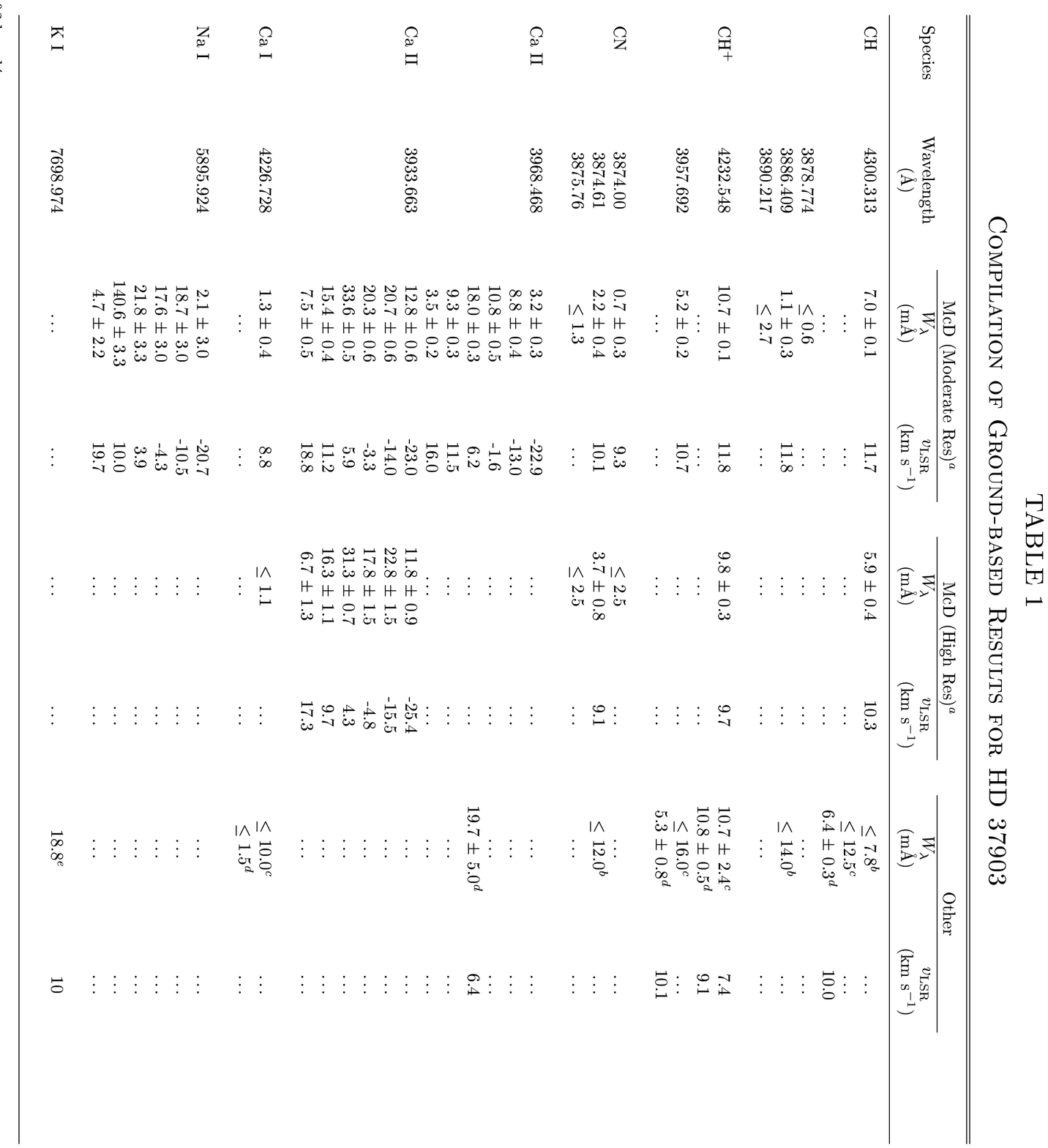}{4.0in}{180}{80}{80}{220}{450}
\end{center}
\end{figure}

\begin{figure}[p]
\begin{center}
\plotfiddle{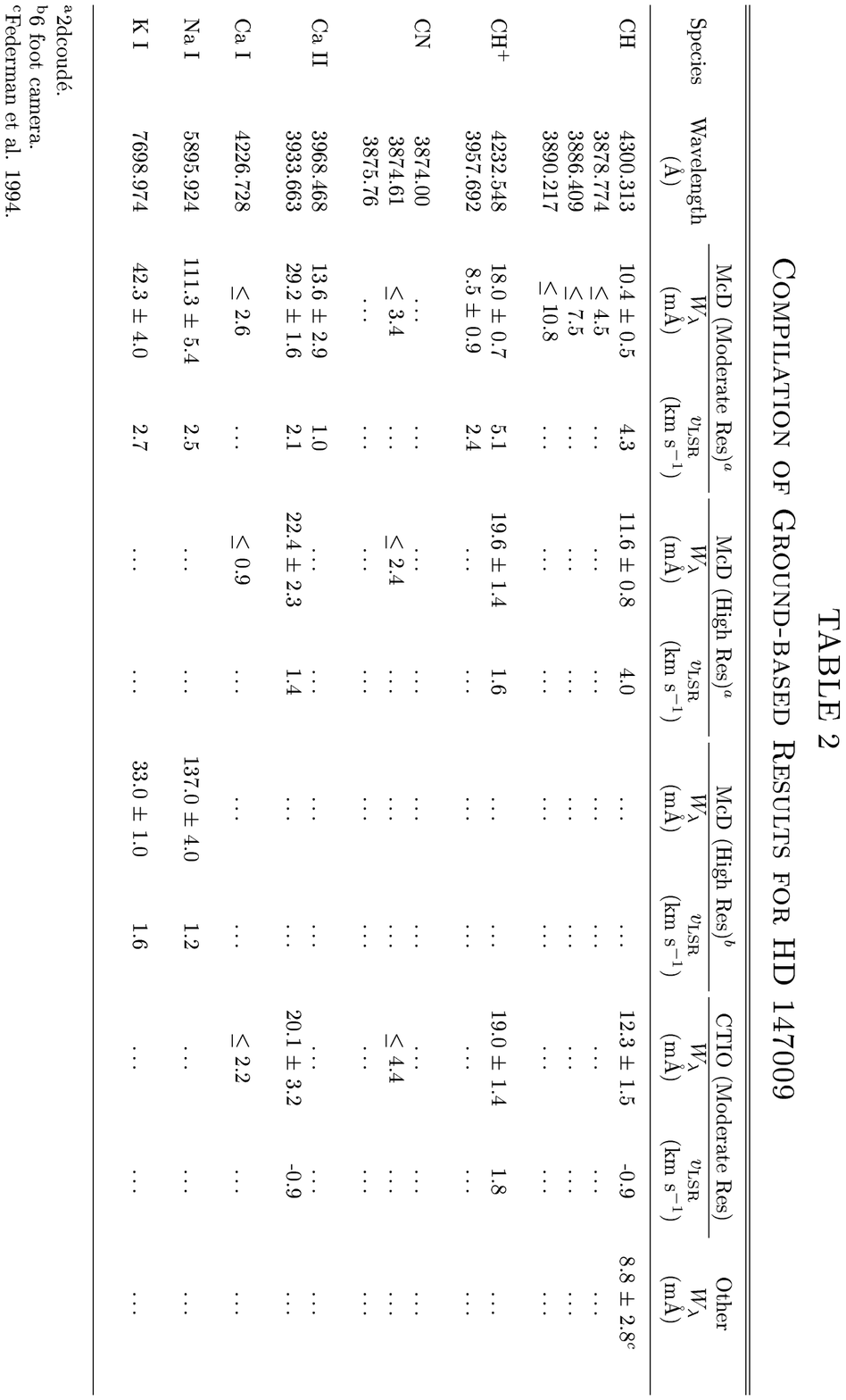}{4.0in}{180}{80}{80}{220}{450}
\end{center}
\end{figure}

\begin{figure}[p]
\begin{center}
\plotfiddle{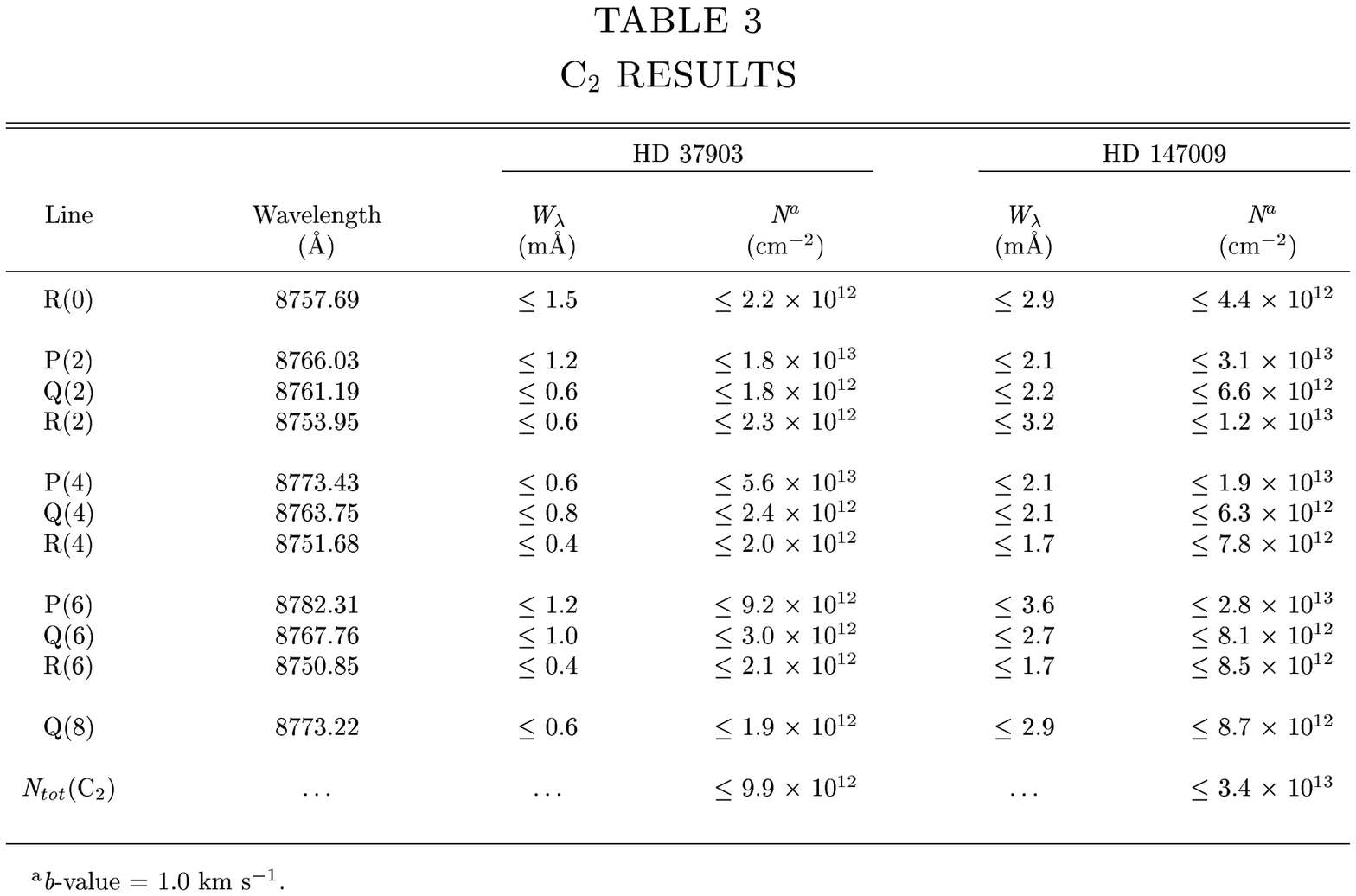}{4.0in}{0}{80}{80}{-220}{-165}
\end{center}
\end{figure}

\begin{figure}[p]
\begin{center}
\plotfiddle{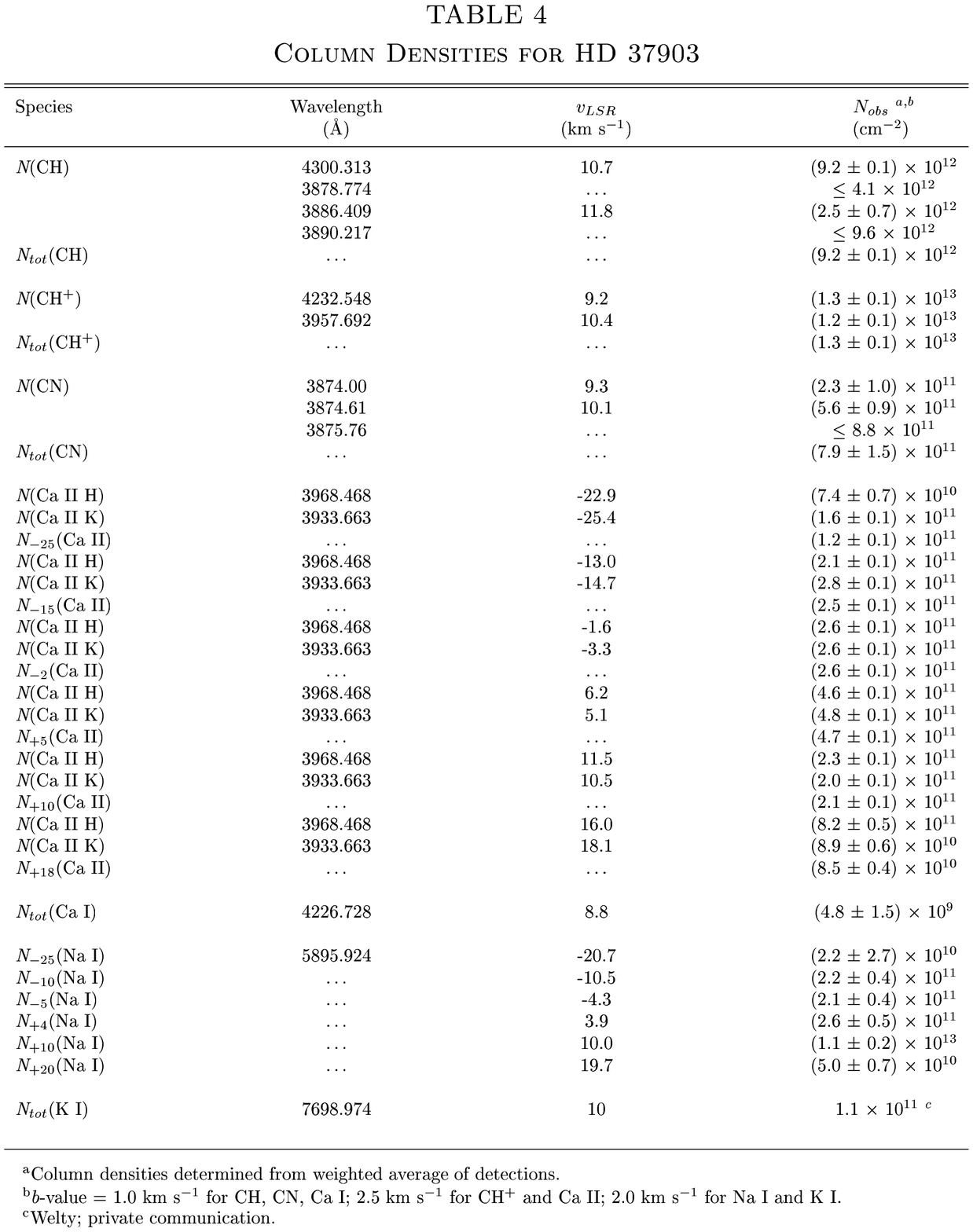}{4.0in}{0}{80}{80}{-220}{-165}
\end{center}
\end{figure}

\newpage
\clearpage

\begin{figure}[p]
\begin{center}
\plotfiddle{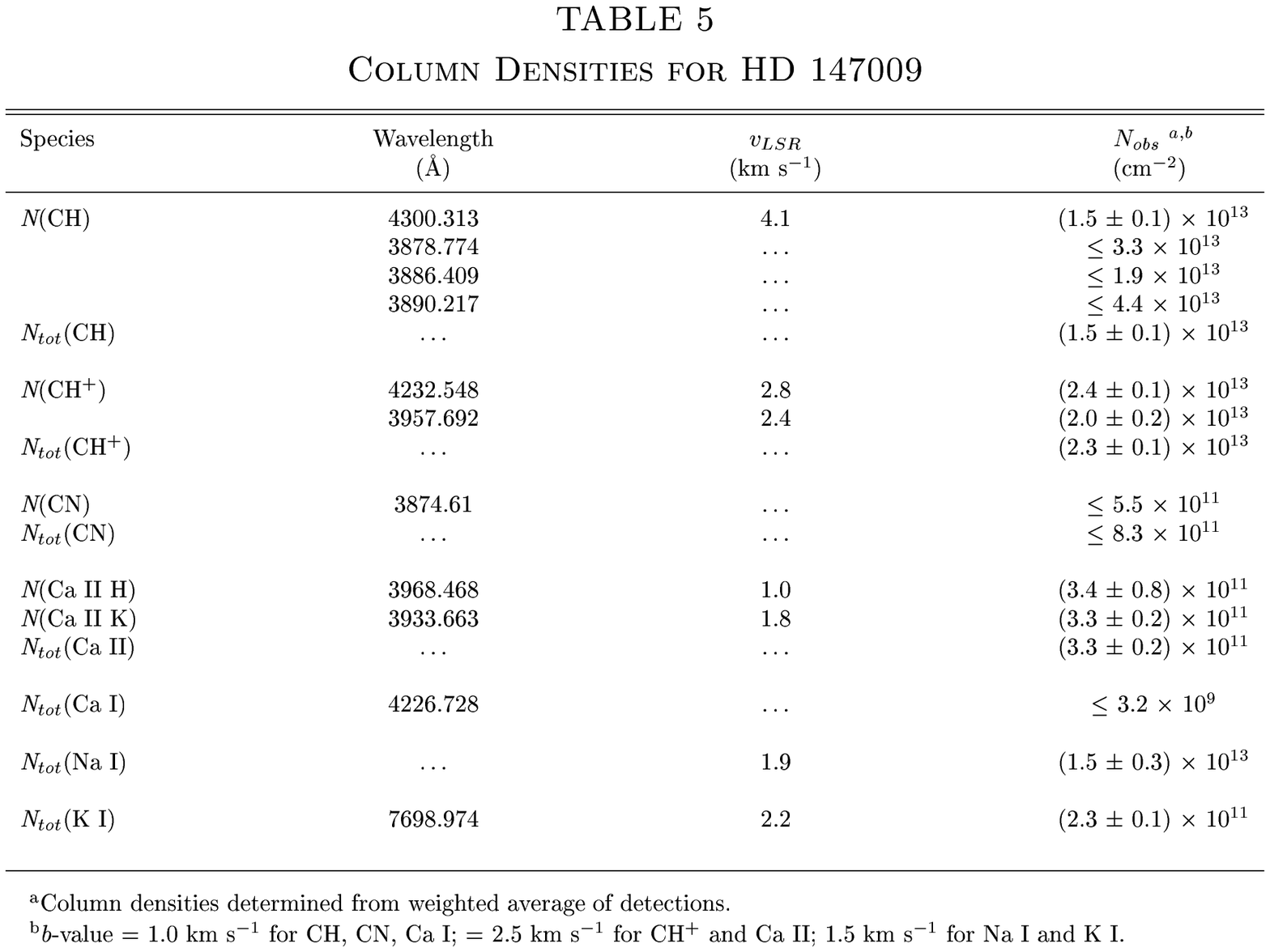}{4.0in}{0}{80}{80}{-220}{-165}
\end{center}
\end{figure}

\begin{figure}[p]
\begin{center}
\plotfiddle{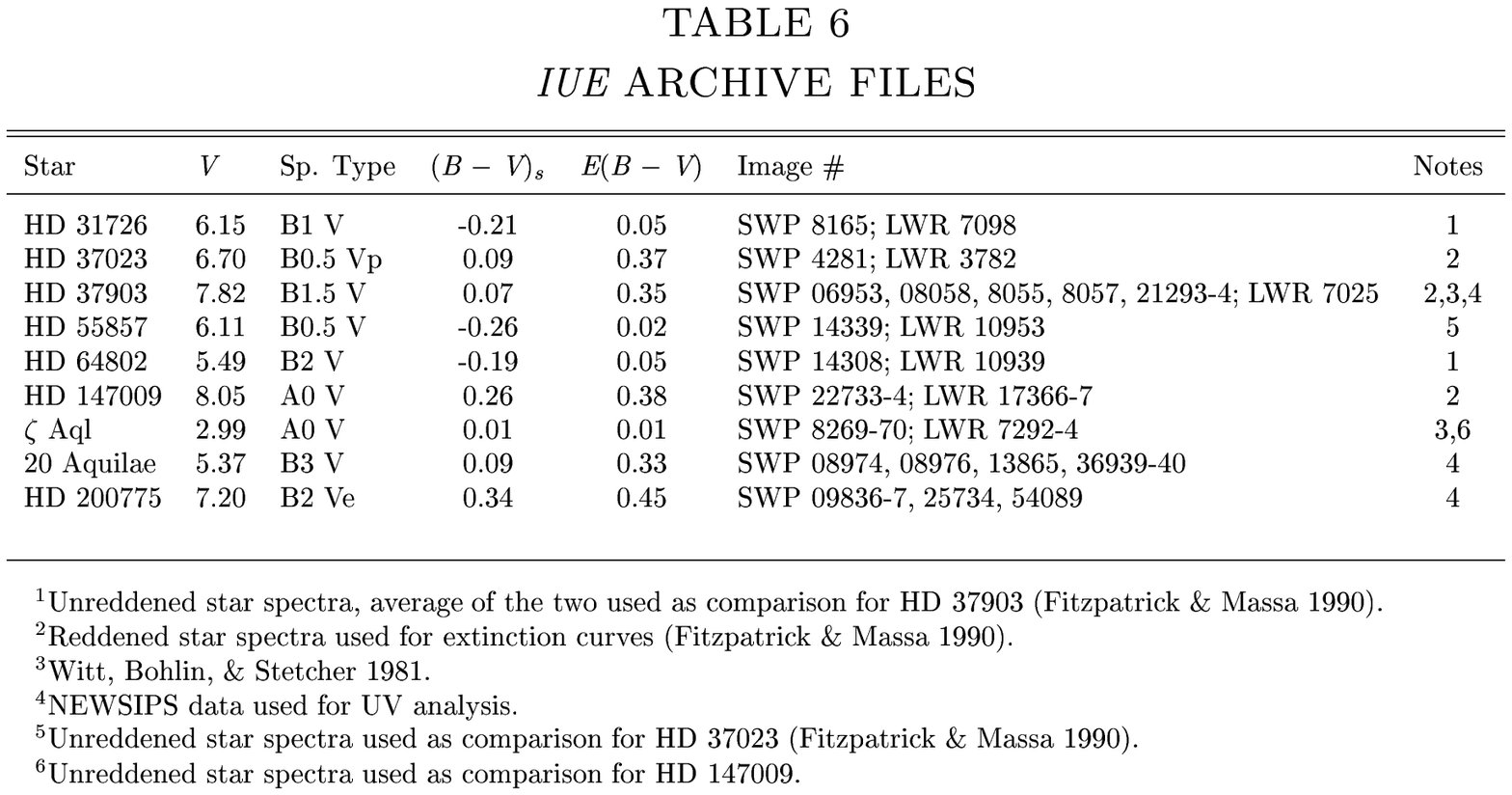}{4.0in}{0}{80}{80}{-220}{-165}
\end{center}
\end{figure}

\begin{figure}[p]
\begin{center}
\plotfiddle{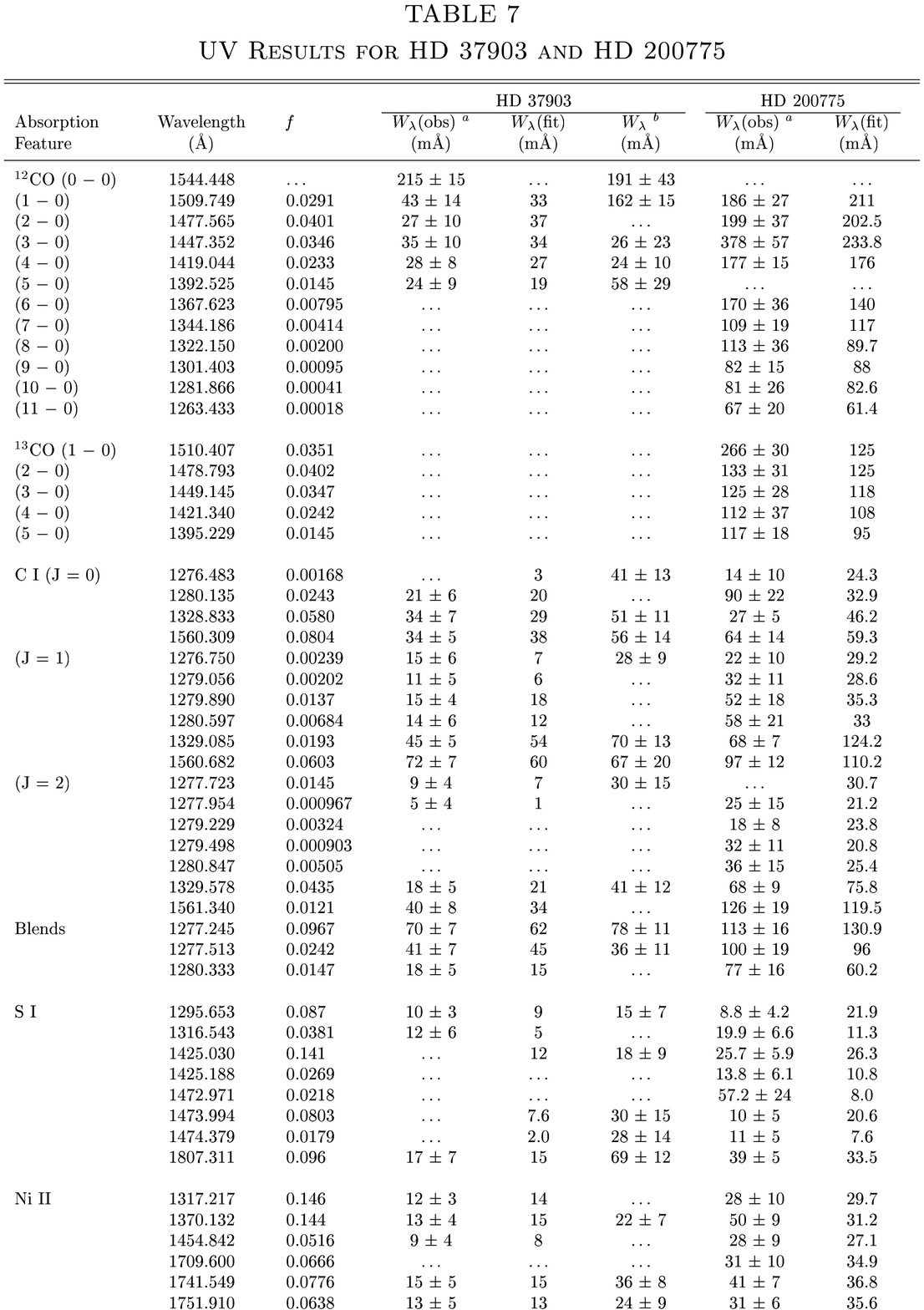}{4.0in}{0}{80}{80}{-220}{-165}
\end{center}
\end{figure}

\begin{figure}[p]
\begin{center}
\plotfiddle{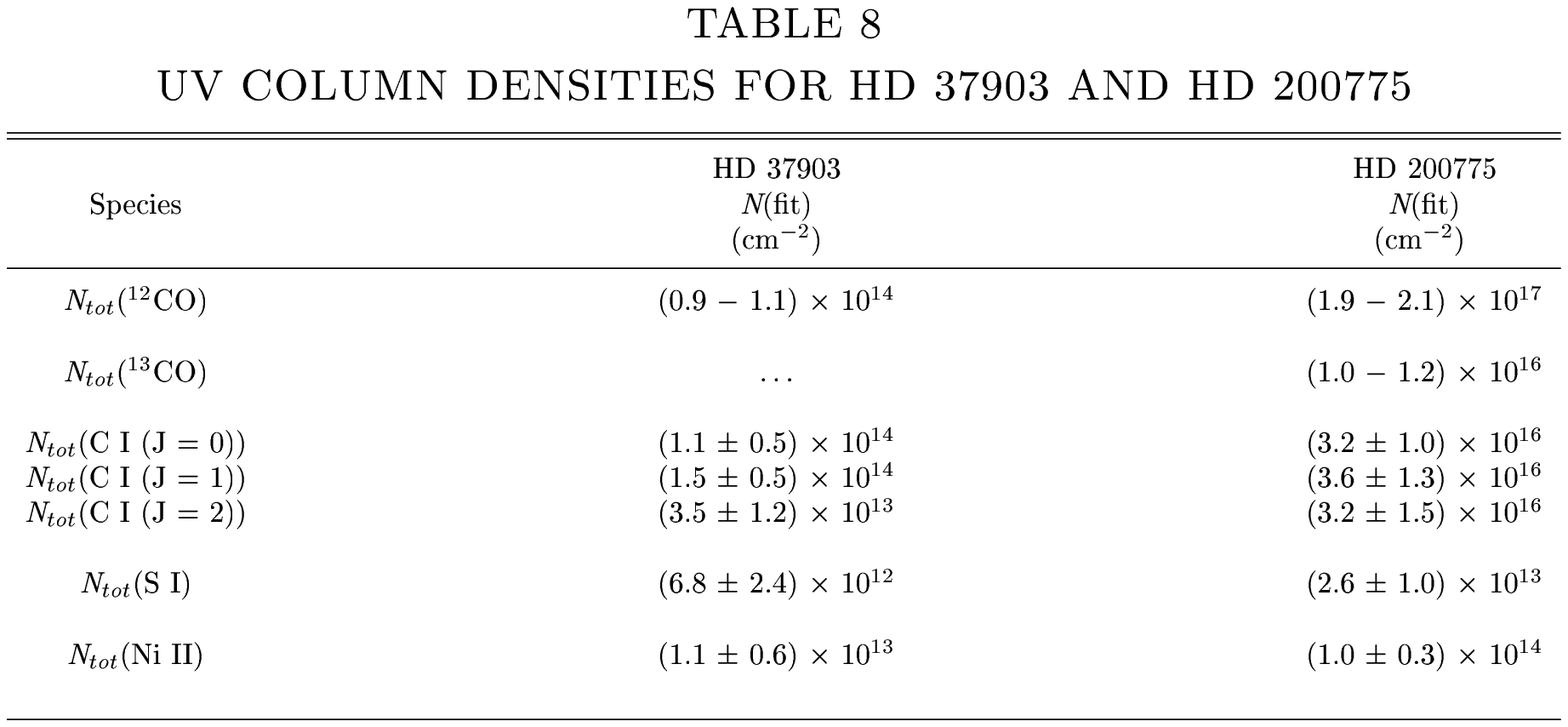}{4.0in}{0}{80}{80}{-220}{-165}
\end{center}
\end{figure}

\begin{figure}[p]
\begin{center}
\plotfiddle{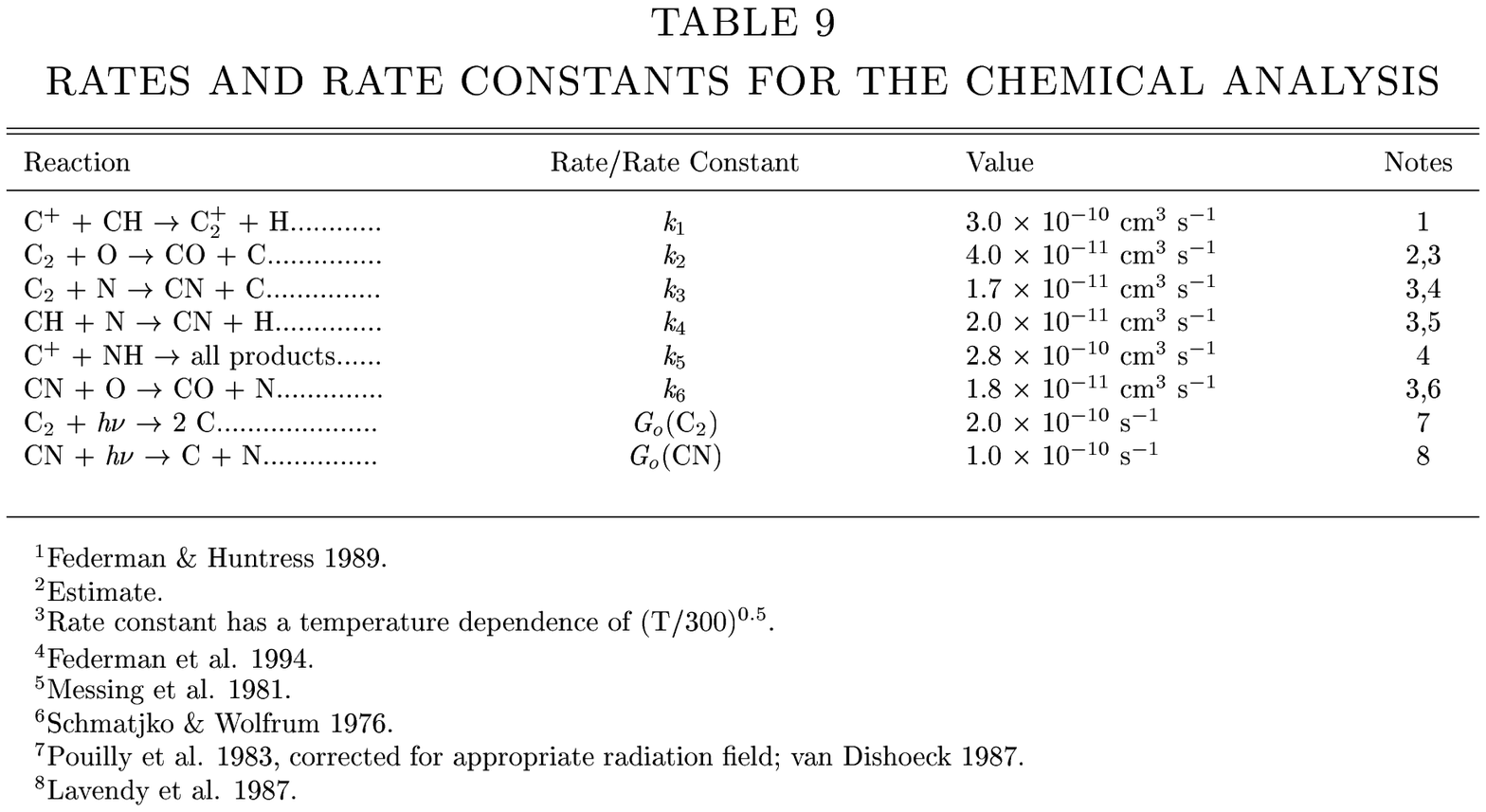}{4.0in}{0}{80}{80}{-220}{-165}
\end{center}
\end{figure}

\begin{figure}[p]
\begin{center}
\plotfiddle{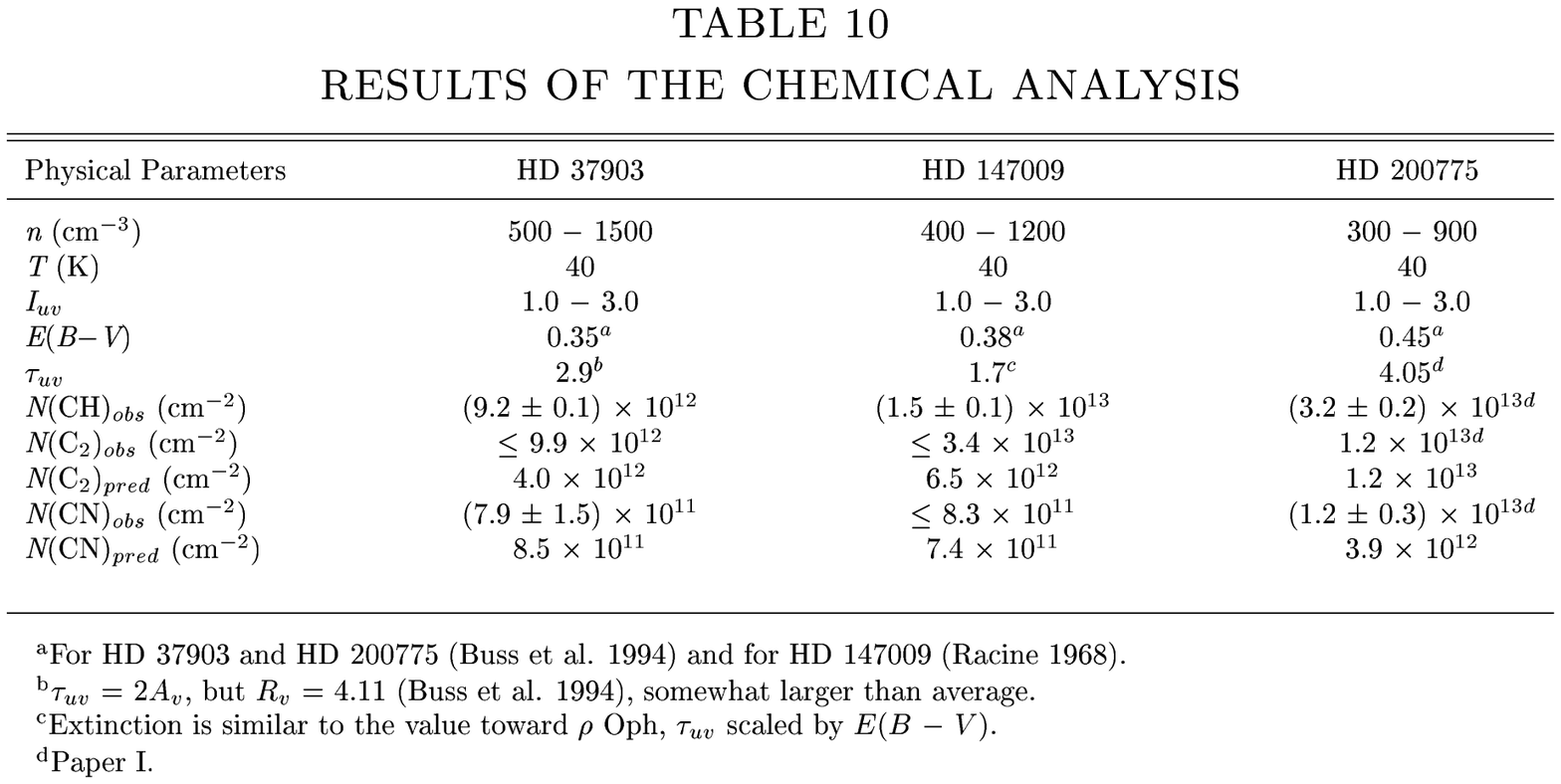}{4.0in}{0}{80}{80}{-220}{-165}
\end{center}
\end{figure}

\begin{figure}[p]
\begin{center}
\plotfiddle{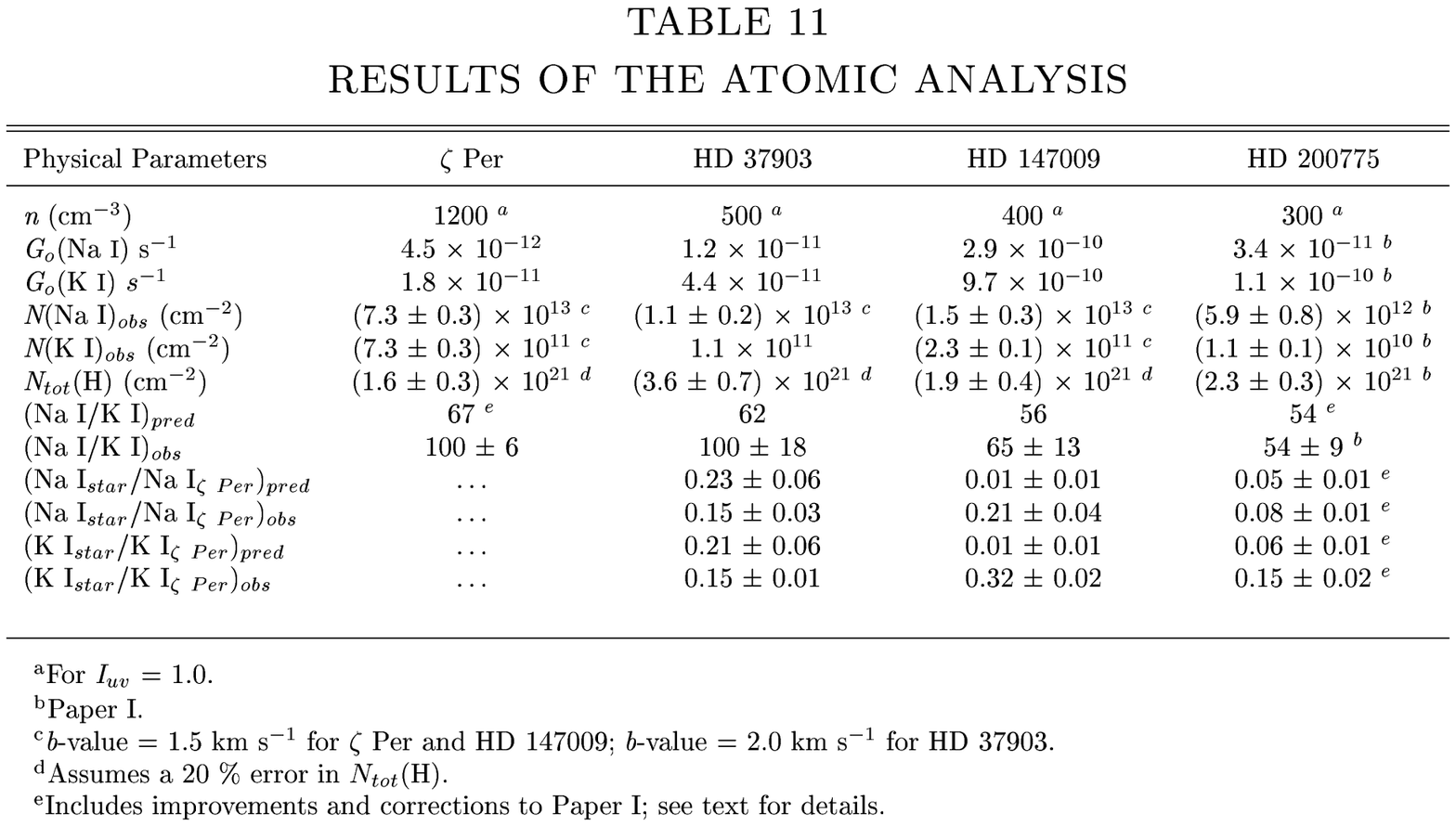}{4.0in}{0}{80}{80}{-220}{-165}
\end{center}
\end{figure}

\begin{figure}[p]
\begin{center}
\plotfiddle{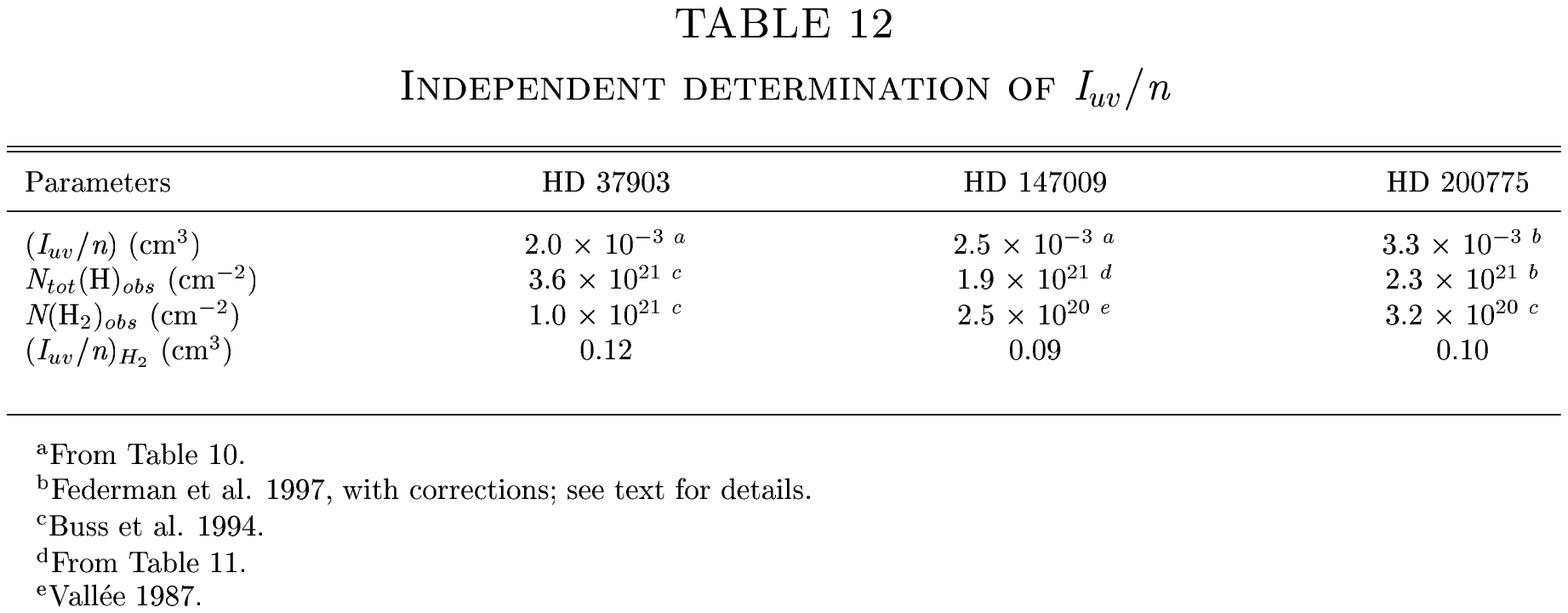}{4.0in}{0}{80}{80}{-220}{-165}
\end{center}
\end{figure}

\begin{figure}[p]
\begin{center}
\plotfiddle{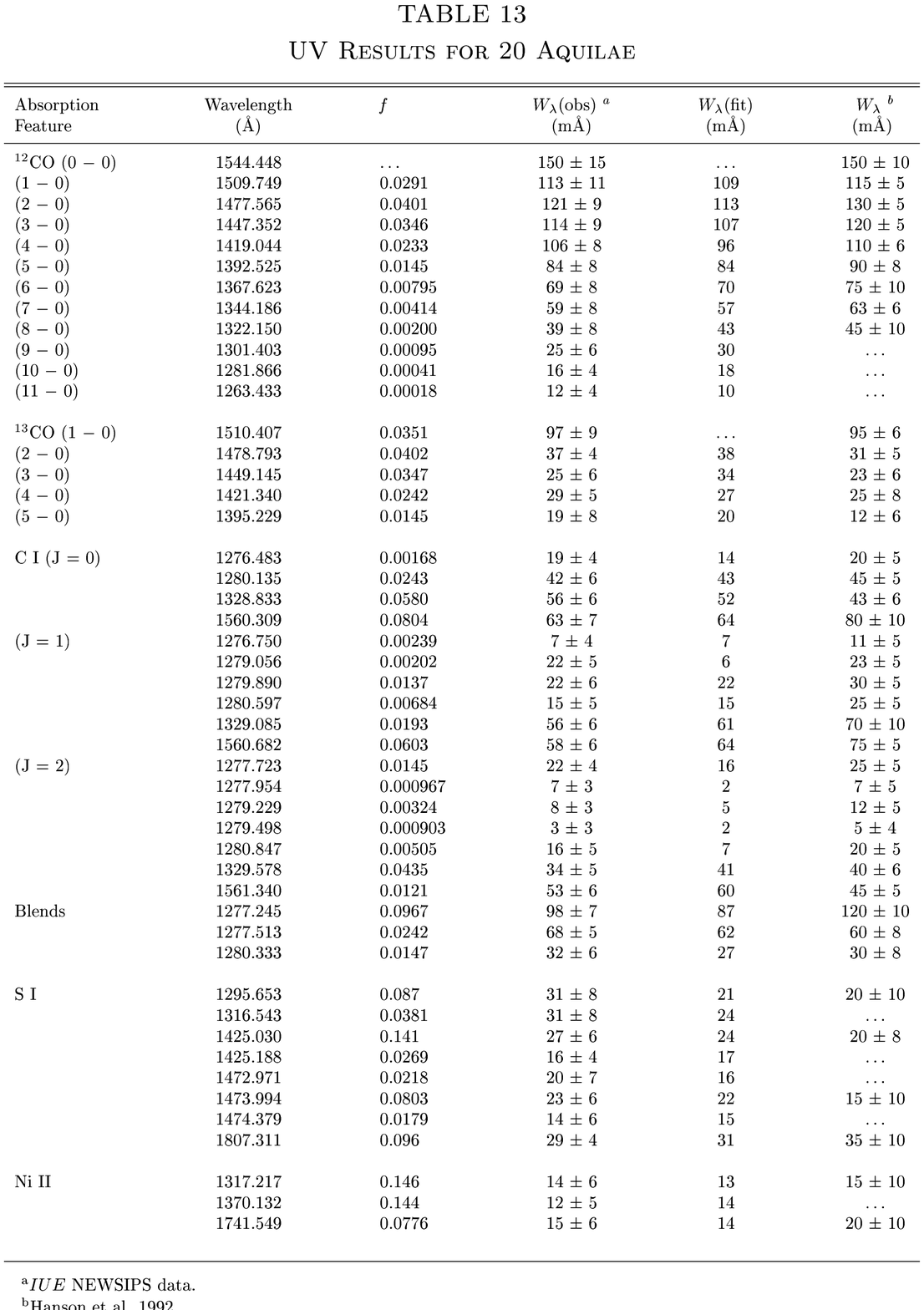}{4.0in}{0}{80}{80}{-220}{-165}
\end{center}
\end{figure}

\newpage
\clearpage

\begin{figure}[p]
\begin{center}
\plotfiddle{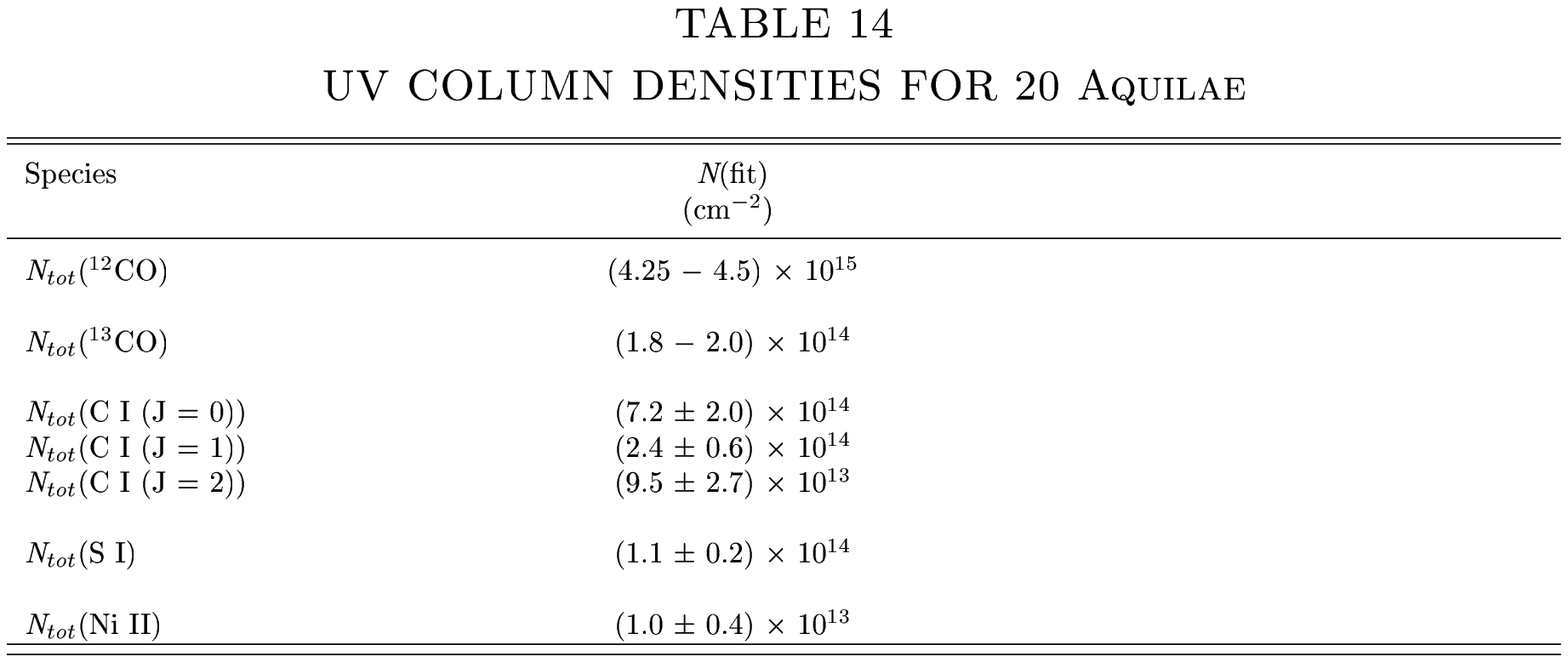}{4.0in}{0}{80}{80}{-220}{-165}
\end{center}
\end{figure}

\begin{figure}[p]
\begin{center}
\plotfiddle{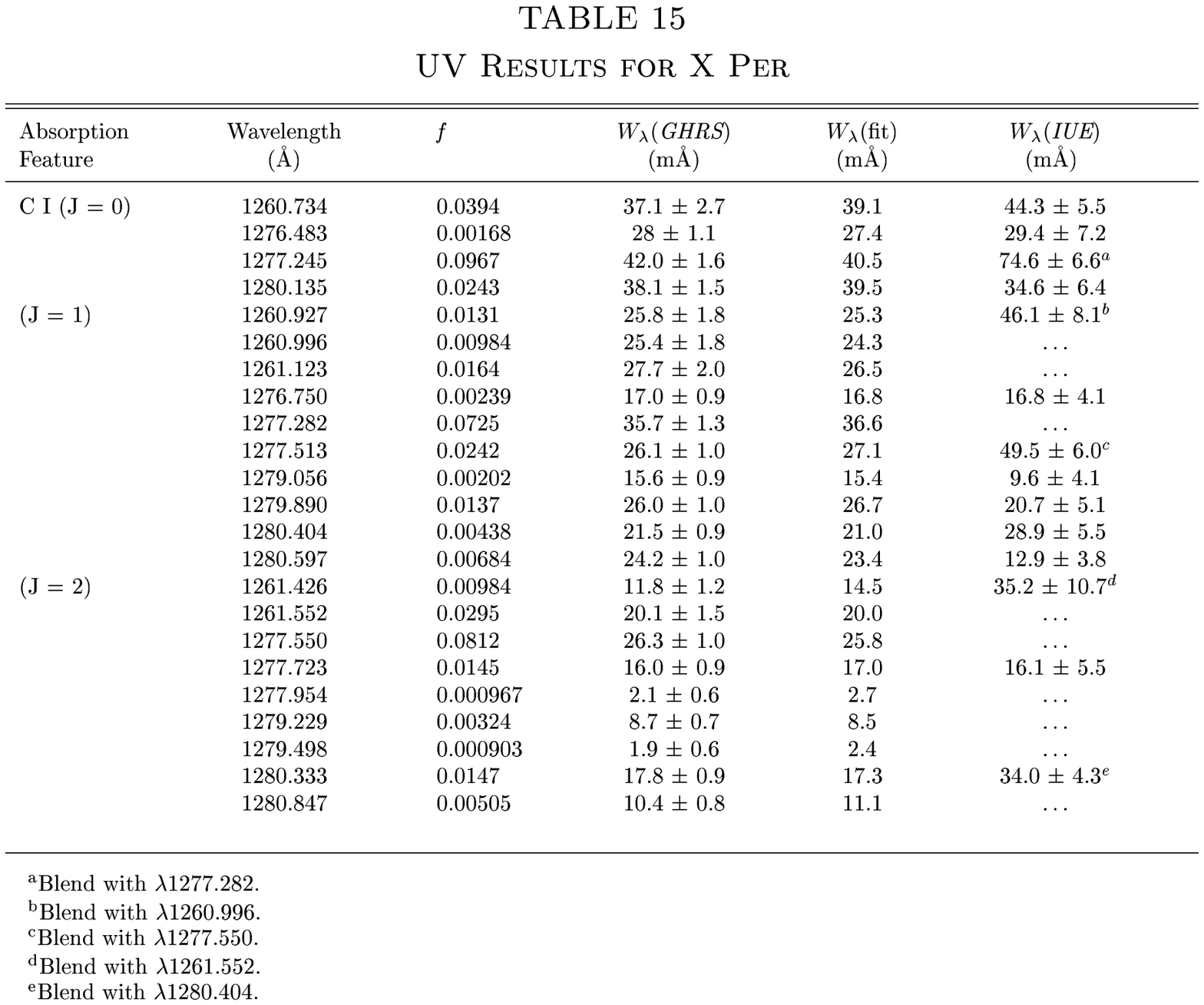}{4.0in}{0}{80}{80}{-220}{-165}
\end{center}
\end{figure}

\begin{figure}[p]
\begin{center}
\plotfiddle{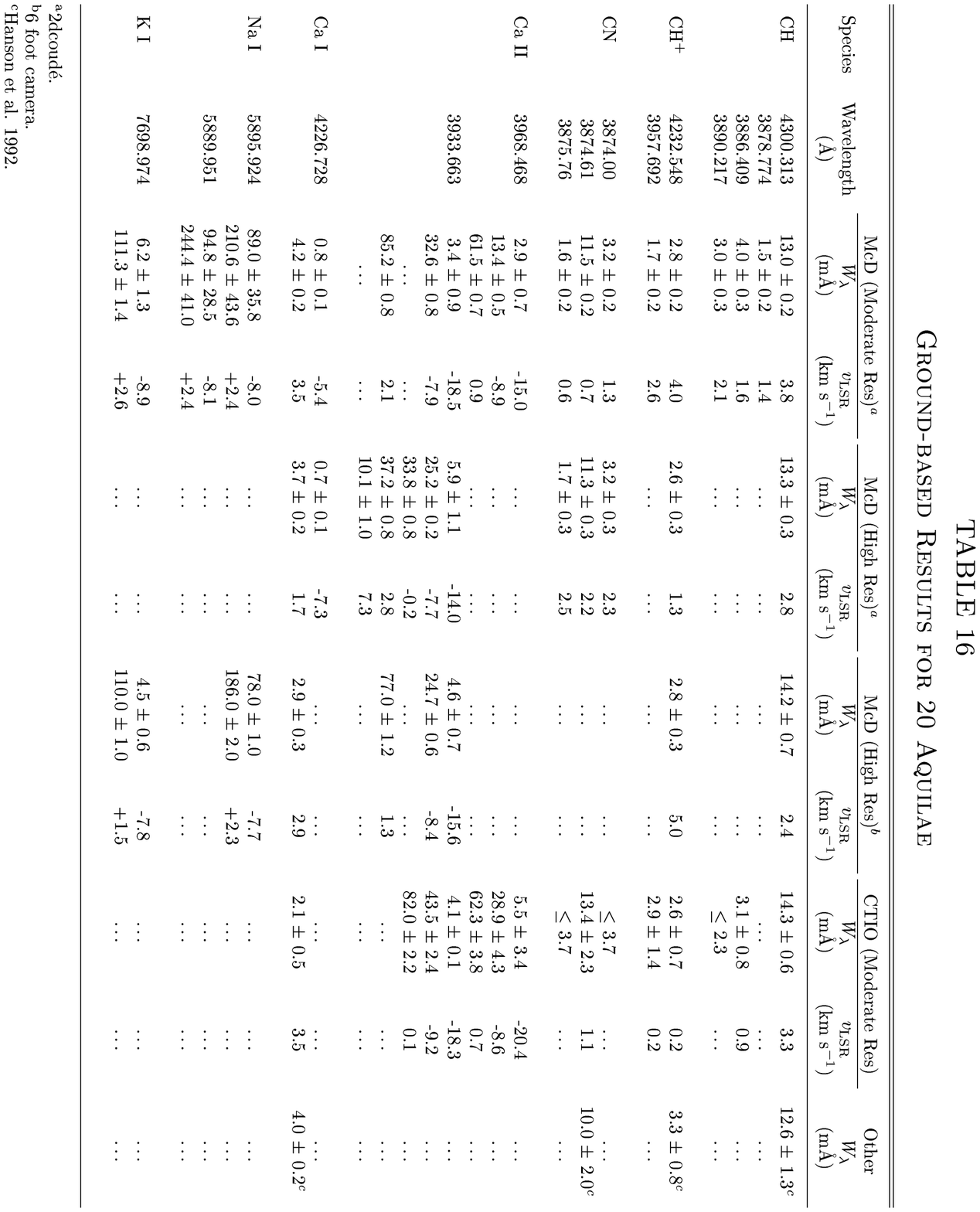}{4.0in}{180}{80}{80}{220}{450}
\end{center}
\end{figure}

\begin{figure}[p]
\begin{center}
\plotfiddle{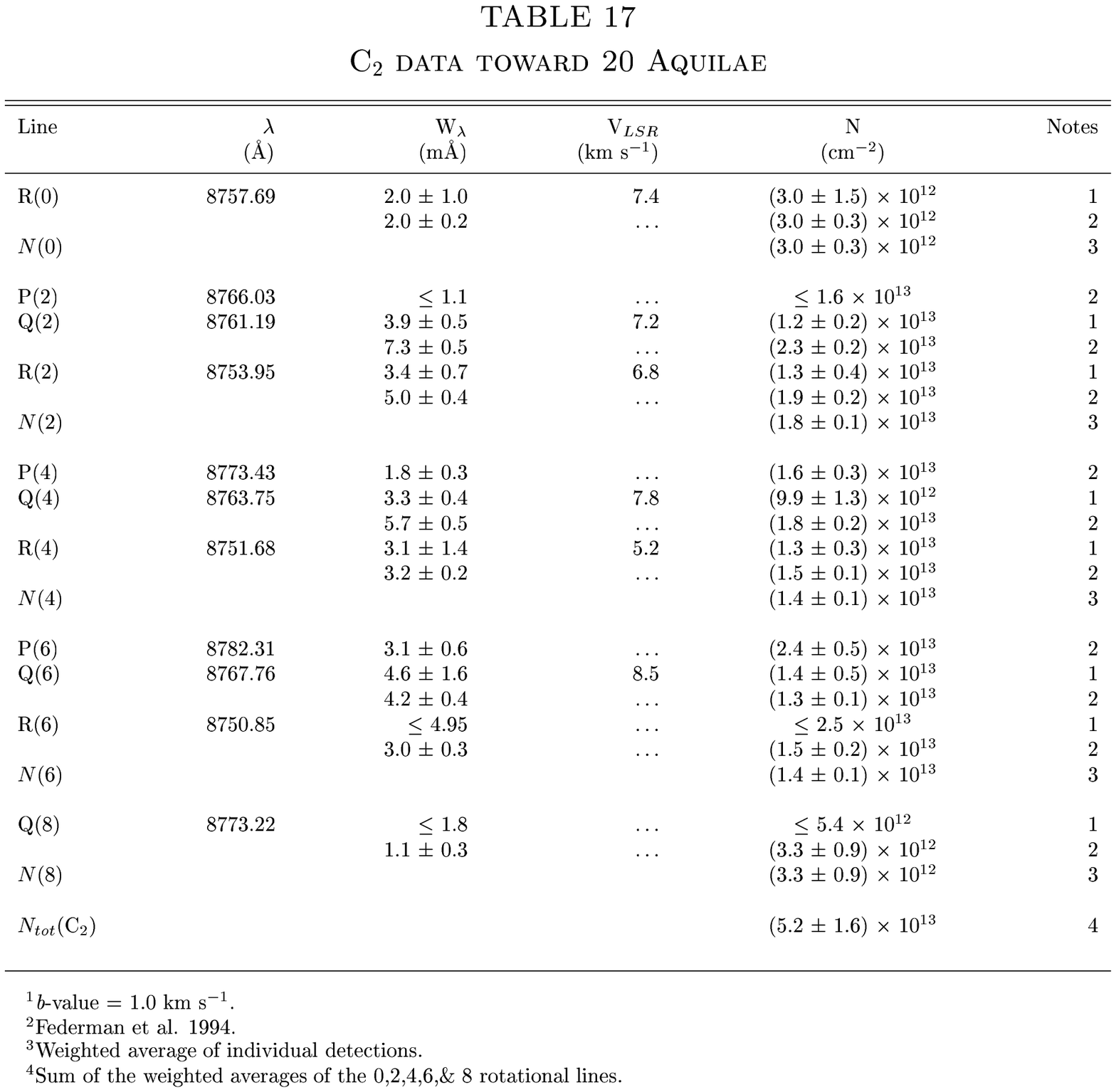}{4.0in}{0}{80}{80}{-220}{-165}
\end{center}
\end{figure}

\begin{figure}[p]
\begin{center}
\plotfiddle{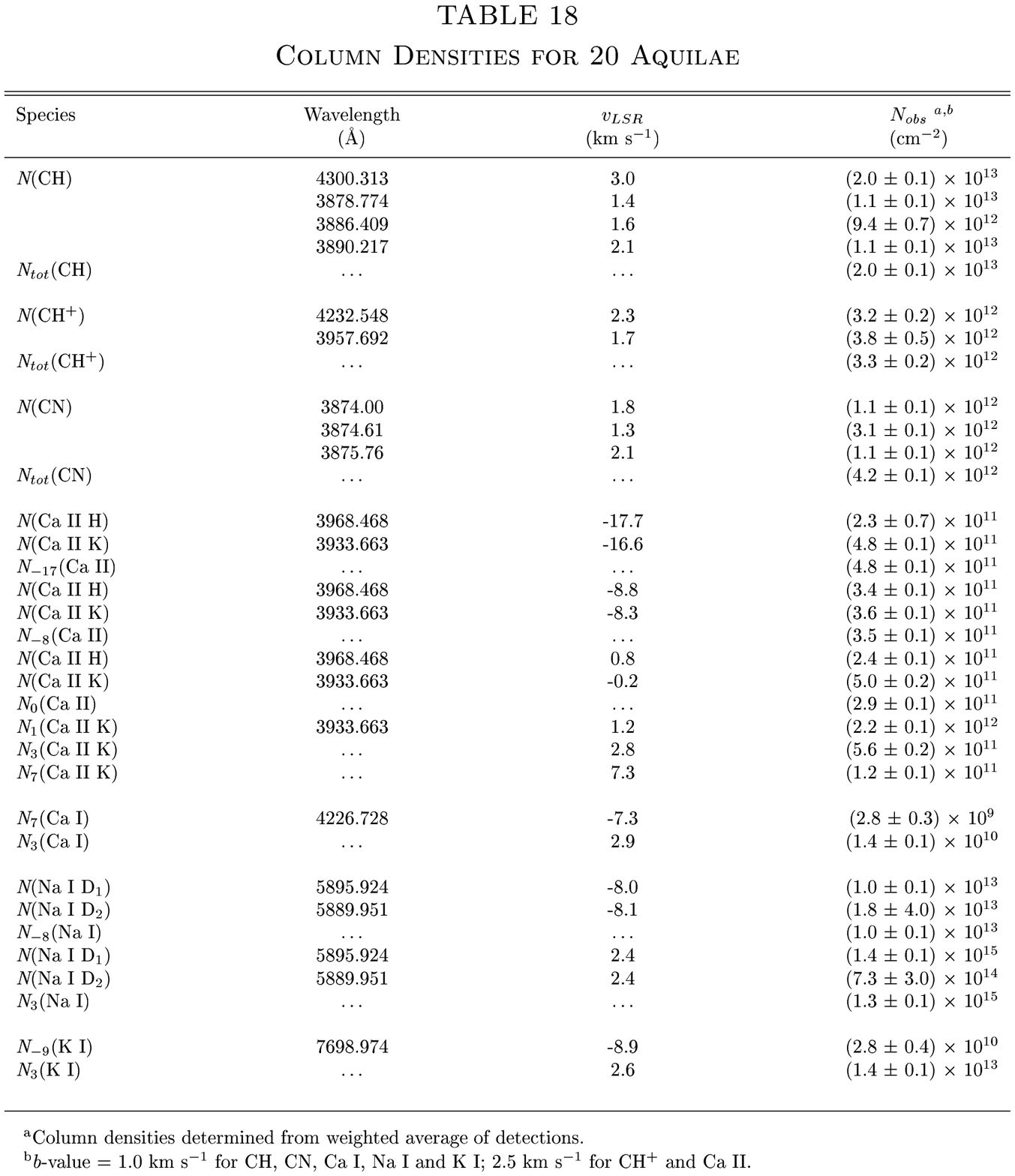}{4.0in}{0}{80}{80}{-220}{-165}
\end{center}
\end{figure}

\newpage
\clearpage

\setcounter{figure}{0}
\begin{figure}[p]
\begin{center}
\plotfiddle{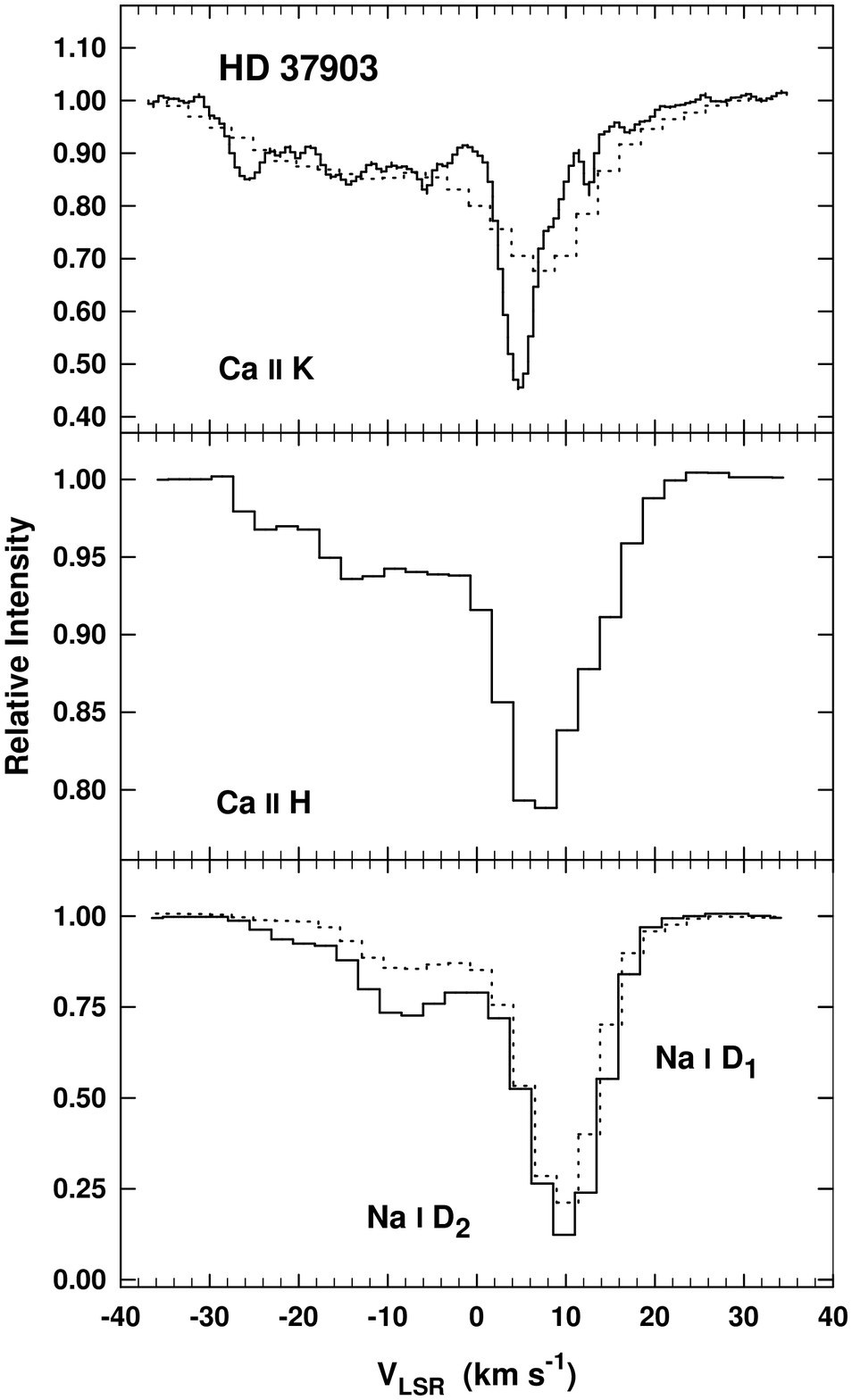}{4.0in}{0}{65}{65}{-180}{-165}
\vspace{2.0in} 
\caption{Representative interstellar atomic spectra toward HD~37903 are shown
here.  Top panel $-$ Ca II K, solid line is high resolution 2dcoud\'{e}
data; dashed line is moderate resolution 2dcoud\'{e} data.  In all
figures the flux is normalized to unity.  There is a nice correspondence between
the two different resolutions in velocity and line shape.  Center panel $-$
moderate resolution 2dcoud\'{e} Ca II H data.  Bottom panel $-$
moderate resolution 2dcoud\'{e} data; dashed line is Na D$_1$ and solid
line is Na D$_2$.  The slight velocity differences are attributed to an
error in wavelength solution.}
\end{center}
\end{figure}

\clearpage
\newpage

\setcounter{figure}{1}
\begin{figure}[p]
\begin{center}
\plotfiddle{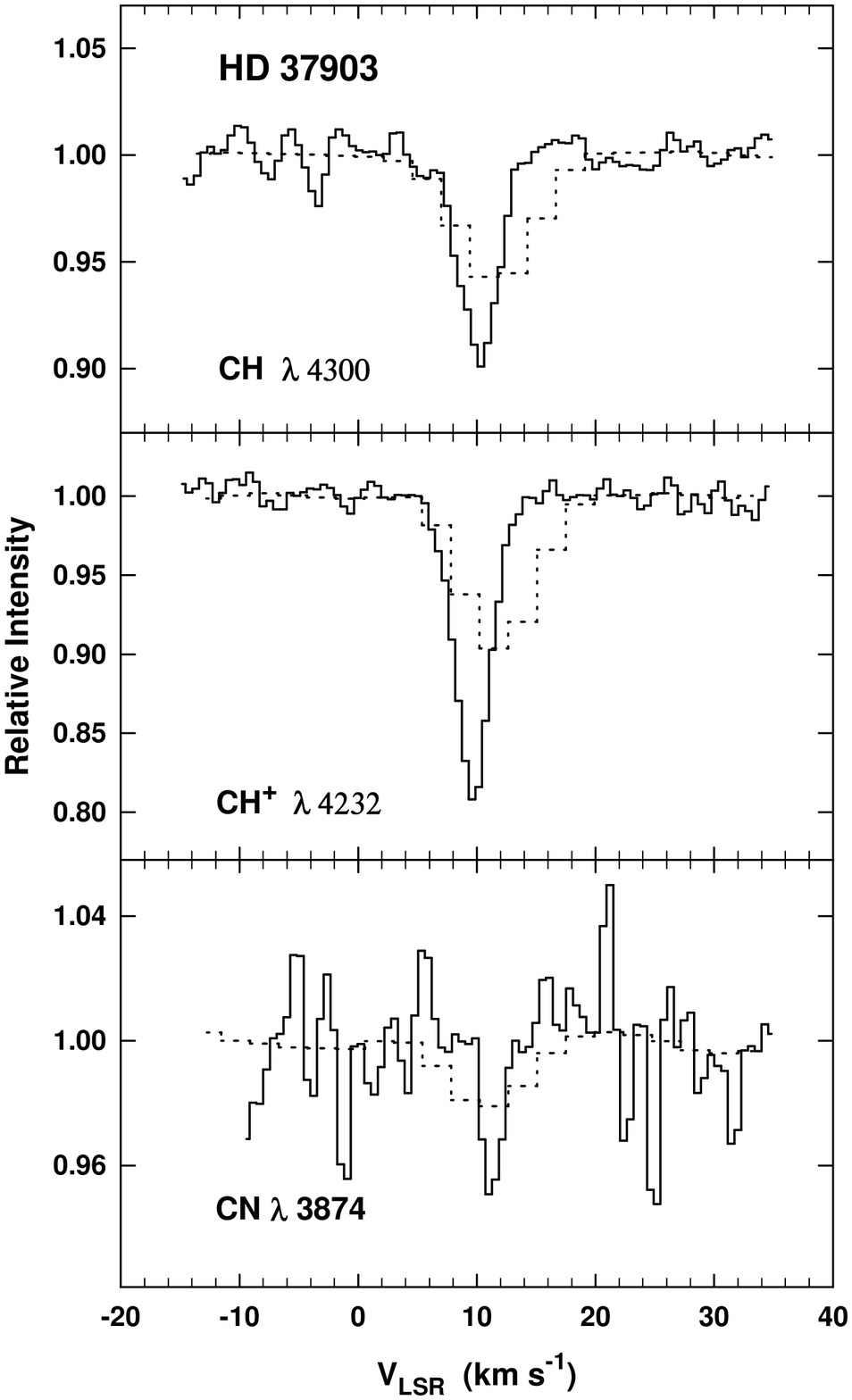}{4.0in}{0}{65}{65}{-180}{-165}
\vspace{2.0in} 
\caption{Representative interstellar molecular spectra toward HD~37903 are
shown.  Top panel $-$ CH, solid line is high resolution and dashed line is
moderate resolution 2dcoud\'{e} data.  Center panel $-$ CH$^+$, same as
top panel. Bottom panel $-$ CN, same as top panel.  There is a general agreement
in velocity and line shape; the slight velocity differences are attributed to an
error in wavelength solution.}
\end{center}
\end{figure}

\clearpage
\newpage

\setcounter{figure}{2}
\begin{figure}[p]
\begin{center}
\plotfiddle{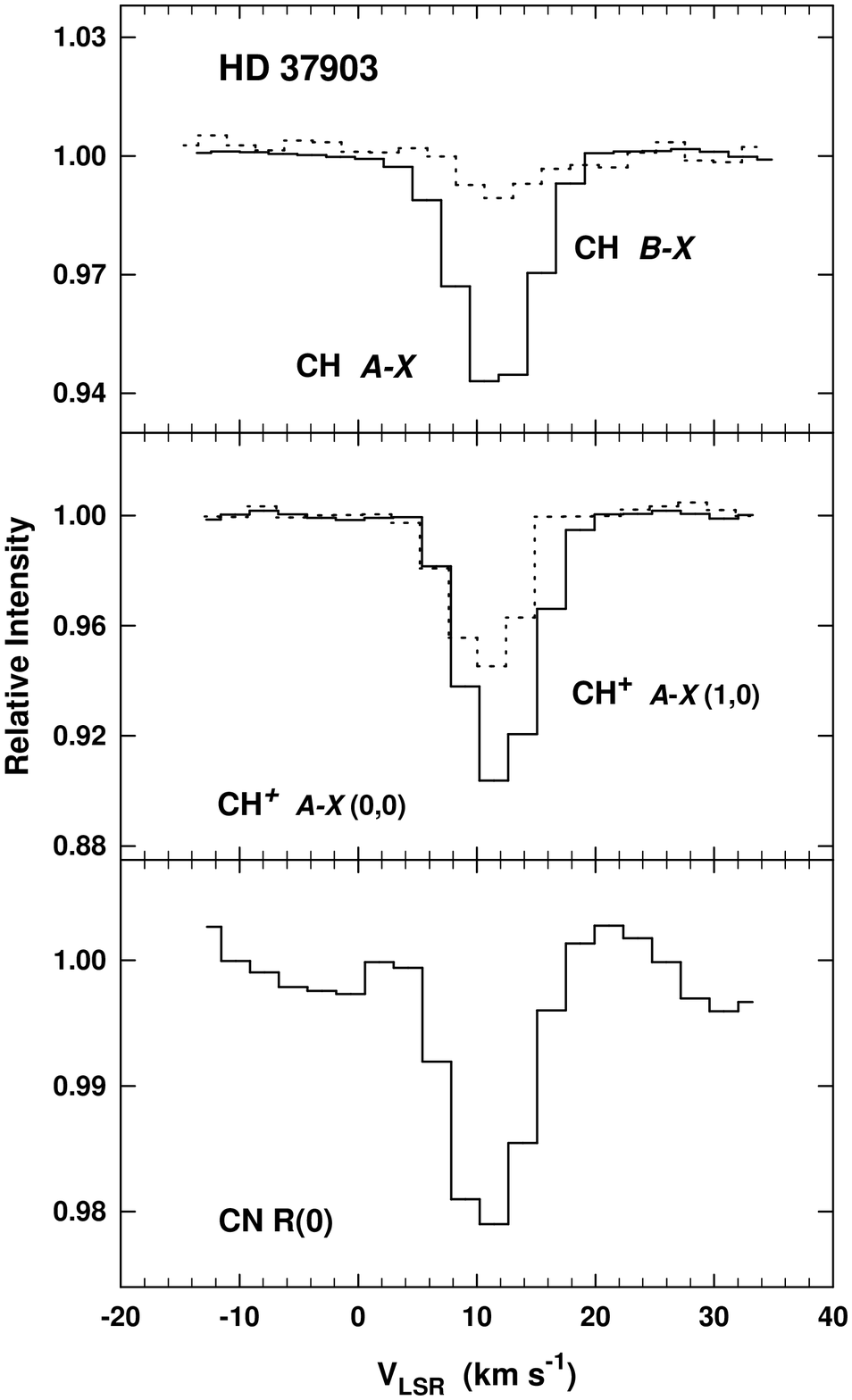}{4.0in}{0}{65}{65}{-180}{-165}
\vspace{2.0in} 
\caption{Moderate resolution 2dcoud\'{e} molecular spectra toward
HD~37903 are shown.  Top panel $-$ CH, solid line is $A-X$ transition and dashed
line is $B-X$ transition.  Center panel $-$ CH$^+$, solid line is $A-X$ (0$-$0)
transition and dashed line is $A-X$ (1$-$0) transition. Bottom panel $-$ CN R(0)
transition is shown.}
\end{center}
\end{figure}

\clearpage
\newpage

\setcounter{figure}{3}
\begin{figure}[p]
\begin{center}
\plotfiddle{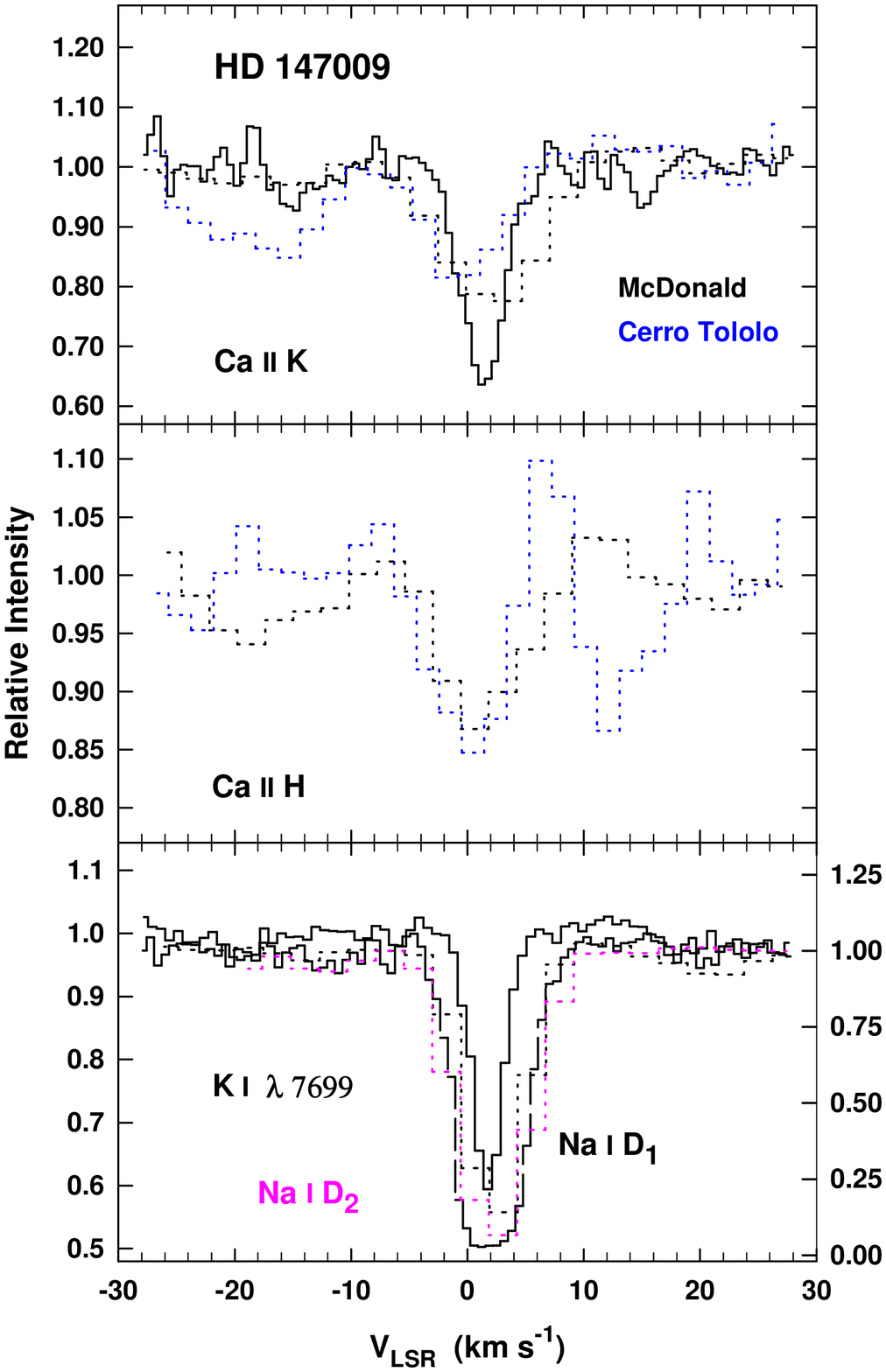}{4.0in}{0}{65}{65}{-180}{-165}
\vspace{2.0in} 
\caption{Representative interstellar atomic spectra toward HD~147009 are shown. 
Top panel $-$ Ca II K, dark solid line is high resolution 2dcoud\'{e}
data; dark dashed line is moderate resolution 2dcoud\'{e} data; and blue 
dashed line is moderate resolution CTIO data.  Center panel $-$ Ca II H, dark
dashed line is moderate resolution 2dcoud\'{e} data; and blue dashed line is
moderate resolution CTIO data.  Apparent minimum in the CTIO data is noise, not 
an additional component.  Bottom panel $-$ shows spectra of several atomic
species taken with different instruments at different resolutions.  There is
excellent agreement among all observations.  Solid line is K I and dark long
dashed line is Na D$_1$ high resolution 6 foot camera data; dark dashed
line is moderate resolution 2dcoud\'{e} Na D$_1$ data; and pink dashed
line is moderate resolution 2dcoud\'{e} Na D$_2$ data.  Use the left scale for 
K I and the right scale for Na I D lines.}
\end{center}
\end{figure}

\clearpage
\newpage

\setcounter{figure}{4}
\begin{figure}[p]
\begin{center}
\plotfiddle{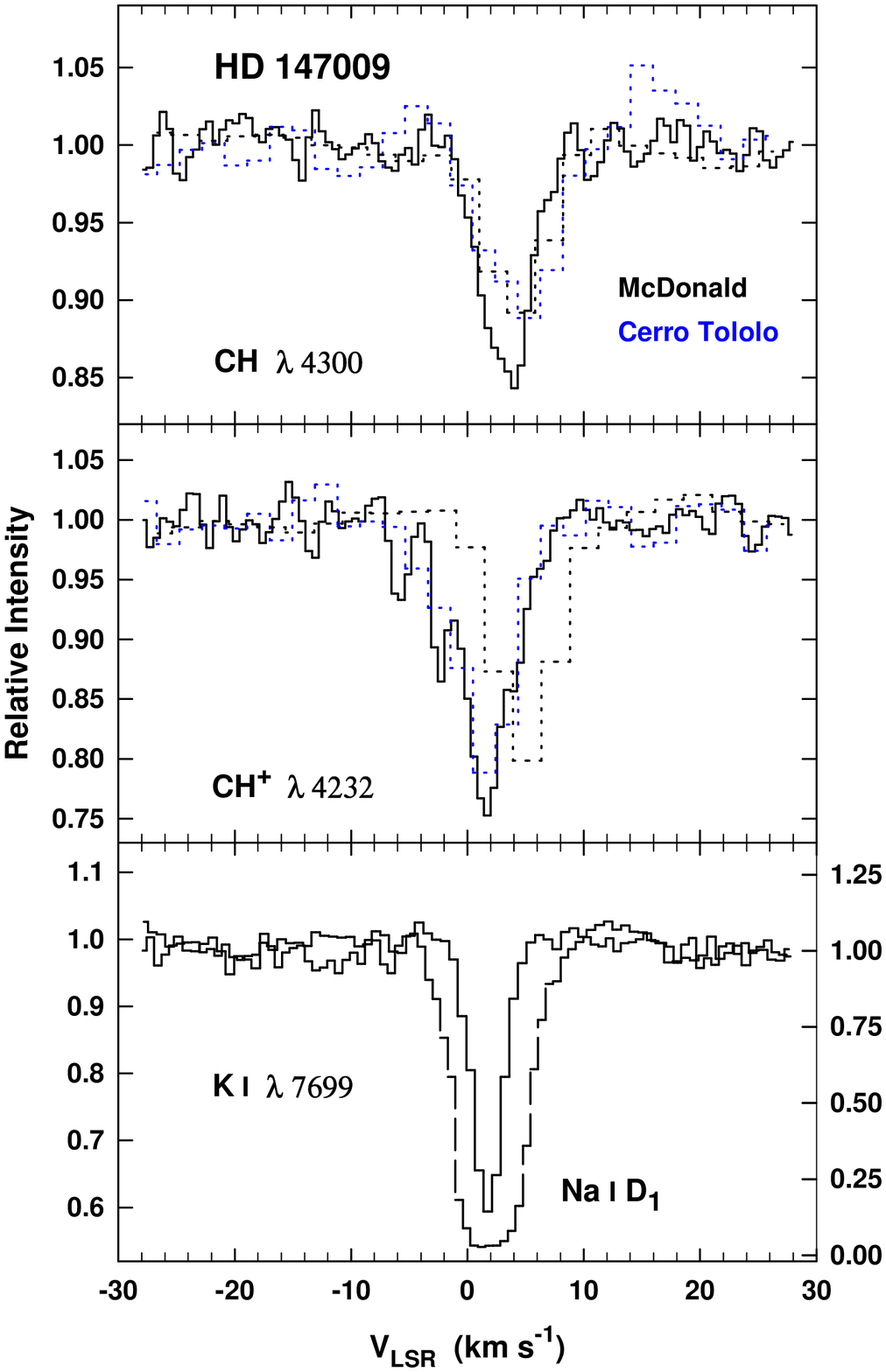}{4.0in}{0}{65}{65}{-180}{-165}
\vspace{2.0in}
\caption{Representative interstellar spectra toward HD~147009 are shown. 
Top panel $-$ CH, dark solid line is high resolution 2dcoud\'{e} data;
dark dashed line is moderate resolution 2dcoud\'{e} data; and blue 
dashed line is moderate resolution CTIO data.  Center panel $-$ CH$^+$,
solid line is high resolution 2dcoud\'{e} data; dark dashed line is
moderate resolution 2dcoud\'{e}; and blue dashed line is moderate
resolution CTIO data.  The slight velocity differences are attributed to an
error in wavelength solution.  Bottom panel $-$ shows spectra of K I, solid line
(left scale), and Na D$_1$, dashed line (right scale), taken with the high 
resolution 6 foot camera.} 
\end{center}
\end{figure}

\clearpage
\newpage

\setcounter{figure}{5}
\begin{figure}[p]
\begin{center}
\plotfiddle{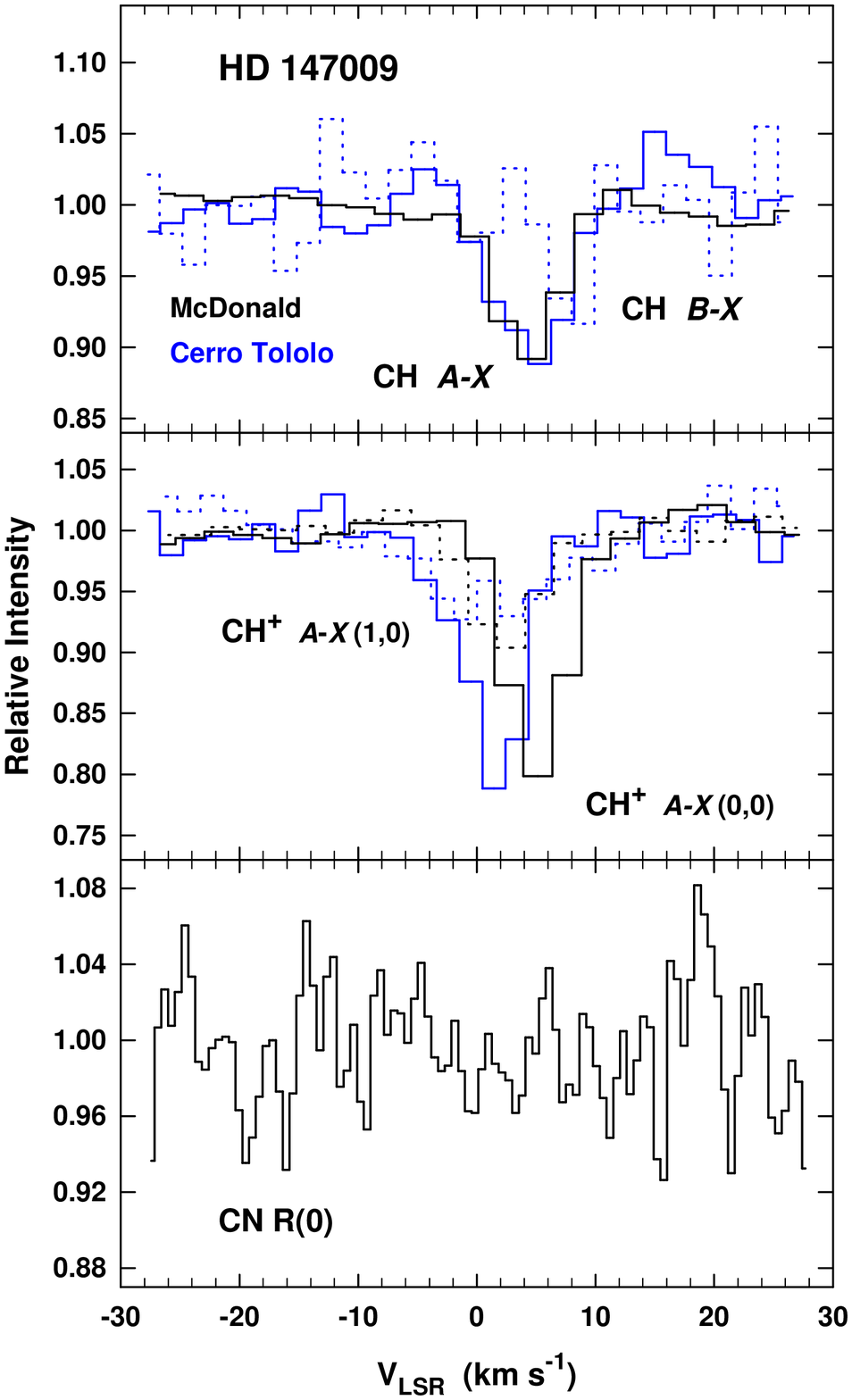}{4.0in}{0}{65}{65}{-180}{-165}
\vspace{2.0in} 
\caption{Interstellar molecular spectra toward HD~147009 are
shown; dark lines are 2dcoud\'{e} data and blue lines are CTIO data.  Top
panel $-$ CH, the solid lines are the $A-X$ transition and the dashed line is
the $B-X$ transition.  Center panel $-$ CH$^+$, the solid lines are the $A-X$
(0$-$0) transition and the dashed lines are the $A-X$ (1$-$0) transition.  The
slight velocity differences are attributed to an error in wavelength solution in
these moderate resolution data.  Bottom panel $-$ shows the high resolution 
2dcoud\'{e} 3874 \AA\ region of the spectrum where CN resides; no detection is
evident.}
\end{center}
\end{figure}

\clearpage
\newpage

\setcounter{figure}{6}
\begin{figure}[p]
\begin{center}
\plotfiddle{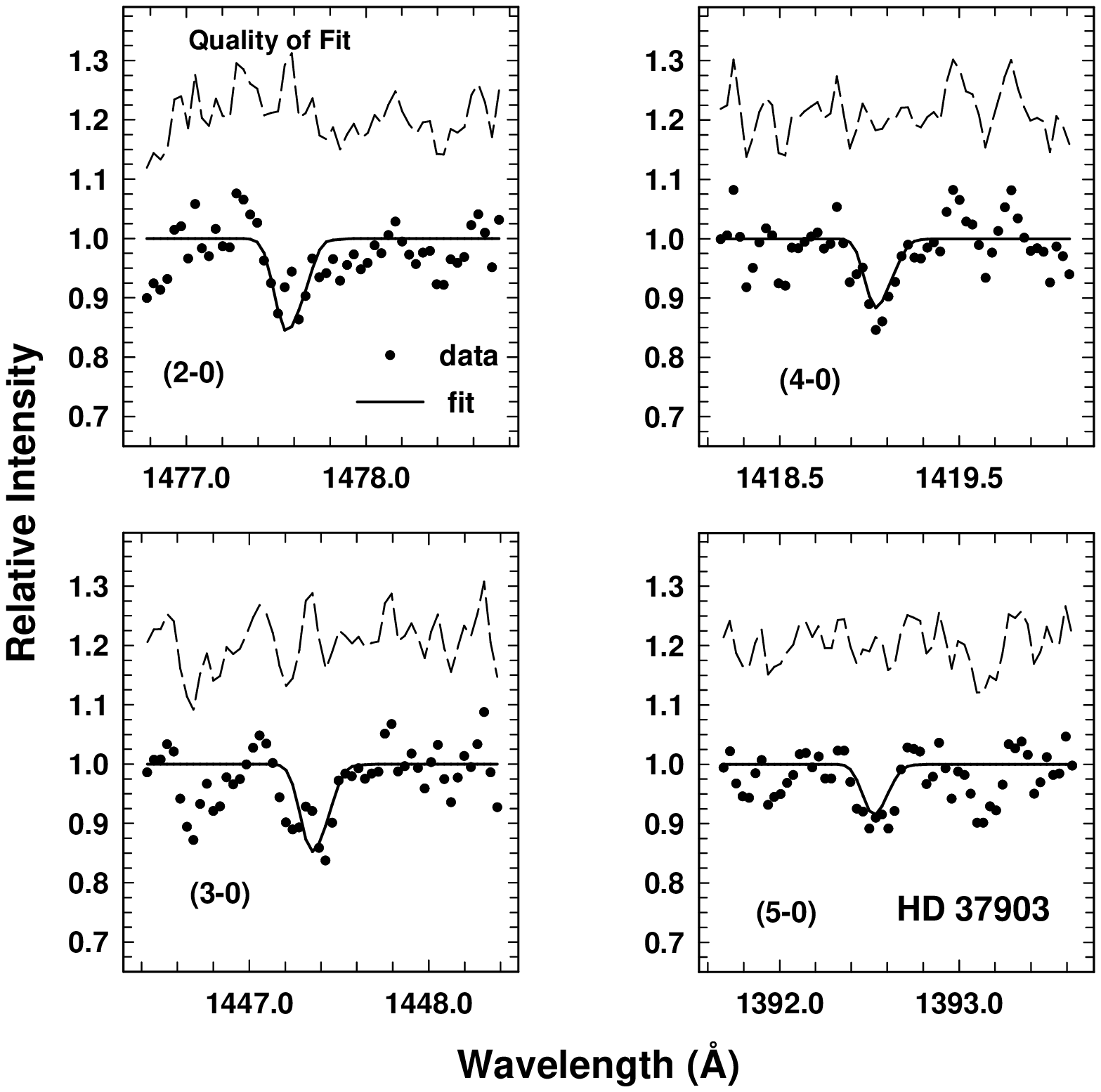}{2.0in}{0}{70}{70}{-280}{-60}
\vspace{0.1in} 
\caption{a.) Representative NEWSIPS $IUE$ $^{12}$CO spectra toward HD~37903 are 
shown.  The data are represented by the filled circles.  Our best fit to the
data (solid line) and the data$-$fit (dashed line, offset to 1.22) are also 
shown.}
\end{center}
\end{figure}

\clearpage
\newpage

\setcounter{figure}{6}
\begin{figure}[p]
\begin{center}
\plotfiddle{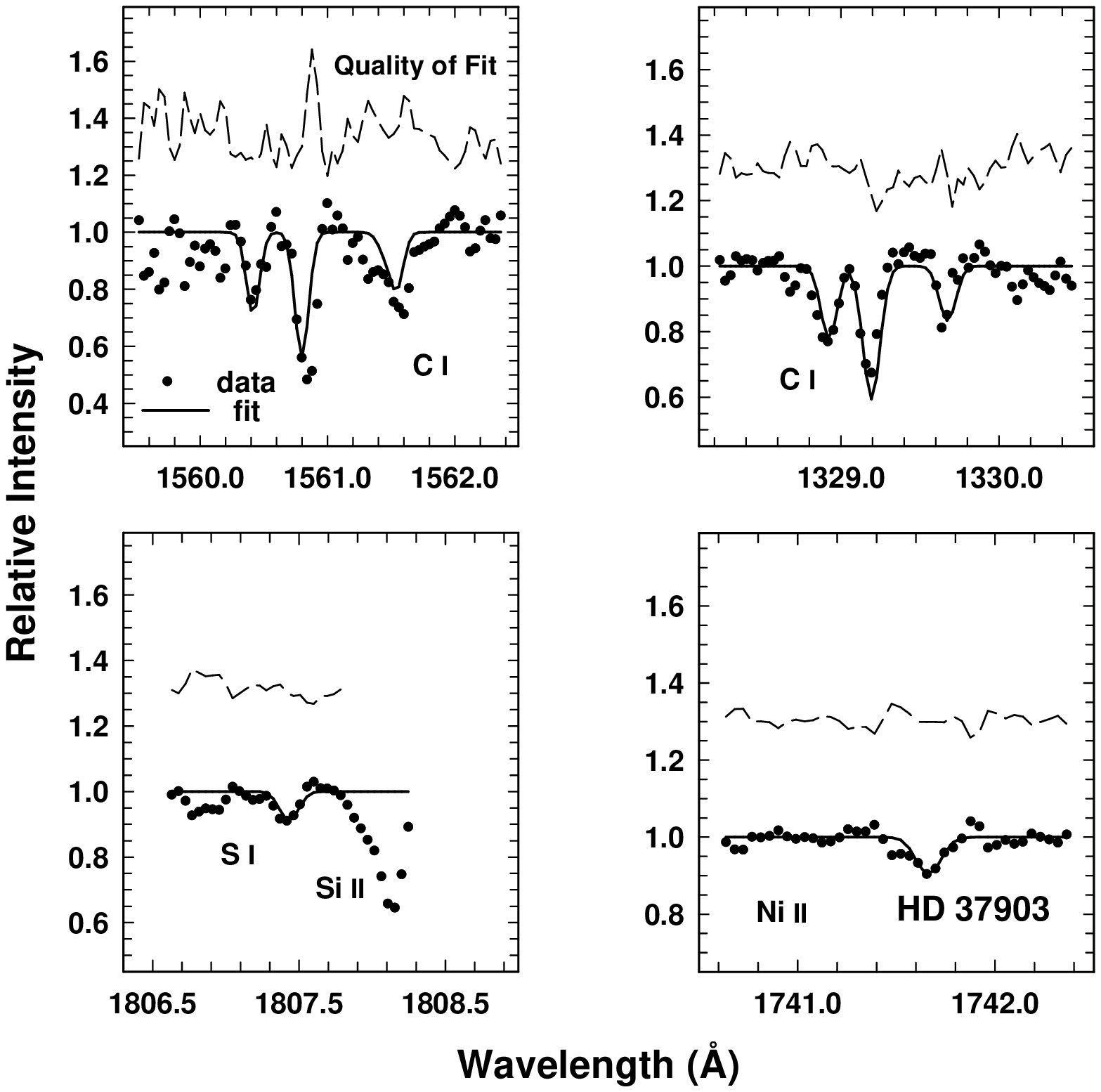}{2.0in}{0}{70}{70}{-280}{-60}
\vspace{0.1in} 
\caption{b.) Same as figure 7a, but spectra of C I, S I and Ni II are shown. 
The Si II line was not fit.  Data$-$fit offset to 1.30.}
\end{center}
\end{figure}

\clearpage
\newpage

\setcounter{figure}{7}
\begin{figure}[p]
\begin{center}
\plotfiddle{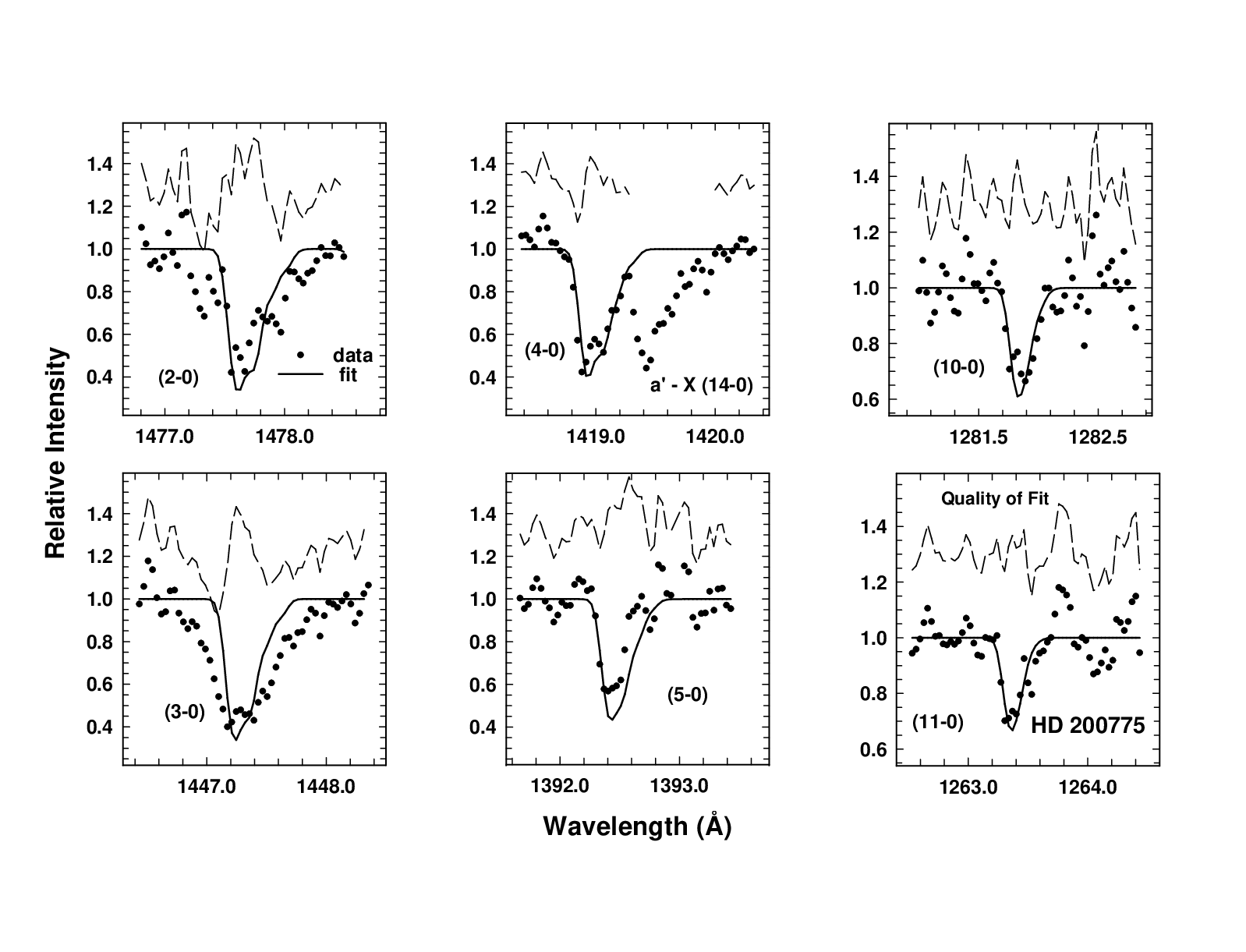}{3.5in}{90}{70}{70}{220}{-240}
\vspace{3.0in} 
\caption{a.) Representative NEWSIPS $IUE$ $^{12}$CO spectra toward HD~200775 
are shown.  The data are represented by the filled circles.  Our best fit to the
data (solid line) and the data$-$fit (dashed line, offset to 1.30) are also 
shown.  The $a^{\prime}-X$ (14$-$0) intersystem band was not fit.}
\end{center}
\end{figure}

\clearpage
\newpage

\setcounter{figure}{7}
\begin{figure}[p]
\begin{center}
\plotfiddle{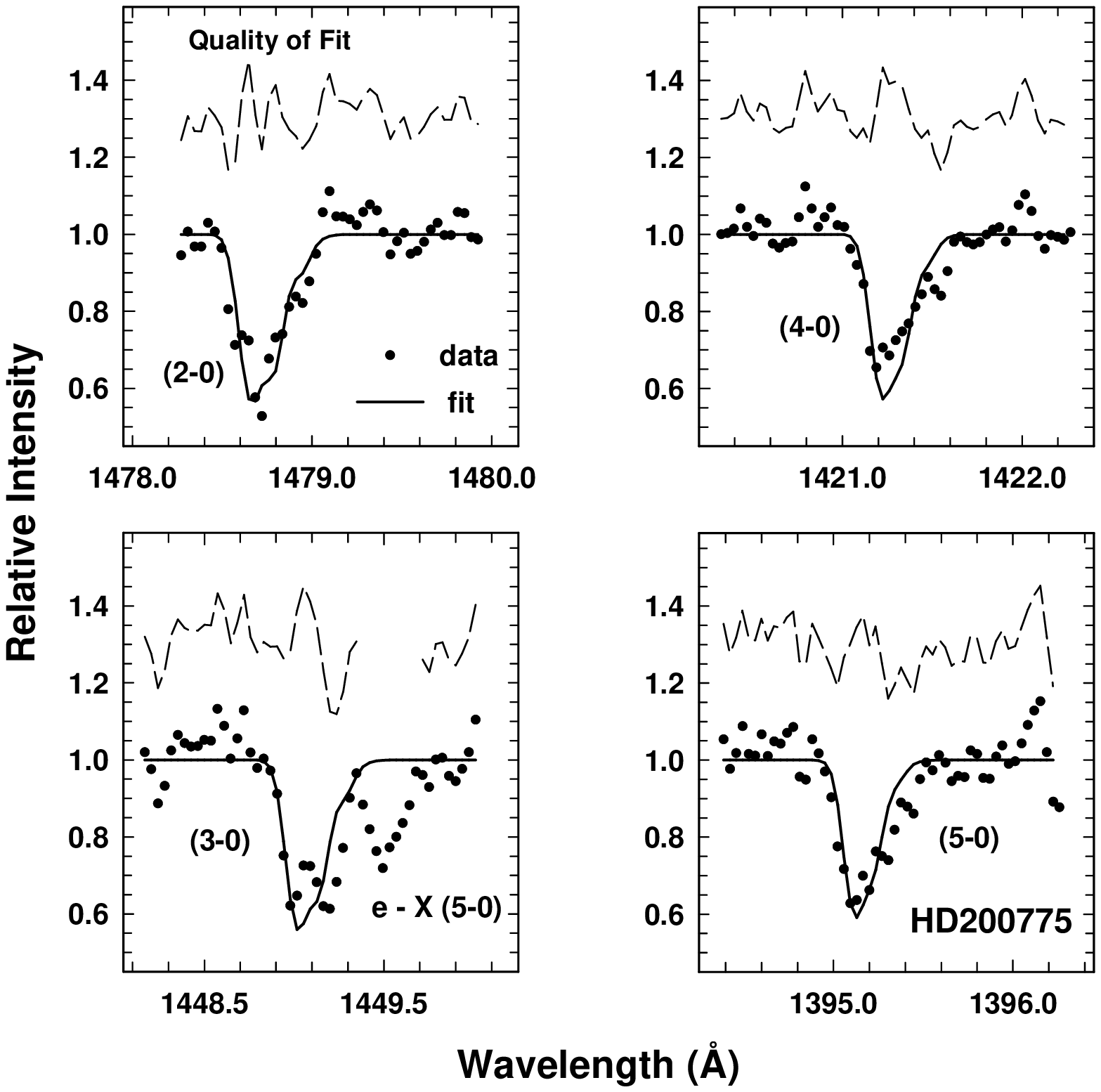}{2.0in}{0}{70}{70}{-280}{-50}
\vspace{0.1in} 
\caption{b.) Same as figure 8a, but spectra of $^{13}$CO are shown.  The $e-X$
(5$-$0) band of $^{12}$CO is clearly seen in the spectrum as well.}
\end{center}
\end{figure}

\clearpage
\newpage

\setcounter{figure}{7}
\begin{figure}[p]
\begin{center}
\plotfiddle{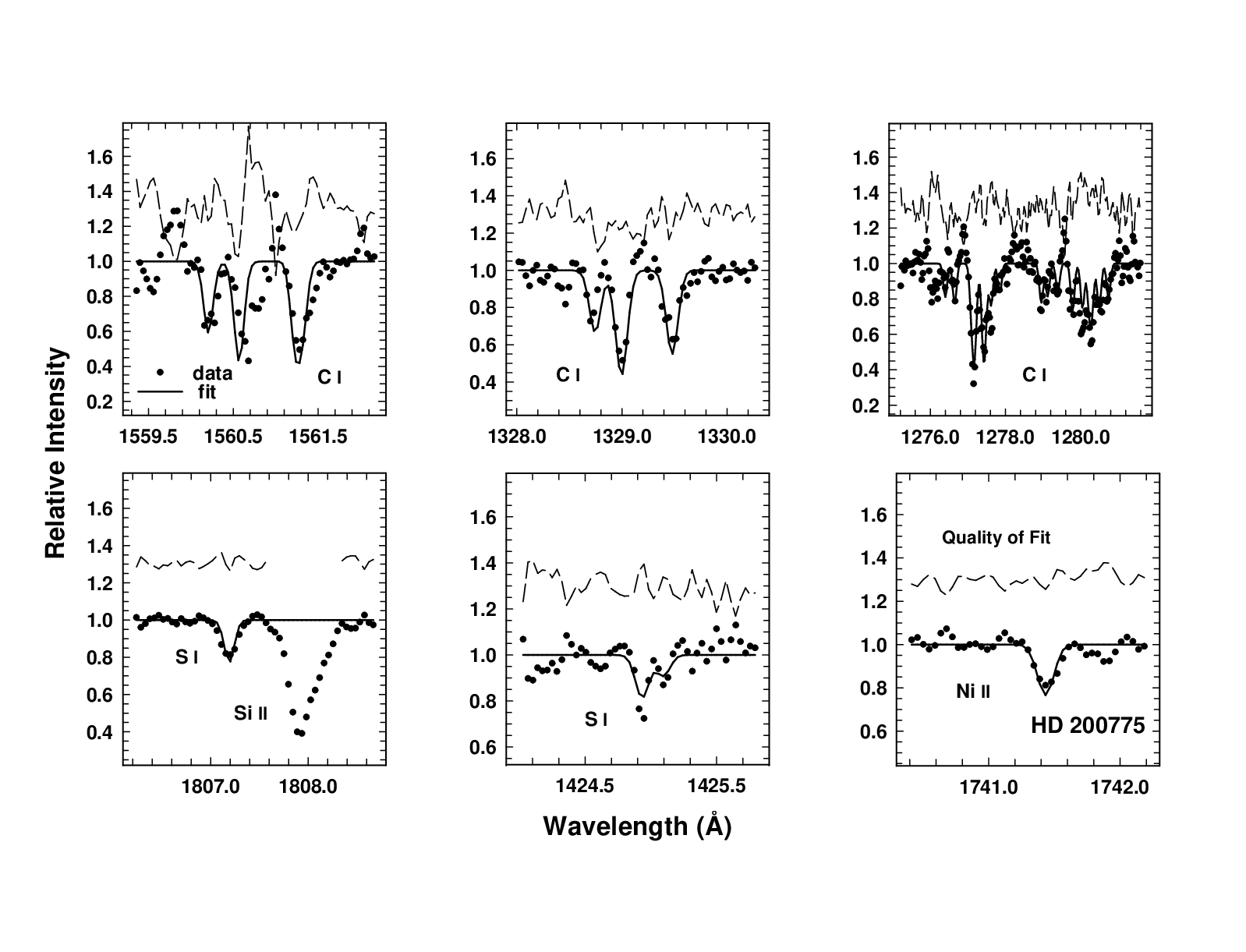}{3.5in}{90}{70}{70}{220}{-200}
\vspace{2.5in} 
\caption{c.) Same as figure 8a, but spectra of C I, S I and Ni II are shown. 
The Si II line was not fit.}
\end{center}
\end{figure}

\clearpage
\newpage

\setcounter{figure}{8}
\begin{figure}[p]
\begin{center}
\plotfiddle{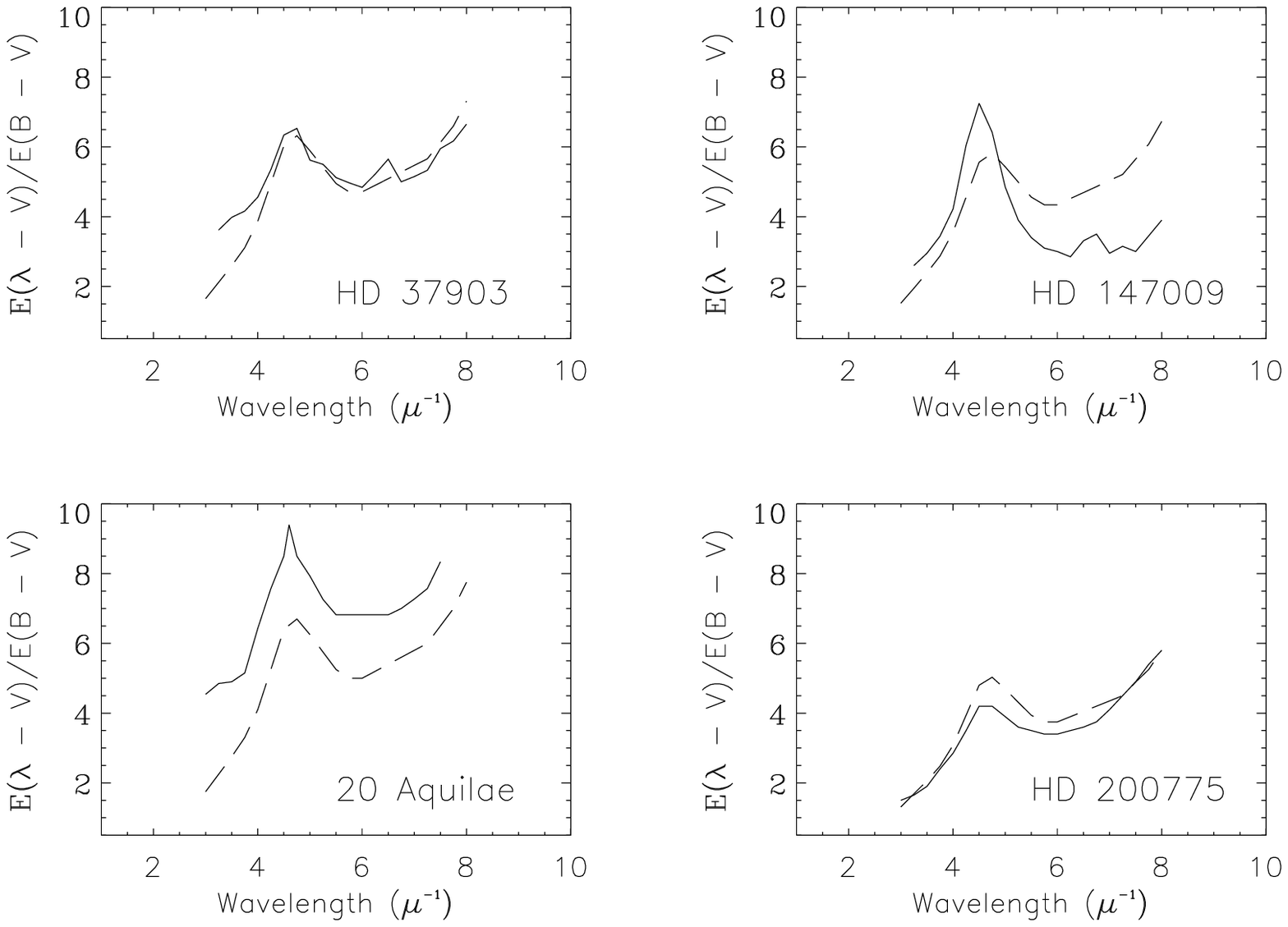}{2.5in}{0}{90}{90}{-230}{-10}
\vspace{0.1in} 
\caption{Shown are the derived extinction curves for the reddened lines
of sights.  The dashed curve is the average interstellar extinction curve
represented by $\zeta$~Per scaled to the $E$($B$ $-$ $V$) of the reddened star. 
Extinction curves for $\zeta$~Per, 20~Aquilae, and HD~200775 were obtained from
Savage \& Mathis 1979, Savage et al. 1985, and Walker et al. 1980,
respectively.}
\end{center}
\end{figure}

\clearpage
\newpage

\setcounter{figure}{9}
\begin{figure}[p]
\begin{center}
\plotfiddle{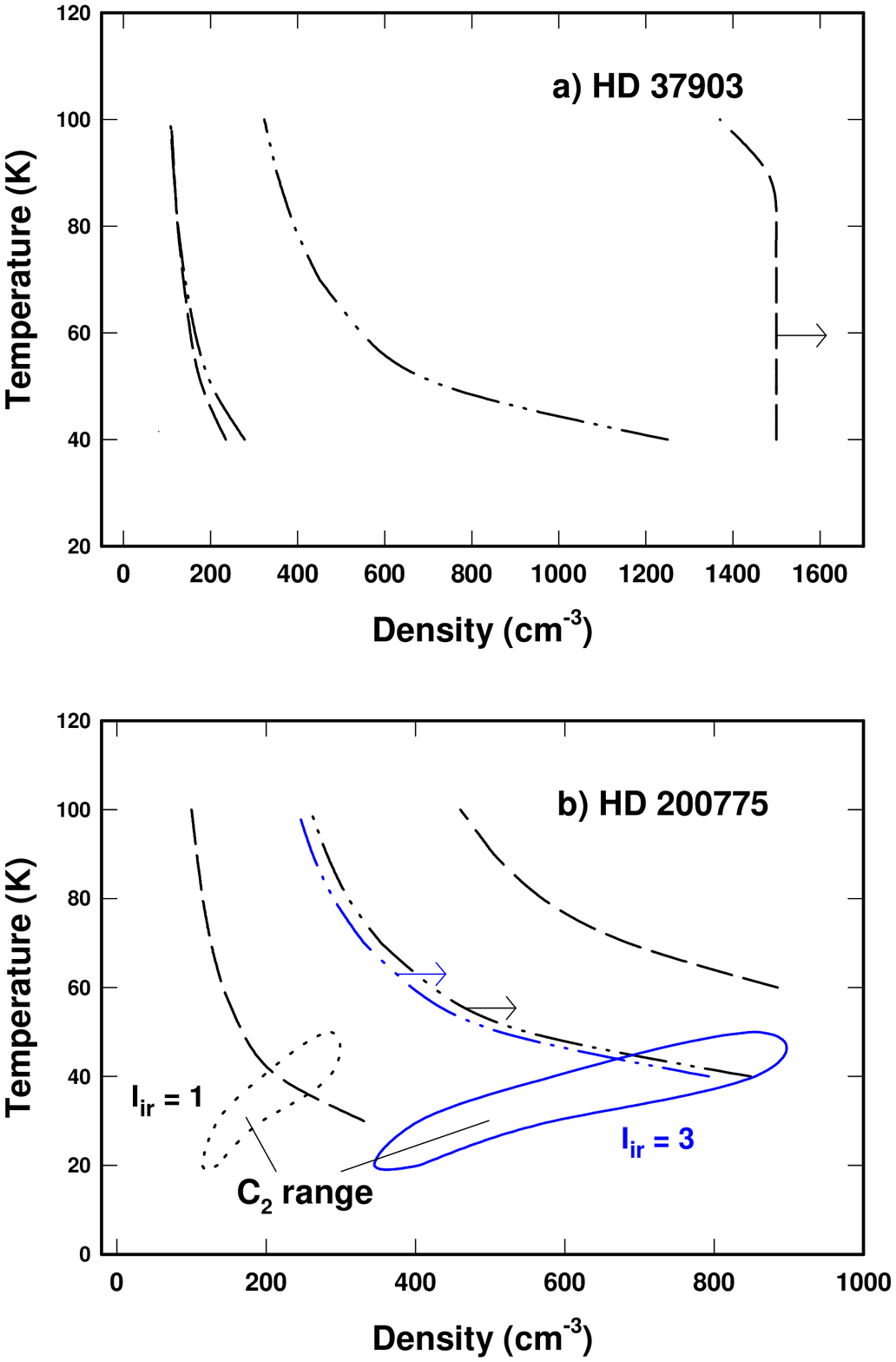}{3.5in}{0}{65}{65}{-180}{-200}
\vspace{2.0in} 
\caption{Results of C I excitation for HD~37903 and HD~200775 are shown. 
Density is for collisional partners, see text.  The dashed lines $-$
$N$(J=1)/$N$(J=0) ratio and the dashed-dotted lines $-$ $N$(J=2)/$N$(J=0) ratio
based on $\pm$ 1 $\sigma$ uncertainty in column densities.  For HD~37903 (top
panel), the arrow indicates upper range for $N$(J=1)/$N$(J=0) ratio is a lower
limit.  Results for $I_{uv}$ = 3 are indistinguishable for $I_{uv}$ =
1.  For HD~200775 (lower panel), upper limit for $N$(J=2)/$N$(J=0) ratio yields
much larger density.  Blue scale shows results for 3-fold enhancement in
radiation field where distinguishable.  Result from C$_2$ excitation is also
shown.}
\end{center}
\end{figure}

\clearpage
\newpage

\setcounter{figure}{10}
\begin{figure}[p]
\begin{center}
\plotfiddle{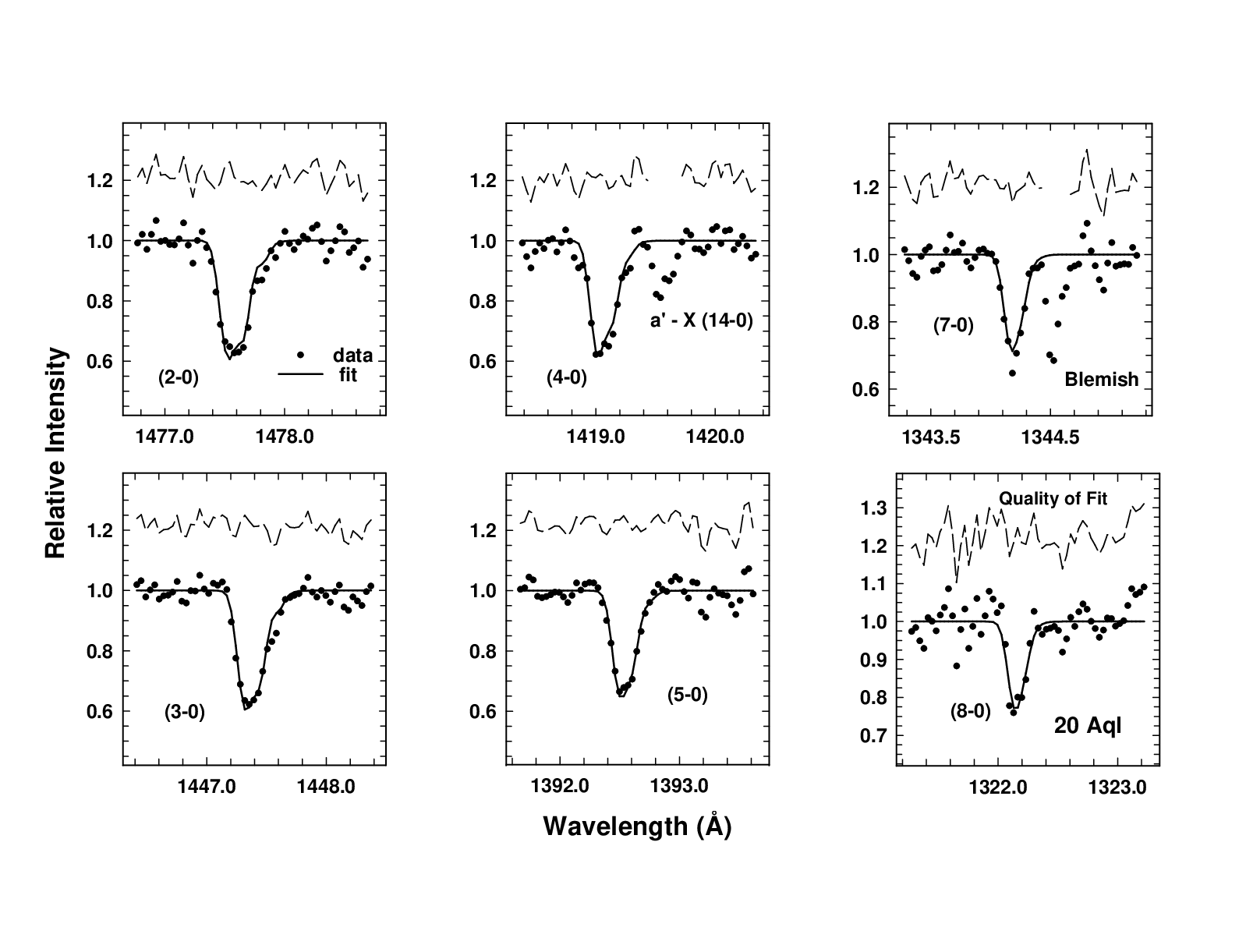}{3.5in}{90}{60}{60}{180}{-160}
\vspace{2.0in} 
\caption{a.)  Representative NEWSIPS $IUE$ $^{12}$CO spectra toward 20~Aquilae
are shown.  The data are represented by the filled circles.  Our best fit to the
data (solid line) and the data$-$fit (dashed line, offset to 1.22) are also
displayed.  The $a^{\prime}-X$ (14$-$0) intersystem band was not fit.}
\end{center}
\end{figure}

\clearpage
\newpage

\setcounter{figure}{10}
\begin{figure}[p]
\begin{center}
\plotfiddle{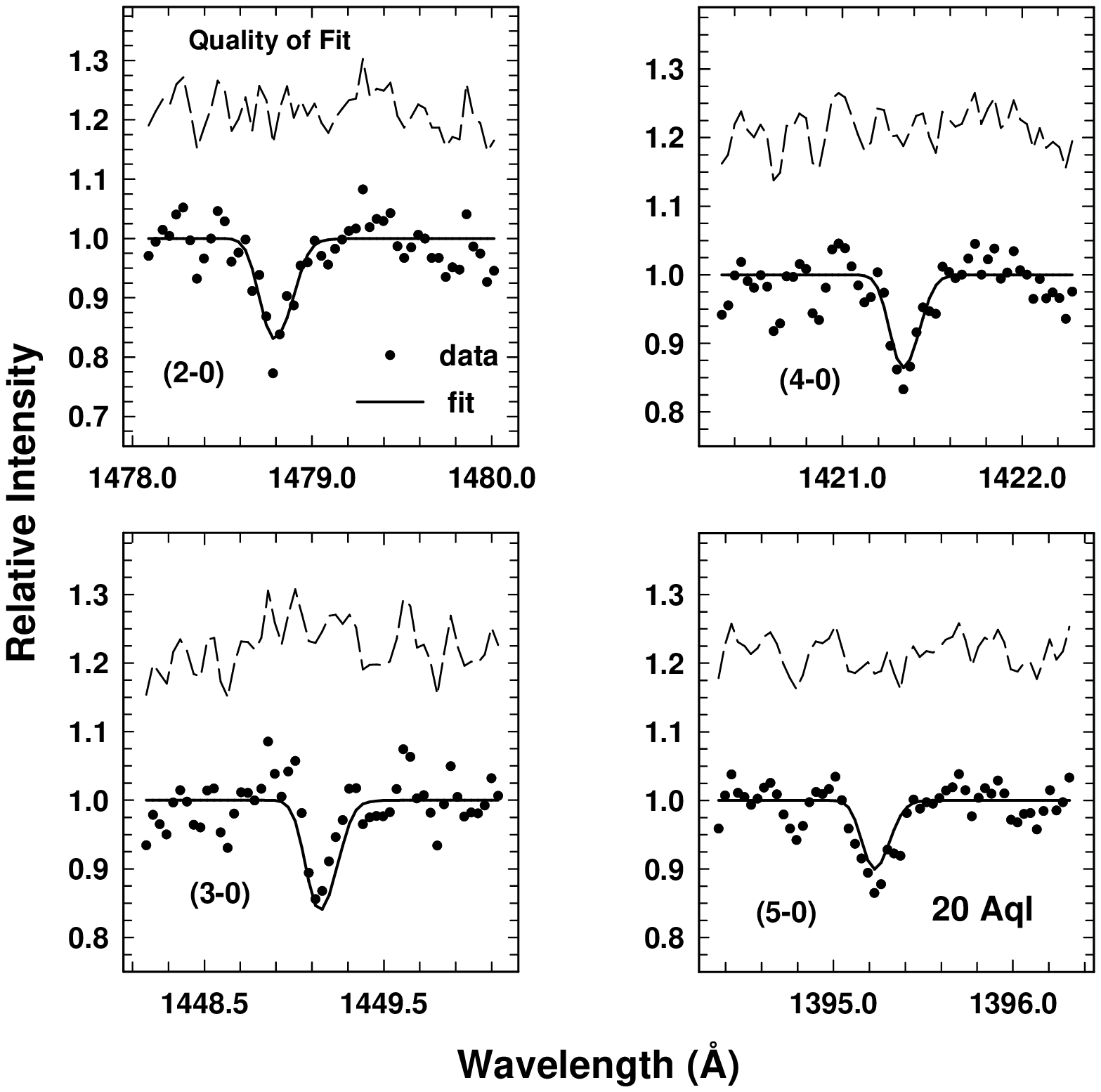}{2.0in}{0}{70}{70}{-280}{-50}
\vspace{0.1in} 
\caption{b.) Same as Figure 11a, but spectra of $^{13}$CO are shown.}
\end{center}
\end{figure}

\clearpage
\newpage

\setcounter{figure}{10}
\begin{figure}[p]
\begin{center}
\plotfiddle{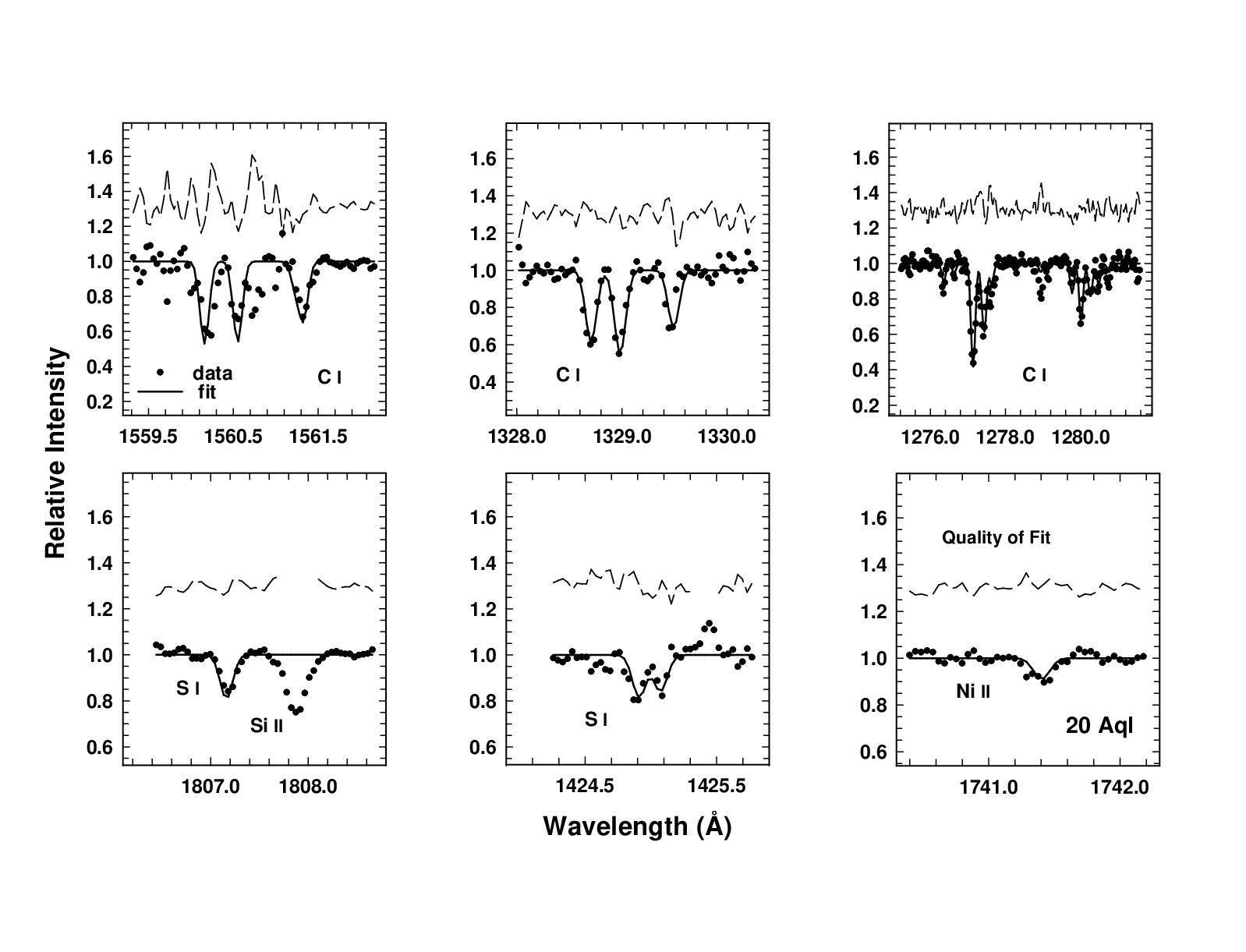}{3.5in}{90}{60}{60}{180}{-160}
\vspace{2.0in} 
\caption{c.) Same as Figure 11a, but spectra of C I, S I and Ni II are shown.
The Si II line at $\lambda$1808 was not fit.  Data$-$fit offset to 1.30.}
\end{center}
\end{figure}

\clearpage
\newpage

\setcounter{figure}{11}
\begin{figure}[p]
\begin{center}
\plotfiddle{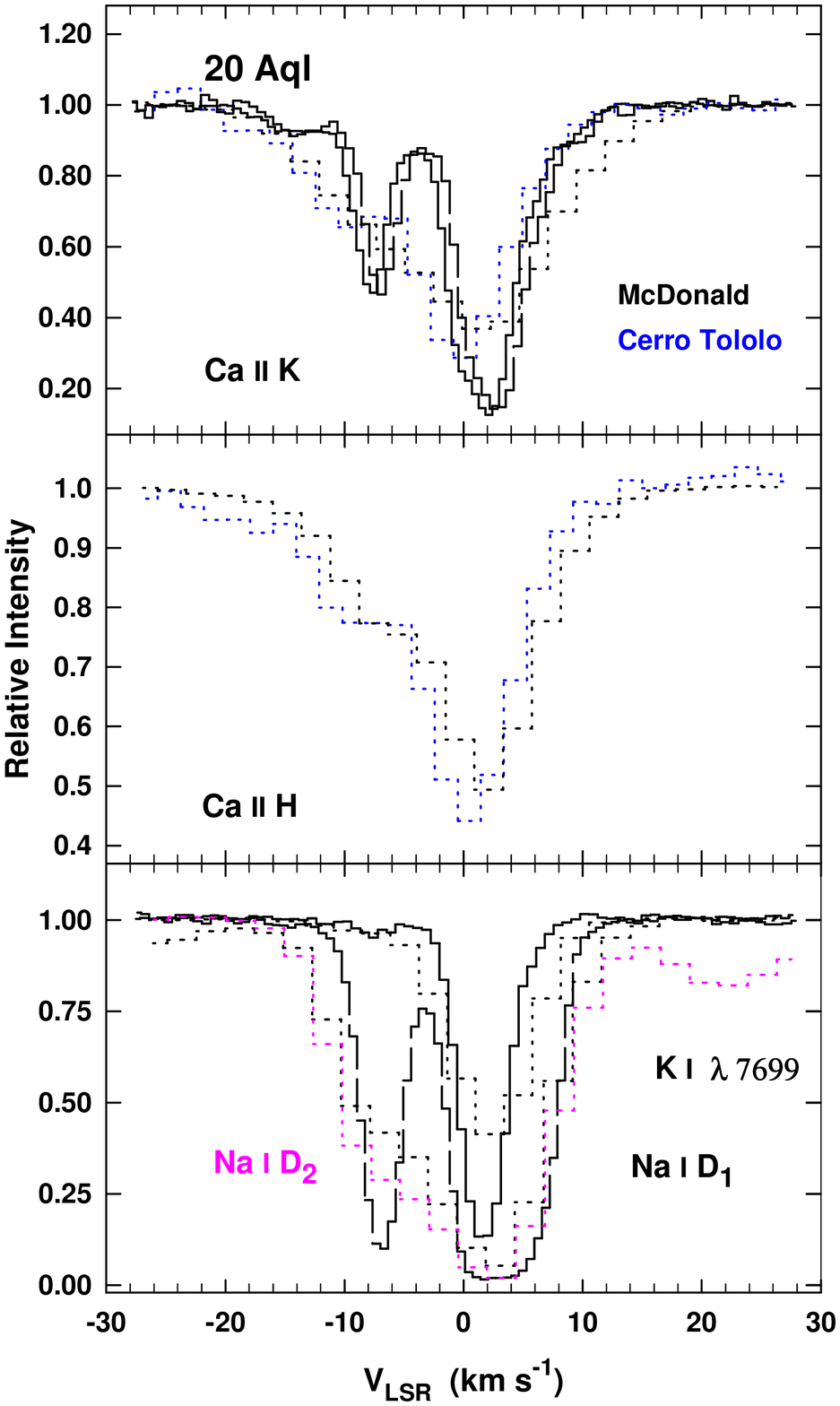}{4.0in}{0}{60}{60}{-180}{-120}
\vspace{1.2in} 
\caption{Representative atomic spectra toward 20~Aquilae are shown.  Top panel
$-$ Ca~II~K, dark solid line is high resolution 6 foot camera data;
dark long-dashed line is high resolution 2dcoud\'{e} data; dark dashed
line is moderate resolution 2dcoud\'{e} data; and the blue dashed line is
moderate resolution CTIO data.  Center panel $-$ Ca II H, dark dashed line is
moderate resolution 2dcoud\'{e}; blue dashed line is moderate
resolution CTIO data.  Bottom panel $-$ shows spectra of several atomic species
taken with different instruments at different resolutions.  There is excellent
agreement among all observations.  Solid line is K I and dark long-dashed line
is Na D$_1$ high resolution 6 foot camera data; saturated dark dashed line is
moderate resolution 2dcoud\'{e} Na D$_1$ data while weaker dark dashed line is K~I; and pink dashed line is moderate resolution 2dcoud\'{e} Na D$_2$ data.}
\end{center}
\end{figure}

\clearpage
\newpage

\setcounter{figure}{12}
\begin{figure}[p]
\begin{center}
\plotfiddle{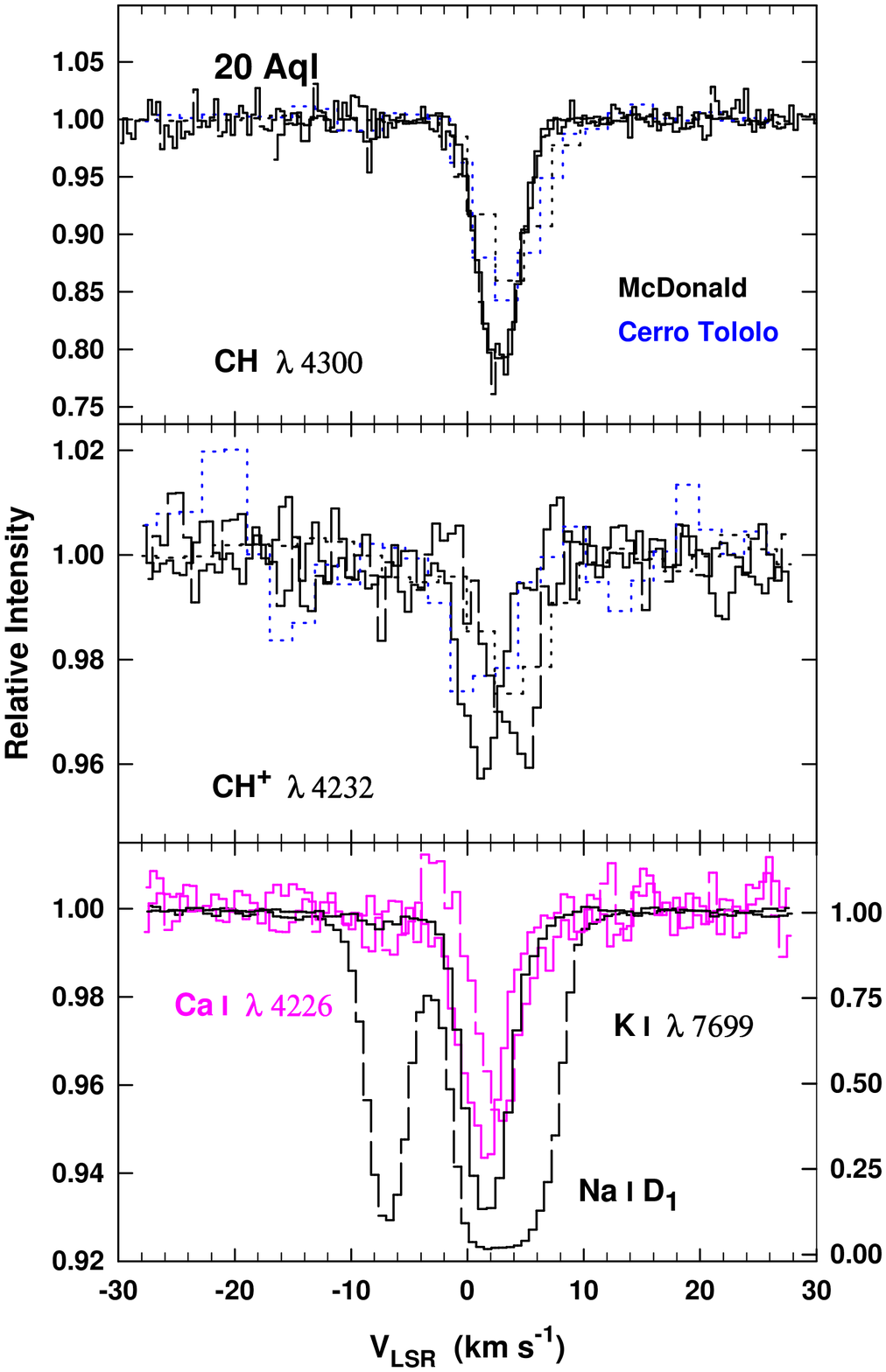}{4.0in}{0}{55}{55}{-180}{-85}
\vspace{0.8in} 
\caption{Representative spectra toward 20~Aquilae are shown. Top panel $-$ CH,
dark solid line is high resolution 2dcoud\'{e} data; dark long-dashed
line shows the high resolution 6 foot camera data; dark dashed line is
moderate resolution 2dcoud\'{e} data; and blue dashed line is moderate
resolution CTIO data.  Center panel $-$ CH$^+$, solid line is high resolution
2dcoud\'{e} data; dark long-dashed line shows the high resolution
6 foot camera data; dark dashed line is moderate resolution 2dcoud\'{e} data;
and blue dashed line is moderate resolution CTIO data.  Velocity offset is
due to error in wavelength solution.  Bottom panel $-$ shows spectra of several
atomic species taken with different instruments at different resolutions.  There
is excellent agreement among all observations.  Solid pink line is Ca I high
resolution 2dcoud\'{e} data; pink and dark long dashed line shows Ca I and Na
D$_1$ high resolution 6 foot camera data; and dark solid line is K I high
resolution 6 foot camera data.  Use the right scale for Na I and K I and the 
left scale for Ca I.}
\end{center}
\end{figure}

\clearpage
\newpage

\setcounter{figure}{13}
\begin{figure}[p]
\begin{center}
\plotfiddle{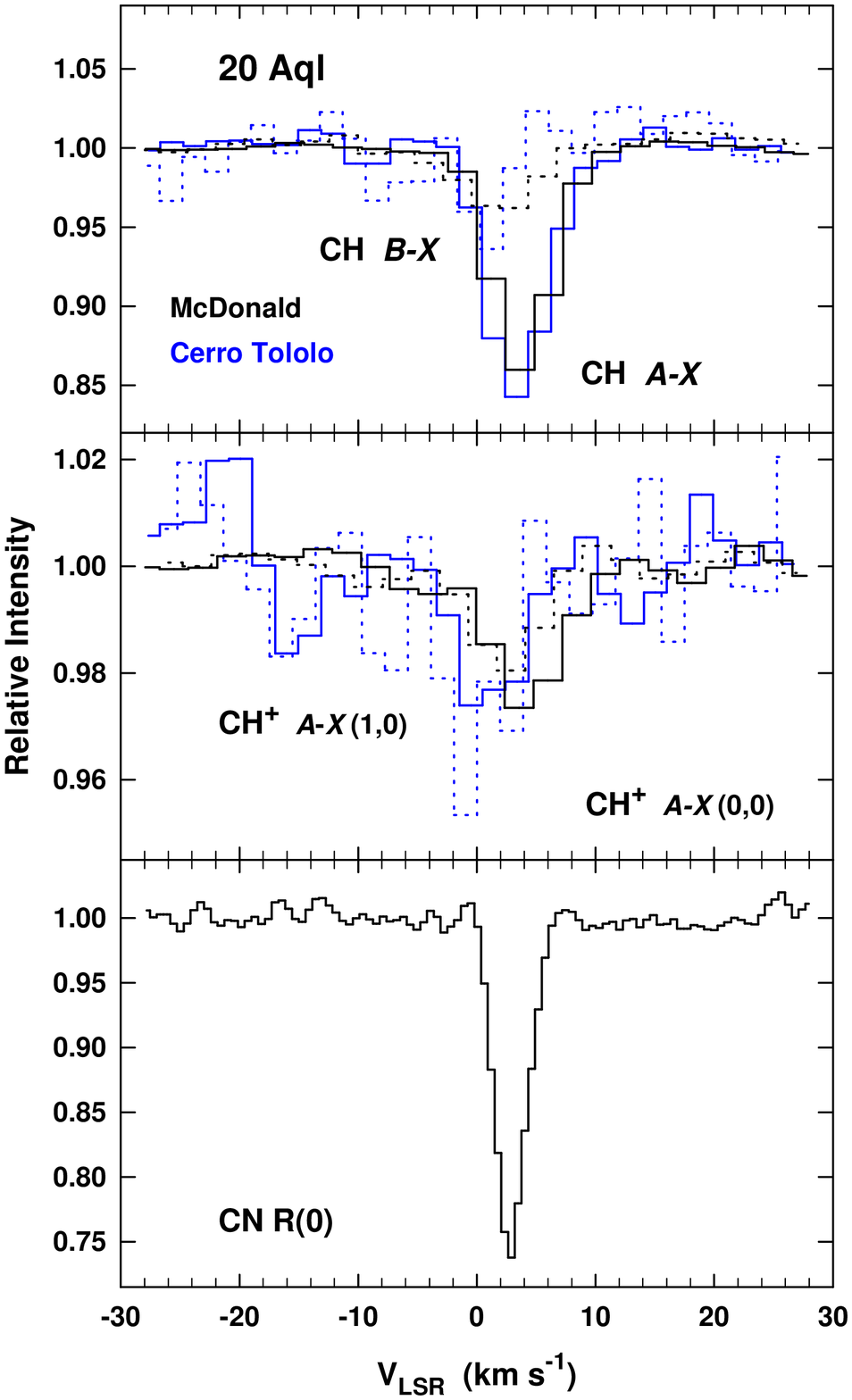}{4.0in}{0}{65}{65}{-180}{-165}
\vspace{2.0in} 
\caption{Moderate resolution molecular spectra toward 20~Aquilae are shown; dark
lines are 2dcou\'{e} data and blue lines are CTIO data.  Top panel $-$ CH, solid
lines are $A-X$ transition and dashed lines are $B-X$ transition.  Center panel
$-$ CH$^+$, solid lines are $A-X$ (0$-$0) transition and dashed lines are $A-X$
(1$-$0) transition. Bottom panel $-$ CN R(0) high resolution 2dcoud\'{e} data 
is shown.}
\end{center}
\end{figure}

\end{document}